\documentclass[12pt,reqno,a4paper]{article}
\usepackage{amsmath,amsfonts,amssymb}
\usepackage{pstricks}
\usepackage[usenames]{color}
\usepackage{pstcol}
\usepackage{hyperref}
\hypersetup{backref,
colorlinks=true,
citecolor=darkblue
}
\numberwithin{equation}{section}
\definecolor{darkblue}{rgb}{0,0,.8}
\definecolor{lightblue}{rgb}{.65,.95,1}
\definecolor{lightlightblue}{rgb}{.85,1,1}
\definecolor{rred}{rgb}{1,0,0}
\definecolor{purple}{rgb}{0.62,0.12,0.94}

\newcommand{\nc}{\newcommand}
\nc{\rnc}{\renewcommand}
\rnc{\title}[1]{{\huge\bf\mbox{}\\\medskip#1\bigskip\medskip\\}}
\rnc{\author}[1]{{\large #1\smallskip\\}}
\nc{\address}[1]{{\em #1\medskip\\}}
\nc{\e}[1]{{\em #1\/}}
\nc{\comment}[1]{}
\nc{\itm}[2]{\\\noindent$\bullet$\ \ \e{#1}. \ #2}
\nc{\ru}[1]{\rule[-#1ex]{0ex}{#1ex}}
\rnc{\baselinestretch}{1.25}
\rnc{\arraystretch}{0.8}
\def\vec#1{\boldsymbol{#1}}
\nc{\spos}[2]{\makebox(0,0)[#1]{$\small{#2}$}}
\nc{\sposb}[2]{\makebox(0,0)[#1]{$ #2 $}}
\nc{\mi}{\!-\!}
\nc{\pl}{\!+\!}
\rnc{\ss}{\scriptstyle}
\nc{\sss}{\scriptscriptstyle}
\nc{\p}[2]{\makebox(0,0)[#1]{$#2$}}
\nc{\pp}[2]{\makebox(0,0)[#1]{$\ss#2$}}
\nc{\ppp}[2]{\makebox(0,0)[#1]{$\sss#2$}}
\nc{\ptext}[6]{\begin{picture}(#1,#2)\put(#3,#4){\p{#5}{\disp#6}}\end{picture}}
\nc{\be}{\begin{equation}}
\nc{\ee}{\end{equation}}
\nc{\bea}{\begin{eqnarray}}
\nc{\eea}{\end{eqnarray}}
\nc{\ba}[1]{\begin{array}{@{}#1@{}}}
\nc{\ea}{\end{array}}
\nc{\bpic}{\begin{picture}}
\nc{\epic}{\end{picture}}
\nc{\disp}{\displaystyle}
\nc{\ade}{\mbox{$A$-$D$-$E$}}
\nc{\calA}{{\cal A}}
\nc{\calH}{{\cal H}}
\nc{\calB}{{\cal B}}
\nc{\calC}{{\cal C}}
\nc{\calE}{{\cal E}}
\nc{\calI}{{\cal I}}
\nc{\calM}{{\cal M}}
\nc{\calN}{{\cal N}}
\nc{\calS}{{\cal S}}
\nc{\calV}{{\cal V}}

\def\hbar{{\overline{h}}}

\def\half {\mbox{$\textstyle {1 \over 2}$}}
\def\gauss#1#2{\left[#1\atop #2\right]}
\def\binom#1#2{\left(#1\atop #2\right)}

\def\ket#1{|#1\rangle}
\def\deriv#1#2{{\partial#1\over \partial #2}}
\def\smat#1{\mbox{\small{\mbox{$\begin{pmatrix}#1\end{pmatrix}$}}}}

\def\({\left(}
\def\){\right)}
\font\tenmsb=msbm10 scaled \magstep1
\font\sevenmsb=msbm7 scaled \magstep1
\font\fivemsb=msbm5 scaled \magstep1
\newfam\msbfam
\textfont\msbfam=\tenmsb
\scriptfont\msbfam=\sevenmsb
\scriptscriptfont\msbfam=\fivemsb
\def\Bbb#1{{\fam\msbfam\relax#1}}
\long\def\omit#1{}
\def\gauss#1#2{\mbox{\small $\left[#1\atop #2\right]$}}

\def\smaller{\small}
\def\punit#1{\hspace{#1\unitlength}}

\def\emptysquare{\hspace{-.11\unitlength}
\begin{pspicture}(1,1)
\pspolygon[linewidth=.25pt](0,0)(1,0)(1,1)(0,1)(0,0)
\end{pspicture}}
\def\loopa{\hspace{-.11\unitlength}
\begin{pspicture}(1,1)
\pspolygon[linewidth=.25pt](0,0)(1,0)(1,1)(0,1)(0,0)
\psarc[linewidth=1.5pt](1,0){.5}{90}{180}
\psarc[linewidth=1.5pt](0,1){.5}{-90}{0}
\end{pspicture}}
\def\loopb{\hspace{-.11\unitlength}
\begin{pspicture}(1,1)
\pspolygon[linewidth=.25pt](0,0)(1,0)(1,1)(0,1)(0,0)
\psarc[linewidth=1.5pt](0,0){.5}{0}{90}
\psarc[linewidth=1.5pt](1,1){.5}{180}{270}
\end{pspicture}}

\def\punit#1{\hspace{#1\unitlength}}

\def\vert#1#2#3{\rule[-4\unitlength]{0in}{8\unitlength}
\begin{picture}(0,0)(-#1,-#2)
\put(0,0){\line(0,1){4}}
\put(0,-1){\makebox(0,0)[t]{\smaller \mbox{$#3$}}}
\end{picture}}

\def\svert#1#2#3{\rule[-4\unitlength]{0in}{8\unitlength}
\begin{picture}(0,0)(-#1,-#2)
\put(0,0){\line(0,1){2}}
\put(0,-1){\makebox(0,0)[t]{\smaller \mbox{$#3$}}}
\end{picture}}

\def\monoid#1#2#3#4{\rule[-4\unitlength]{0in}{8\unitlength}
\begin{picture}(0,0)(-#1,-#2)
\put(2,0){\oval(4,3.5)[t]}
\put(2,4){\oval(4,3.5)[b]}
\put(0,-1){\makebox(0,0)[t]{\smaller \mbox{$#3$}}}
\put(4,-1){\makebox(0,0)[t]{\smaller \mbox{$#4$}}}
\end{picture}}

\def\downmonoid#1#2#3#4{\rule[-0\unitlength]{0in}{1\unitlength}
\begin{picture}(0,0)(-#1,-#2)
\put(2,0){\oval(4,3.5)[t]}
\put(0,-1){\makebox(0,0)[t]{\smaller \mbox{$#3$}}}
\put(4,-1){\makebox(0,0)[t]{\smaller \mbox{$#4$}}}
\end{picture}}

\def\down2monoid#1#2#3#4{\rule[-4\unitlength]{0in}{8\unitlength}
\begin{picture}(0,0)(-#1,-#2)
\put(2,0){\oval(12,10)[t]}
\put(-4,-1){\makebox(0,0)[t]{\smaller \mbox{$#3$}}}
\put(8,-1){\makebox(0,0)[t]{\smaller \mbox{$#4$}}}
\end{picture}}

\def\boxi#1#2#3{
\setlength{\unitlength}{.1in}
\begin{picture}(0,0)(-#2,-#3)
\multiput(0,0)(1,0){4}{\line(1,0){.5}}
\multiput(0,0)(0,1){4}{\line(0,1){.5}}
\multiput(0,4)(1,0){4}{\line(1,0){.5}}
\multiput(4,0)(0,1){4}{\line(0,1){.5}}
\put(2,2){\makebox(0,0)[c]{\smaller \mbox{$#1$}}}
\thicklines
\put(0,2){\line(1,1){2}}
\put(2,0){\line(1,1){2}}
\end{picture}}
\def\boxe#1#2#3{
\setlength{\unitlength}{.1in}
\begin{picture}(0,0)(-#2,-#3)
\multiput(0,0)(1,0){4}{\line(1,0){.5}}
\multiput(0,0)(0,1){4}{\line(0,1){.5}}
\multiput(0,4)(1,0){4}{\line(1,0){.5}}
\multiput(4,0)(0,1){4}{\line(0,1){.5}}
\put(2,2){\makebox(0,0)[c]{\smaller \mbox{$#1$}}}
\thicklines
\put(2,4){\line(1,-1){2}}
\put(0,2){\line(1,-1){2}}
\end{picture}}

\def\pdiamond#1{
\begin{pspicture}[.45](0,0)(0,8)
\pspolygon[linewidth=.25pt](0,4)(2,0)(4,4)(2,8)
\rput(2,4){\small $#1$}
\psarc(0,4){.35}{-63.4}{63.4}
\end{pspicture}}
\def\phexbr#1{\punit{-2}
\begin{pspicture}[.45](0,0)(0,8)
\pspolygon[linewidth=.25pt](2,0)(6.5,0)(8.5,4)(4,4)
\rput(5.25,2){\small $#1$}
\psarc(2,0){.35}{0}{63.4}
\end{pspicture}}
\def\phextr#1{\punit{-2}
\begin{pspicture}[.45](0,0)(0,8)
\pspolygon[linewidth=.25pt](4,4)(8.5,4)(6.5,8)(2,8)
\rput(5.25,6){\small $#1$}
\psarc(4,4){.35}{0}{126.8}
\end{pspicture}}
\def\phextl#1{\punit{-2}
\begin{pspicture}[.45](0,0)(0,8)
\pspolygon[linewidth=.25pt](2,4)(6.5,4)(8.5,8)(4,8)
\rput(5.25,6){\small $#1$}
\psarc(2,4){.35}{0}{63.4}
\end{pspicture}}
\def\phexbl#1{\punit{-2}
\begin{pspicture}[.45](0,0)(0,8)
\pspolygon[linewidth=.25pt](4,0)(8.5,0)(6.5,4)(2,4)
\rput(5.25,2){\small $#1$}
\psarc(4,0){.35}{0}{126.8}
\end{pspicture}}

\def\pdiamonda{
\begin{pspicture}[.45](0,0)(0,8)
\pspolygon[linewidth=.25pt](0,4)(2,0)(4,4)(2,8)
\psbezier[linewidth=1pt](1,2)(2.1,2.5)(2.1,5.5)(1,6)
\psbezier[linewidth=1pt](3,2)(1.9,2.5)(1.9,5.5)(3,6)
\end{pspicture}}
\def\pdiamondb{
\begin{pspicture}[.45](0,0)(0,8)
\pspolygon[linewidth=.25pt](0,4)(2,0)(4,4)(2,8)
\psarc[linewidth=1pt](2,0){2.236}{63.4}{116.5}
\psarc[linewidth=1pt](2,8){2.236}{-116.5}{-63.4}
\end{pspicture}}
\def\phexbra{\punit{-2}
\begin{pspicture}[.45](0,0)(0,8)
\pspolygon[linewidth=.25pt](2,0)(6.5,0)(8.5,4)(4,4)
\psbezier[linewidth=1pt](3,2)(4.5,1.25)(6.25,2.5)(6.25,4)
\psbezier[linewidth=1pt](4.25,0)(4.25,1.5)(6,2.75)(7.5,2)
\end{pspicture}}
\def\phexbrb{\punit{-2}
\begin{pspicture}[.45](0,0)(0,8)
\pspolygon[linewidth=.25pt](2,0)(6.5,0)(8.5,4)(4,4)
\psarc[linewidth=1pt](2,0){2.236}{0}{63.4}
\psarc[linewidth=1pt](8.5,4){2.236}{-180}{-116.5}
\end{pspicture}}
\def\phextra{\punit{-2}
\begin{pspicture}[.45](0,0)(0,8)
\pspolygon[linewidth=.25pt](4,4)(8.5,4)(6.5,8)(2,8)
\psbezier[linewidth=1pt](3,6)(4.5,6.75)(6.25,5.5)(6.25,4)
\psbezier[linewidth=1pt](4.25,8)(4,6.5)(6,5.25)(7.5,6)
\end{pspicture}}
\def\phextrb{\punit{-2}
\begin{pspicture}[.45](0,0)(0,8)
\pspolygon[linewidth=.25pt](4,4)(8.5,4)(6.5,8)(2,8)
\psarc[linewidth=1pt](8.5,4){2.236}{116.5}{180}
\psarc[linewidth=1pt](2,8){2.236}{-63.4}{0}
\end{pspicture}}

\def\phextla{\punit{-2}
\begin{pspicture}[.45](0,0)(0,8)
\pspolygon[linewidth=.25pt](2,4)(6.5,4)(8.5,8)(4,8)
\psbezier[linewidth=1pt](3,6)(4.5,5.25)(6.25,6.5)(6.25,8)
\psbezier[linewidth=1pt](4.25,4)(4.25,5.5)(6,6.75)(7.5,6)
\end{pspicture}}
\def\phextlb{\punit{-2}
\begin{pspicture}[.45](0,0)(0,8)
\pspolygon[linewidth=.25pt](2,4)(6.5,4)(8.5,8)(4,8)
\psarc[linewidth=1pt](2,4){2.25}{0}{63.4}
\psarc[linewidth=1pt](8.5,8){2.25}{-180}{-116.5}
\end{pspicture}}

\def\phexblb{\punit{-2}
\begin{pspicture}[.45](0,0)(0,8)
\pspolygon[linewidth=.25pt](4,0)(8.5,0)(6.5,4)(2,4)
\psarc[linewidth=1pt](8.5,0){2.25}{116.5}{180}
\psarc[linewidth=1pt](2,4){2.25}{-63.4}{0}
\end{pspicture}}

\def\ptri#1{
\begin{pspicture}[.45](0,0)(4,8)
\pspolygon[linewidth=.25pt](0,4)(4,0)(4,8)(0,4)
\rput(2.3,4){\small $#1$}
\end{pspicture}}
\def\ptriarc#1{
\begin{pspicture}[.45](0,0)(4,8)
\pspolygon[linewidth=.25pt](0,4)(4,0)(4,8)(0,4)
\psarc(0,4){2.8}{-45}{45}
\rput(3.5,4){\small $#1$}
\end{pspicture}}
\def\psq#1{
\begin{pspicture}[.45](0,0)(4,4)
\pspolygon[linewidth=.25pt](0,0)(4,0)(4,4)(0,4)
\rput(2,2){\small $#1$}
\psarc(0,0){.35}{0}{90}
\end{pspicture}}
\def\psqbr#1{
\begin{pspicture}[.45](0,0)(4,4)
\pspolygon[linewidth=.25pt](0,0)(4,0)(4,4)(0,4)
\rput(2,2){\small $#1$}
\psarc(4,0){.35}{90}{180}
\end{pspicture}}
\def\psqtl#1{
\begin{pspicture}[.45](0,0)(4,4)
\pspolygon[linewidth=.25pt](0,0)(4,0)(4,4)(0,4)
\rput(2,2){\small $#1$}
\psarc(0,4){.35}{-90}{0}
\end{pspicture}}
\def\psqtr#1{
\begin{pspicture}[.45](0,0)(4,4)
\pspolygon[linewidth=.25pt](0,0)(4,0)(4,4)(0,4)
\rput(2,2){\small $#1$}
\psarc(4,4){.35}{-180}{-90}
\end{pspicture}}
\def\psqa#1{
\begin{pspicture}[.45](0,0)(4,4)
\pspolygon[linewidth=.25pt](0,0)(4,0)(4,4)(0,4)
\psarc[linewidth=1pt](4,0){2}{90}{180}
\psarc[linewidth=1pt](0,4){2}{-90}{0}
\rput(2,2){\small $#1$}
\end{pspicture}}
\def\psqb#1{
\begin{pspicture}[.45](0,0)(4,4)
\pspolygon[linewidth=.25pt](0,0)(4,0)(4,4)(0,4)
\psarc[linewidth=1pt](0,0){2}{0}{90}
\psarc[linewidth=1pt](4,4){2}{180}{270}
\rput(2,2){\small $#1$}
\end{pspicture}}

\nc{\botrightrefnodots}[4]{\begin{picture}(1.5,2.5)
\put(1.5,0){\line(0,1){2}}
\put(1.5,1){\line(-1,1){0.5}}
\put(1.5,1){\line(-1,-1){1}}
\put(1.5,0){\line(-1,1){1}}
\put(1.5,2){\line(-1,-1){1.0}}
\put(0.5,0){\line(1,-1){0.5}}
\put(1.0,-0.5){\line(1,1){0.5}}
\put(1.45,0.5){\spos{r}{#1}}
\put(1.45,1.5){\spos{r}{#2}}
\put(1,0){\spos{}{#3}}
\put(1,1){\spos{}{#4}}\end{picture}}
\nc{\toprightrefnodots}[4]{\begin{picture}(1.5,2.5)
\put(1.5,0){\line(0,1){2}}
\put(1.5,1){\line(-1,1){1}}
\put(1.5,1){\line(-1,-1){0.5}}
\put(1.5,0){\line(-1,1){1.0}}
\put(1.5,2){\line(-1,-1){1}}
\put(0.5,2.0){\line(1,1){0.5}}
\put(1.0,2.5){\line(1,-1){0.5}}
\put(1.45,0.5){\spos{r}{#1}}
\put(1.45,1.5){\spos{r}{#2}}
\put(1,1){\spos{}{#3}}
\put(1,2){\spos{}{#4}}\end{picture}}

\def\verta{
\begin{pspicture}[.45](0,0)(4,4)
\pspolygon[linewidth=.25pt](0,0)(4,0)(4,4)(0,4)
\psline[linewidth=1pt,arrowsize=6pt]{->}(0,2)(4,2)
\psline[linewidth=1pt,arrowsize=6pt]{->}(2,0)(2,4)
\psline[linewidth=1pt,arrowsize=6pt]{->}(0,2)(2,2)
\psline[linewidth=1pt,arrowsize=6pt]{->}(2,0)(2,2)
\psarc(0,0){.5}{0}{90}
\end{pspicture}}
\def\vertb{
\begin{pspicture}[.45](0,0)(4,4)
\pspolygon[linewidth=.25pt](0,0)(4,0)(4,4)(0,4)
\psline[linewidth=1pt,arrowsize=6pt]{->}(4,2)(0,2)
\psline[linewidth=1pt,arrowsize=6pt]{->}(2,4)(2,0)
\psline[linewidth=1pt,arrowsize=6pt]{->}(4,2)(2,2)
\psline[linewidth=1pt,arrowsize=6pt]{->}(2,4)(2,2)
\psarc(0,0){.5}{0}{90}
\end{pspicture}}
\def\vertc{
\begin{pspicture}[.45](0,0)(4,4)
\pspolygon[linewidth=.25pt](0,0)(4,0)(4,4)(0,4)
\psline[linewidth=1pt,arrowsize=6pt]{->}(4,2)(0,2)
\psline[linewidth=1pt,arrowsize=6pt]{->}(2,0)(2,4)
\psline[linewidth=1pt,arrowsize=6pt]{->}(4,2)(2,2)
\psline[linewidth=1pt,arrowsize=6pt]{->}(2,0)(2,2)
\psarc(0,0){.5}{0}{90}
\end{pspicture}}
\def\vertd{
\begin{pspicture}[.45](0,0)(4,4)
\pspolygon[linewidth=.25pt](0,0)(4,0)(4,4)(0,4)
\psline[linewidth=1pt,arrowsize=6pt]{->}(0,2)(4,2)
\psline[linewidth=1pt,arrowsize=6pt]{->}(2,4)(2,0)
\psline[linewidth=1pt,arrowsize=6pt]{->}(0,2)(2,2)
\psline[linewidth=1pt,arrowsize=6pt]{->}(2,4)(2,2)
\psarc(0,0){.5}{0}{90}
\end{pspicture}}
\def\verte{
\begin{pspicture}[.45](0,0)(4,4)
\pspolygon[linewidth=.25pt](0,0)(4,0)(4,4)(0,4)
\psline[linewidth=1pt,arrowsize=6pt]{->}(2,2)(0,2)
\psline[linewidth=1pt,arrowsize=6pt]{->}(2,2)(4,2)
\psline[linewidth=1pt,arrowsize=6pt]{->}(2,4)(2,2)
\psline[linewidth=1pt,arrowsize=6pt]{->}(2,0)(2,2)
\psarc(0,0){.5}{0}{90}
\end{pspicture}}
\def\vertf{
\begin{pspicture}[.45](0,0)(4,4)
\pspolygon[linewidth=.25pt](0,0)(4,0)(4,4)(0,4)
\psline[linewidth=1pt,arrowsize=6pt]{->}(0,2)(2,2)
\psline[linewidth=1pt,arrowsize=6pt]{->}(4,2)(2,2)
\psline[linewidth=1pt,arrowsize=6pt]{->}(2,2)(2,4)
\psline[linewidth=1pt,arrowsize=6pt]{->}(2,2)(2,0)
\psarc(0,0){.5}{0}{90}
\end{pspicture}}

\def\vertbdya{
\begin{pspicture}[.45](0,0)(4,8)
\pspolygon[linewidth=.25pt](0,4)(4,0)(4,8)(0,4)
\psarc(0,4){2.8}{-45}{35}
\psline[linewidth=1pt,arrowsize=6pt]{->}(2.6,5)(1.95,6)
\end{pspicture}}
\def\vertbdyb{
\begin{pspicture}[.45](0,0)(4,8)
\pspolygon[linewidth=.25pt](0,4)(4,0)(4,8)(0,4)
\psarc(0,4){2.8}{-35}{45}
\psline[linewidth=1pt,arrowsize=6pt]{->}(2.6,3)(1.95,2)
\end{pspicture}}

\begin{document}
%\begin{titlepage}
%\vspace*{\fill}
\rnc{\baselinestretch}{1.1}
\begin{center}
\title{Logarithmic Minimal Models}
\medskip
\author{Paul A. Pearce, J\o rgen Rasmussen}
\address{Department of Mathematics and Statistics\\
University of Melbourne\\Parkville, Victoria 3010, Australia}
\medskip
\author{Jean-Bernard Zuber}
\address{
LPTHE Tour 24-25 5\`eme \'etage, 
Universit\'e Pierre et Marie Curie -- Paris6\\
4 Place Jussieu,
F 75252 Paris Cedex 5, France}
\bigskip\medskip

\begin{abstract}
\noindent  Working in the dense loop representation, we use the planar
Temperley-Lieb \mbox{algebra} to build integrable lattice
models called \mbox{logarithmic} minimal models ${\cal LM}(p,p')$. 
Specifically, we construct Yang-Baxter integrable Temperley-Lieb models on the strip acting on link states and consider their associated Hamiltonian limits. 
These models and their associated representations of the Temperley-Lieb algebra are inherently non-local and not (time-reversal) symmetric. 
We argue that, in the continuum scaling limit, they yield logarithmic conformal field theories with central charges
$c=1-{6(p-p')^2\over pp'}$ where $p,p'=1,2,\ldots$ are coprime. The first few 
 members of the principal series ${\cal LM}(m,m+1)$ are critical dense polymers 
\mbox{($m=1$, $c\!=\!-2$)}, critical percolation \mbox{($m=2$, $c\!=\!0$)} and 
logarithmic Ising model \mbox{($m=3$, $c=1/2$)}. 
For the principal series, we find an infinite family of integrable and conformal
boundary conditions organized in an extended Kac table with conformal weights
$\Delta_{r,s}={((m+1)r-ms)^2-1\over 4m(m+1)}$, $r,s=1,2,\ldots$. The associated
conformal partition functions are given in terms of Virasoro characters of
highest-weight representations. Individually, these characters decompose into a finite number of
characters of irreducible representations. We show with examples how indecomposable representations arise from fusion. 
\end{abstract}
\end{center}

%%%%%%%%%%%%%%%%%%%%%%%%%%%%%%%%%%%%%%%%%%%%%%%%%%%%

\section{Introduction}

There is much current 
interest~\cite{Gurarie93,Roh96,Flohr,GabK99, Gaberdiel01}
in Logarithmic
Conformal Field Theories (LCFTs) including LCFTs in the presence of
boundaries~\cite{bdylcft}. 
The present paper aims at studying a family of lattice integrable models, for which, 
it is believed, the associated conformal field theories are logarithmic. 
The two primary signatures of LCFTs are first the appearance of logarithmic branch cuts
in correlation functions, and second and perhaps
more fundamentally, the appearance of indecomposable representations 
of the underlying conformal algebra (Virasoro or one of its extensions)
and their accompanying Jordan cells. 
Throughout this paper, we reserve the term indecomposable representation for a
representation exhibiting a Jordan-cell structure.
 On the lattice, the transfer matrix or the Hamiltonian on a strip
 are the precursors of the Virasoro generator $L_0$. 
To find lattice realizations of an LCFT, it is thus necessary to consider 
systems in which the transfer matrix is not diagonalizable and admits
Jordan cells. 
For simple lattice models, such as the six-vertex model or RSOS models, the transfer matrices are (time-reversal) symmetric. Since these transfer matrices are real, this implies that they are diagonalizable so  something different is needed. 

Indecomposable representations and their associated 
Jordan matrices have been shown to occur in a variety of 
algebras: Temperley-Lieb algebra and quantum groups at roots of 
unity~\cite{PaSa90} and superalgebras~\cite{Marcu,RozSa92,Maas}. This has led to 
supersymmetric and fermionic models~\cite{Sa92,Kausch95,KoMa97,Iva99,ReSa01}.
In the present paper, we make use of {\it non-local} degrees of freedom. 

Usually in statistical mechanics, one works with {\it local} degrees of freedom, 
such as spins or heights. In contrast, in other classes of physical
problems~\cite{Sa87,DuplantierSaleur} 
such as percolation (see Figure~1) and polymers, one needs to keep track 
of connectivities or some other degrees of freedom
which are inherently {\em non-local}. This shift in paradigm has a dramatic effect on the
physical properties of these models. Specifically, for the models 
considered here, we confirm that the set of exponents extends beyond~\cite{Sa87} the ``minimal Kac table''
and that  their associated conformal field theories
are in fact logarithmic~\cite{Kausch95,Ca99,GuLu99}.  
Indeed, it is demonstrated that the transfer matrices,
although real, in some cases are not diagonalizable and hence lead to Jordan cells.

\begin{figure}[t]
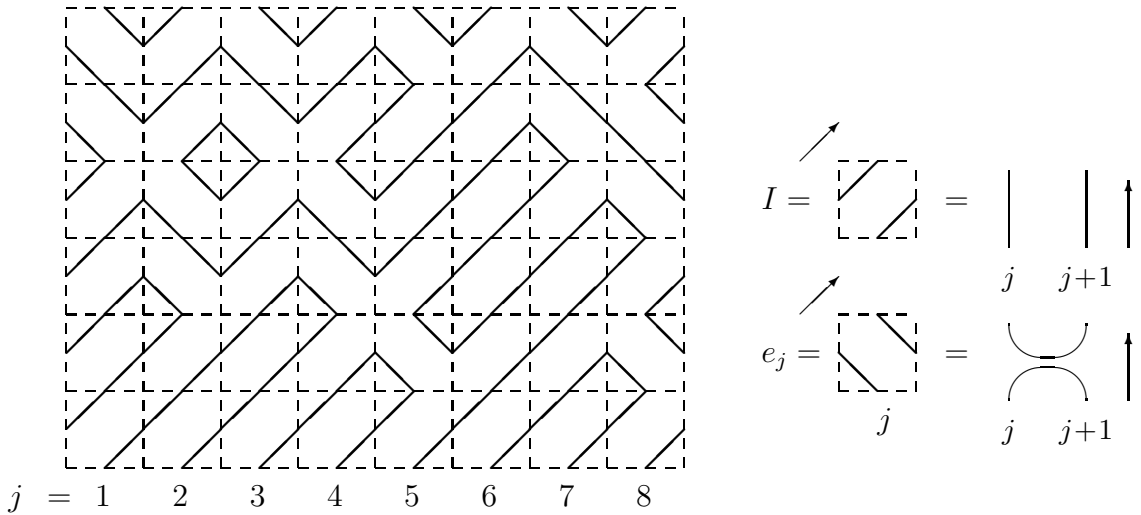

\mbox{}\vspace{0.8in}
$$
\mbox{}\hspace{-3.0in}
\boxe{}{-12}{20}\boxi{}{-8}{20}\boxe{}{-4}{20}\boxi{}0{20}\boxe{}4{20}\boxi{}8{20}\boxe{}{12}{20}
\boxi{}{16}{20}
\boxe{}{-12}{16}\boxi{}{-8}{16}\boxe{}{-4}{16}\boxi{}0{16}\boxi{}4{16}\boxi{}8{16}\boxe{}{12}{16}
\boxe{}{16}{16}
\boxi{}{-12}{12}\boxe{}{-8}{12}\boxi{}{-4}{12}\boxe{}0{12}\boxi{}4{12}\boxi{}8{12}\boxi{}{12}{12}
\boxe{}{16}{12}
\boxi{}{-12}8\boxe{}{-8}8\boxi{}{-4}8\boxe{}08\boxi{}48\boxi{}88\boxi{}{12}8\boxi{}{16}{8}
\boxi{}{-12}4\boxi{}{-8}4\boxi{}{-4}4\boxi{}04\boxe{}44\boxi{}84\boxi{}{12}4\boxe{}{16}{4}
\boxi{}{-12}0\boxi{}{-8}0\boxi{}{-4}0\boxi{}00\boxi{}40\boxi{}80\boxi{}{12}0\boxi{}{16}{0}
\put(-15,-2){$j$}
\put(-13,-2){$=$}
\put(-10.5,-2){$1$}
\put(-6.5,-2){$2$}
\put(-2.5,-2){$3$}
\put(1.5,-2){$4$}
\put(5.5,-2){$5$}
\put(9.5,-2){$6$}
\put(13.5,-2){$7$}
\put(17.5,-2){$8$}
\put(26,16){\vector(1,1){2.0}}
\put(24,13.5){$I=\mbox{}$}\boxi{}{28}{12}
\put(33,13.5){$\;=\quad\vert 0{-2}{j}\punit4\vert 0{-2}{j\!+\!1}$}
\put(43,11.5){\vector(0,1){3.5}}
\put(26,8){\vector(1,1){2.0}}
\put(24,5.5){$e_j=\mbox{}$}\boxe{}{28}{4}
\put(30,5.5){$\punit3\;=\quad\monoid 0{-2}{j}{j\!+\!1}$}
\put(43,3.5){\vector(0,1){3.5}}
\put(30,2){$j$}
$$
\caption{A simple model of critical percolation on the square lattice. Each face in
a column $j$ can be in one of two equally probable configurations which are associated
with the identity $I$ or the Temperley-Lieb generator $e_j$. }
\label{connectivities}
\end{figure}

There is some evidence~\cite{Kausch91,Roh96,Flohr96aug,Caselle,JR,EberleFlohr06} to suggest that there is an LCFT associated with each minimal model ${\cal M}(p,p')$. These LCFTs
are in some sense the simplest LCFTs. 
In this paper, we develop ideas involving non-local connectivities, the planar Temperley-Lieb (TL) algebra~\cite{TempLieb,Jones} and its action on states of planar link diagrams, to build integrable lattice models which we call logarithmic minimal models ${\cal LM}(p,p')$. These models might play a similar role for logarithmic theories as the Andrews-Baxter-Forrester RSOS models~\cite{ABFFB} do for rational theories. 
The isotropic critical percolation model ${\cal LM}(2,3)$ is illustrated in Figure~1. 
The idea to use transfer matrices acting on connectivity states dates back to the early eighties~\cite{BloteEtAl}. The role of planar link diagrams as ideals of the TL algebra was emphasized in \cite{PRGN}. The approach developed here has its roots in the loop version of the $O(n)$ model~\cite{Nienhuis}. 
Presumably, an alternative approach could be developed by using cluster transfer matrices~\cite{ChangSchrockJacobsen}.

We assert that the continuum scaling limit of the ${\cal LM}(p,p')$ lattice models define logarithmic 
CFTs which we also call logarithmic minimal models and also denote by ${\cal LM}(p,p')$. 
These theories offer a laboratory for studying 
LCFTs further by opening up new approaches to this important class of problems. In particular, these 
theories are amenable to study by the use of functional equations, Bethe ansatz, $T$-systems, $Y$-systems and Thermodynamic Bethe Ansatz.

We believe the main originality of the present paper lies in the 
use of boundary conditions in the loop model, 
 that are consistent with integrability.
It has been known for a long time~\cite{Ca84, SaBa89, Ca89} that 
boundary conditions are  suitable to expose the representation content 
of a CFT and to study the fusion of these representations. 
Here we borrow from the work by Behrend and Pearce~\cite{BP01} the construction 
of boundary states that are solutions of the Boundary Yang-Baxter Equations 
(BYBEs). These boundary conditions in the conformal continuum limit are
expected to give rise to representations of the Virasoro algebra. Specifically, 
the boundary conditions that we consider are labelled by a pair of integers $(r,s)$ 
with $(1,1)$ playing the role of the vacuum 
boundary condition. Imposing the boundary conditions $(1,1)$ and $(r,s)$ on the two sides of the strip
gives rise to a certain representation $(r,s)$, to be defined below, of the
Virasoro algebra and enables us to write an explicit form of the corresponding Hamiltonian.
Imposing $(r,s)$ and $(r',s')$ boundary conditions gives access~\cite{Ca89} to the 
fusion of representations $(r,s)$ and $(r',s')$. Our model thus provides a
practical tool to study the fusion of representations and to see the 
generation of indecomposable representations.

The layout of this paper is as follows. We start in Section~2 by summarising the spectral conformal data obtained for our logarithmic minimal models. 
We recall  the various types of representations of the Virasoro algebra, namely, irreducible, reducible and indecomposable representations. In Appendix~A, we show how the 
characters of the reducible representations can be written as a sum of finitely many characters of irreducible representations.
In Section~3, we use the planar TL algebra to define 
integrable lattice realizations of the minimal LCFTs. We review the definition of the
planar TL algebra~\cite{Jones}. We show that the lattice 
logarithmic minimal models are integrable in the sense that the 
local face operators $X(u)$, where $u$ is the spectral parameter, satisfy the Yang-Baxter Equations (YBEs) and Boundary Yang-Baxter Equations (BYBEs). 
We also use the construction of Behrend and Pearce~\cite{BP01} to obtain an infinite hierarchy of solutions to the BYBE labelled by extended Kac labels $(r,s)$ with $r,s=1,2,\ldots$.
In Section~4, we discuss the relation between the planar TL algebra and the more usual linear TL algebra. We introduce link diagrams which are the non-local states that keep track of connectivities. We also specialize the inversion, YBE and BYBEs to their appropriate forms in the linear TL algebra.
In Section~5, we set up commuting double-row transfer matrices $\vec D(u)$ and argue that there exist integrable and conformal boundary conditions labelled by the entries of the infinite Kac table
$(r,s)$ with $r,s=1,2,3,\ldots$. Next we obtain explicit expressions for the integrable Temperley-Lieb link Hamiltonians by taking the logarithmic derivative of the commuting double-row transfer matrices $\vec D(u)$ at $u=0$. 
In Section~6, we discuss the relation of the logarithmic minimal models to the six-vertex model.
We also present analytic expressions for the bulk and boundary free energies including the forms applicable in the Hamiltonian limit.
In Section~7, we turn to the conformal spectra on the strip obtained in the continuum scaling limit, first 
restricting ourselves to the case where one boundary is the vacuum. In these cases, we find that the transfer matrices are diagonalizable and the spectrum generating functions 
are given by a single conformal character corresponding to a 
quasi-rational quotient module of the Virasoro algebra. We present numerical evidence to support this assertion.
In Section~8, by considering non-trivial boundaries on both sides of the strip, we show how indecomposable representations are generated by fusion of  the $(r,s)$ representations. 
This observation is crucial in the claim that our models are logarithmic. 
We leave a more detailed discussion of the fusion algebras to a subsequent paper.  
Section~9 contains a brief discussion.

\section{Logarithmic Minimal CFT}

\subsection{Spectral Data}

%  Extended Kac tables
\def \st#1{\raisebox{-6pt}{\rule{0pt}{18pt}}\makebox[16pt]{\small ${#1}$}}
\def\vvdots{\mathinner{\mkern1mu\raise1pt\vbox{\kern7pt\hbox{.}}\mkern2mu
 \raise4pt\hbox{.}\mkern2mu\raise7pt\hbox{.}\mkern1mu}}
\begin{table}[p]
\begin{center}
$m=1,\ c=-2$
\hspace{5.8cm}
$m=2,\ c=0$\\ \mbox{}\\
\begin{pspicture}(0,0)(7,8)
%\psgrid
\rput[bl](0.15,0){\color{lightlightblue}{\rule{1cm}{7.25cm}}}
\rput[bl](0,0){
\begin{tabular}{|c|c|c|c|c|c|c|c|c|}
\hline
\st{\vdots}&\st{\vdots}&\st{\vdots}&\st{\vdots}&\st{\vdots}&\st{\vdots}&\st{\vvdots}\\ \hline
\st{63\over 8}&\st{35\over 8}&\st{15\over 8}&\st{3\over 8}&\st{-{1\over 8}}&\st{3\over 8}&\st{\cdots}\\ \hline
\st{6}&\st{3}&\st{1}&\st{0}&\st{0}&\st{1}&\st{\cdots}\\ \hline
\st{35\over 8}&\st{15\over 8}&\st{3\over 8}&\st{-{1\over 8}}&\st{3\over 8}&\st{15\over 8}&\st{\cdots}\\ \hline
\st{3}&\st{1}&\st{0}&\st{0}&\st{1}&\st{3}&\st{\cdots}\\ \hline
\st{15\over 8}&\st{3\over 8}&\st{-{1\over 8}}&\st{3\over 8}&\st{15\over 8}&\st{35\over 8}&\st{\cdots}\\ \hline
\st{1}&\st{0}&\st{0}&\st{1}&\st{3}&\st{6}&\st{\cdots}\\ \hline
\st{3\over 8}&\st{-{1\over 8}}&\st{3\over 8}&\st{15\over 8}&\st{35\over 8}&\st{63\over 8}&\st{\cdots}\\ \hline
\st{0}&\st{0}&\st{1}&\st{3}&\st{6}&\st{10}&\st{\cdots}\\ \hline
\st{-{1\over 8}}&\st{3\over 8}&\st{15\over 8}&\st{35\over 8}&\st{63\over 8}&\st{99\over 8}&\st{\cdots}\\ \hline
\st{0}&\st{1}&\st{3}&\st{6}&\st{10}&\st{15}&\st{\cdots}\\ \hline
\end{tabular}}
\end{pspicture}
\mbox{}\ \ \ \ \ \ \ \ \ \ \mbox{}
\begin{pspicture}(0,0)(7,8)
%\psgrid
\rput[bl](0.15,0){\color{lightlightblue}{\rule{1.97cm}{7.25cm}}}
\rput[bl](0.15,0){\color{lightblue}{\rule{1cm}{1.3cm}}}
\rput[bl](0,0){
\begin{tabular}{|c|c|c|c|c|c|c|c|c|}
\hline
\st{\vdots}&\st{\vdots}&\st{\vdots}&\st{\vdots}&\st{\vdots}&\st{\vdots}&\st{\vvdots}\\ \hline
\st{12}&\st{65\over 8}&\st{5}&\st{21\over 8}&\st{1}&\st{1\over 8}&\st{\cdots}\\ \hline
\st{28\over 3}&\st{143\over 24}&\st{10\over 3}&\st{35\over 24}&\st{1\over 3}&\st{-{1\over 24}}&\st{\cdots}\\ \hline
\st{7}&\st{33\over 8}&\st{2}&\st{5\over 8}&\st{0}&\st{1\over 8}&\st{\cdots}\\ \hline
\st{5}&\st{21\over 8}&\st{1}&\st{1\over 8}&\st{0}&\st{5\over 8}&\st{\cdots}\\ \hline
\st{10\over 3}&\st{35\over 24}&\st{1\over 3}&\st{-{1\over 24}}&\st{1\over 3}&\st{35\over 24}&\st{\cdots}\\ \hline
\st{2}&\st{5\over 8}&\st{0}&\st{1\over 8}&\st{1}&\st{21\over 8}&\st{\cdots}\\ \hline
\st{1}&\st{{1\over 8}}&\st{0}&\st{5\over 8}&\st{2}&\st{33\over 8}&\st{\cdots}\\ \hline
\st{1\over 3}&\st{-{1\over 24}}&\st{1\over 3}&\st{35\over 24}&\st{10\over 3}&\st{143\over 24}&\st{\cdots}\\ \hline
\st{0}&\st{1\over 8}&\st{1}&\st{21\over 8}&\st{5}&\st{65\over 8}&\st{\cdots}\\ \hline
\st{0}&\st{5\over 8}&\st{2}&\st{33\over 8}&\st{7}&\st{85\over 8}&\st{\cdots}\\ \hline
\end{tabular}}
\end{pspicture}
\\ \mbox{}\\ \mbox{}\\
$m=3,\ c=1/2$
\hspace{5.3cm}
$m=4,\ c=7/10$\\ \mbox{}\\
\begin{pspicture}(0,0)(7,8)
%\psgrid
\rput[bl](0.15,0){\color{lightlightblue}{\rule{2.95cm}{7.25cm}}}
\rput[bl](0.15,0){\color{lightblue}{\rule{1.95cm}{1.94cm}}}
\rput[bl](0,0){
\begin{tabular}{|c|c|c|c|c|c|c|c|c|}
\hline
\st{\vdots}&\st{\vdots}&\st{\vdots}&\st{\vdots}&\st{\vdots}&\st{\vdots}&\st{\vvdots}\\ \hline
\st{225\over 16}&\st{161\over 16}&\st{323\over 48}&\st{65\over 16}&\st{33\over 16}&\st{35\over 48}&\st{\cdots}\\ \hline
\st{11}&\st{15\over 2}&\st{14\over 3}&\st{5\over 2}&\st{1}&\st{1\over 6}&\st{\cdots}\\ \hline
\st{133\over 16}&\st{85\over 16}&\st{143\over 48}&\st{21\over 16}&\st{5\over 16}&\st{-{1\over 48}}&\st{\cdots}\\ \hline
\st{6}&\st{7\over 2}&\st{5\over 3}&\st{1\over 2}&\st{0}&\st{1\over 6}&\st{\cdots}\\ \hline
\st{65\over 16}&\st{33\over 16}&\st{35\over 48}&\st{1\over 16}&\st{1\over 16}&\st{35\over 48}&\st{\cdots}\\ \hline
\st{5\over 2}&\st{1}&\st{1\over 6}&\st{0}&\st{1\over 2}&\st{5\over 3}&\st{\cdots}\\ \hline
\st{21\over 16}&\st{{5\over 16}}&\st{-{1\over 48}}&\st{5\over 16}&\st{21\over 16}&\st{143\over 48}&\st{\cdots}\\ \hline
\st{1\over 2}&\st{0}&\st{1\over 6}&\st{1}&\st{5\over 2}&\st{14\over 3}&\st{\cdots}\\ \hline
\st{{1\over 16}}&\st{1\over 16}&\st{35\over 48}&\st{33\over 16}&\st{65\over 16}&\st{323\over 48}&\st{\cdots}\\ \hline
\st{0}&\st{1\over 2}&\st{5\over 3}&\st{7\over 2}&\st{6}&\st{55\over 6}&\st{\cdots}\\ \hline
\end{tabular}}
\end{pspicture}
\mbox{}\ \ \ \ \ \ \ \ \ \ \mbox{}
\begin{pspicture}(0,0)(7,8)
%\psgrid
\rput[bl](0.15,0){\color{lightlightblue}{\rule{3.9cm}{7.25cm}}}
\rput[bl](0.15,0){\color{lightblue}{\rule{2.9cm}{2.56cm}}}
\rput[bl](0,0){
\begin{tabular}{|c|c|c|c|c|c|c|c|c|}
\hline
\st{\vdots}&\st{\vdots}&\st{\vdots}&\st{\vdots}&\st{\vdots}&\st{\vdots}&\st{\vvdots}\\ \hline
\st{153\over 10}&\st{899\over 80}&\st{39\over 5}&\st{399\over 80}&\st{14\over 5}&\st{99\over 80}&\st{\cdots}\\ \hline
\st{12}&\st{135\over 16}&\st{11\over 2}&\st{51\over 16}&\st{3\over 2}&\st{7\over 16}&\st{\cdots}\\ \hline
\st{91\over 10}&\st{483\over 80}&\st{18\over 5}&\st{143\over 80}&\st{3\over 5}&\st{3\over 80}&\st{\cdots}\\ \hline
\st{33\over 5}&\st{323\over 80}&\st{21\over 10}&\st{63\over 80}&\st{1\over 10}&\st{3\over 80}&\st{\cdots}\\ \hline
\st{9\over 2}&\st{39\over 16}&\st{1}&\st{3\over 16}&\st{0}&\st{7\over 16}&\st{\cdots}\\ \hline
\st{14\over 5}&\st{99\over 80}&\st{3\over 10}&\st{-{1\over 80}}&\st{3\over 10}&\st{99\over 80}&\st{\cdots}\\ \hline
\st{3\over 2}&\st{{7\over 16}}&\st{{0}}&\st{3\over 16}&\st{1}&\st{39\over 16}&\st{\cdots}\\ \hline
\st{3\over 5}&\st{3\over 80}&\st{1\over 10}&\st{63\over 80}&\st{21\over 10}&\st{323\over 80}&\st{\cdots}\\ \hline
\st{{1\over 10}}&\st{3\over 80}&\st{3\over 5}&\st{143\over 80}&\st{18\over 5}&\st{483\over 80}&\st{\cdots}\\ \hline
\st{0}&\st{7\over 16}&\st{3\over 2}&\st{51\over 16}&\st{11\over 2}&\st{135\over 16}&\st{\cdots}\\ \hline
\end{tabular}}
\end{pspicture}
\end{center}
\caption{Lower left corner of the extended Kac table of conformal weights $\Delta_{r,s}$ for 
$m=1,2,3,4$ corresponding, respectively,
to critical dense polymers ($c=-2$), critical percolation ($c=0$),
the logarithmic Ising model ($c=1/2$), and the logarithmic tricritical Ising model 
($c=7/10$). All of the distinct conformal weights occur in the first $m$ columns.}
\end{table}

The usual rational minimal models are constructed on a finite set of irreducible
highest-weight representations which arise as quotients of Verma modules 
of the Virasoro algebra.
The value of the central charge is specified by two coprime integers
$p, p'$, with $1<p<p'$ and
\be
c=c(p,p'):=1-{6(p-p')^2\over pp'} 
\ee
The conformal weights, which label the irreducible representations,
 are given by the Kac formula
\begin{equation}
\Delta_{r,s}=\Delta_{p-r,p'-s}=
{(p'r-ps)^2-(p-p')^2\over 4pp'},
\quad\qquad 1\le r \le p-1\ ,\quad 1\le s \le p'-1
\end{equation}

In contrast, logarithmic CFTs are constructed on representations of 
the Virasoro algebra
which are {\it not}\/ all irreducible highest-weight representations:
some of the representations are indecomposable.
In the simple class of such theories considered here, 
the central charges and conformal 
weights are as in the usual minimal models (up to the 
bounds on the labels $r, s$), but the 
Virasoro generators act on some representations  through  Jordan cells.

To be more precise, the CFTs that will appear in the 
continuum limit of our lattice models have central charges and conformal weights
\be
c=1-{6\lambda^2\over \pi(\pi-\lambda)},\quad 0<\lambda<\pi;\qquad
\Delta_{r,s}={[\pi r-(\pi-\lambda)s]^2-\lambda^2\over 4\pi(\pi-\lambda)}, \quad r,s=1,2,3,\ldots
\label{ctsc}
\ee
where $\lambda$ is the {\it crossing parameter} of the 
lattice model. Whenever $\lambda/\pi$ is rational of the form  $\lambda={(p'-p)\pi\over p'}$, where  $p,p'$ are two coprime integers with $0<p<p'$, this central charge coincides with $c(p,p')$. 
The corresponding CFT is not rational nor unitary and 
will be denoted by ${\cal LM}(p,p')$.  
The conformal weights lie in an {\it infinitely extended} Kac table
\begin{equation}
\Delta_{r,s}={(p'r-ps)^2-(p-p')^2\over 4pp'}, 
\qquad r,s=1,2,3,\ldots
\end{equation}
While $s$ varies over an infinite range, it may be necessary to restrict the values of $r$ according to the model. 

The most studied LCFTs so far are models with central charges $c=c(1,p')$ or $c=c(2,p')$~\cite{Kausch91,Roh96,Flohr96aug,FjelstadEtAl,EberleFlohr06}. In the present paper, 
our primary focus is the series ${\cal LM}(m,m+1)$ with central charges
\begin{equation}
c=1-{6\over m(m+1)},\qquad m=1,2,3,\ldots
\end{equation}
called the {\it principal series} and we restrict $r$ to the range $1\le r\le m$.
The first few members of the principal series are of particular significance since they include critical dense polymers 
\mbox{($m=1$, $c\!=\!-2$)} and critical percolation \mbox{($m=2$, $c\!=\!0$)}. The other members of the series are new lattice models. Borrowing nomenclature from the usual rational models, we call them the
logarithmic Ising model \mbox{($m=3$, $c=1/2$)}, the logarithmic tricritical Ising model \mbox{($m=4$, $c=7/10$)} and so on. The conformal weights of these models are shown in Table~1. Despite the nomenclature, the properties of these models are actually very different from their rational cousins.

It is observed that all of the distinct conformal weights fall in the first $m$ columns of the extended Kac table of ${\cal LM}(m,m+1)$. This follows from a simple combination of the symmetries $\Delta_{r+kp,s+kp'}=\Delta_{r,s}=\Delta_{p-r,p'-s}$ with $k\in{\Bbb Z}$. 
In the logarithmic theories, these symmetries merely express the 
{\it coincidence}\/ of conformal weights and do {\it not}\/ indicate the identification of representations.

Specifying the central charge and conformal weights may not uniquely determine a LCFT. It is conceivable that two LCFTs could have the same spectral data but differ in their Jordan cell structures. 
Thus, we do not claim any exhaustive classification of LCFTs, nor do we claim that the logarithmic minimal models exhaust the LCFTs of any given central charge.
Instead, we pragmatically define minimal LCFTs as the continuum scaling limits of our logarithmic minimal models, which are well-defined integrable lattice models, and then study their conformal properties.

%%%%%%%%%%%%%%%%%%%%%%%%%%%%%%%%%%%%%%%%%%%%%%%%%%%%

\subsection{Quasi-Rational Representations and Characters}

The concepts of rational CFT, with 
its finite number of representations of the
chiral algebra, and of the fusion of these representations are quite familiar. 
The logarithmic minimal models, 
on the other hand, possess a countably infinite number of representations.
We anticipate that the logarithmic minimal models are quasi-rational in the  
sense that the fusion of any two representations  
produces only a finite number of such representations. 
The representations of such a theory will be called quasi-rational, 
following Nahm~\cite{Nahm94}, who gave a criterion for 
quasi-rationality. 

For any {\it rational}\/ or {\it irrational}\/ value of 
$\lambda/\pi $ and for any
positive integers $r, s$, the module
(representation) $V_{\Delta_{r,s}}$ of the Virasoro algebra
of highest weight $\Delta_{r,s}$ given by (2.4) 
is reducible; it has a submodule $V_{\Delta_{r,-s}}$ 
of highest weight  $\Delta_{r,-s}=\Delta_{r,s}+rs$.
The character of the quotient module 
$Q_{r,s}:=V_{\Delta_{r,s}}/V_{\Delta_{r,-s}}$ is
\begin{equation} 
\chi_{r,s}(q)  
=q^{-c/24}\,{q^{\Delta_{r,s}}-q^{\Delta_{r,-s}}
\over \prod_{n=1}^\infty (1-q^n)}
=q^{-c/24}\,{q^{\Delta_{r,s}}(1-q^{rs})
\over \prod_{n=1}^\infty (1-q^n)}
\end{equation}
Such quotients $Q_{r,s}$ are irreducible for (generic) irrational values of 
$\lambda/\pi$, while they are not  necessarily irreducible if   
${\lambda\over\pi}={p'-p\over p'}$ is rational. In our 
construction, the spectrum depends on the free parameter $\lambda$ and we find it varies continuously with $\lambda$.
This supports our assertion that the characters $\chi_{r,s}(q)$ above 
are appropriate building blocks to describe the conformal 
spectra of the logarithmic models  ${\cal LM}(p,p')$, 
even though ${\lambda\over\pi}={p'-p\over p'}$ is rational and the associated characters are not irreducible. We denote by $(r,s)$ the corresponding representations. It is stressed that we
are only equating the {\em characters} of the representations $(r,s)$ and $Q_{r,s}$, not the
representations themselves, since we are only probing the action of $L_0$.
This means that we at this point are leaving open the possibility that the representation $(r,s)$,
unlike $Q_{r,s}$, is fully reducible.
In Appendix~A, we show how the characters of the representations  
$Q_{r,s}$ decompose into a 
finite number of characters of irreducible representations of the Virasoro algebra. 

The characters $\chi_{r,s}(q)$ arise as the 
limit of finitized characters for a lattice strip of $N$ columns
\bea
\chi_{r,s}(q)\;=\;\lim_{N\to\infty}\chi_{r,s}^{(N)}(q)
\label{quasi}
\eea
where
\bea
\chi_{r,s}^{(N)}(q)\;=\;q^{-c/24+\Delta_{r,s}}
\Big(\gauss{N}{(N-s+r)/2}_q-
q^{rs}\gauss{N}{(N-s-r)/2}_q\Big)
\label{finitizedchar}
\eea
Here $\gauss{N}{M}_q$ is a $q$-binomial (Gaussian polynomial) and $N=r\!-\!s$ mod $2$. The dimension of the vector space of states is given by $\mbox{dim}\,{\cal V}=\chi_{r,s}^{(N)}(1)$. The characters $\chi_{r,s}(q)=q^{-c/24+\Delta_{r,s}}\sum_E q^E$ are the spectrum generating functions for the integer energies $E$ of an infinite system. A {\em finitized} character~\cite{Melzer,Berkovich} is obtained by a consistent truncation of the space of states of the infinite system. The energies of a {\em finite} system therefore do not precisely coincide with integer energies of the finitized character but they converge to them as $N\to\infty$. 

The fusion of the
representations $(r,s)$ generates new representations that may be 
indecomposable. For example, we will confirm in Section~8 that for critical dense polymers  ($m=1$, $c=-2$) 
the fusion of $(1,2)$ with itself is
\be
(1,2) \otimes_f (1,2) = (1,1) \oplus_i (1,3)
\ee
As indicated by the subscript $i$, the right side is not a direct sum of representations but rather an {\it indecomposable}\/ combination exhibiting Jordan cells.
Of course, since the character of the indecomposable representation is insensitive to the off-diagonal terms,  it is simply $\chi_{1,1}(q)+\chi_{1,3}(q)$. 

By construction, the set of (irreducible, reducible or indecomposable) representations generated by this fusion prescription is closed and constitutes a set of quasi-rational representations.

\section{Planar Temperley-Lieb Algebra} 
\psset{unit=.1in}
\setlength{\unitlength}{.1in}

{}From a simple perspective, a planar algebra~\cite{Jones} is a closed algebra of diagrams known as planar tangles. The diagrams can be interpreted (by selecting in- and out- states) as giving rise to  
multiplications in different directions corresponding to a consistent action on a collection of vector spaces. Here we only consider the planar Temperley-Lieb (TL) algebra. 

Given a planar algebra admitting local face operators satisfying the Yang-Baxter equation, one can build an integrable lattice model.  
To define integrable lattice models~\cite{BaxBook} on the square lattice which realize
the minimal LCFTs ${\cal LM}(p,p')$ in the continuum scaling limit, we use solutions of
the Yang-Baxter Equation (YBE) built from the planar TL algebra~\cite{Jones} 
${\cal T}={\cal T}(\lambda)$ with crossing parameter $\lambda\in {\Bbb R}$ which for ${\cal LM}(p,p')$ is specialized to
\begin{equation}
\lambda={(p'-p)\pi\over p'},\qquad \mbox{$p,p'$ coprime}
\end{equation}
We introduce a complex spectral parameter $u\in{\Bbb C}$ and set
\bea
s_r(u)=\frac{\sin(u\!+\!r\lambda)}{\sin\!\lambda},\qquad r\in{\Bbb Z}
\eea
The local face operators are defined as linear combinations of elementary 2-boxes (monoids~\cite{Kauf}) by
\be
X(u)\;=\;\psq u\;=\;s_1(-u)\;\psqa{}\ +\ s_0(u)\;\psqb{}
\ee
Consequently, we have the local crossing relation
\be
X(\lambda-u)\;=\;\psq{\lambda-u}\;=\;s_0(u)\;\psqa{}\ +\ s_1(-u)\;\psqb{}
\;=\;\psqbr u
\ee
The 2 in 2-box refers to the fact that there are 2 connectivities in and 2 connectivities out.
The lower-left corner of a lattice face is marked to fix which monoid gets the weight $s_1(-u)$ and which gets the weight $s_0(u)$. 
Internally, the nodes at the centers of the edges of a face can be connected in pairs in one of two ways as specified by the two elementary 2-boxes. 
%The weights, or unnormalized probabilities, assigned to these elementary 2-boxes are
%\be
%W\left(\psqa u\right)\;=\;s_1(-u),\qquad\qquad W\left(\psqb u\right)\;=\;s_0(u)
%\ee
%We often omit the spectral parameters in the elementary 2-boxes when they are clear from context.

The usual physical requirement is that these weights are positive but it is useful here to relax this constraint.
{}From the diagonal reflection symmetries and crossing symmetries, we have
\psset{unit=.1in}
\setlength{\unitlength}{.1in}
\be
X(u)\;=\;\psq u\;=\;\psqtr u\;=\;\psqbr{\lambda\!-\!u}\;=\;\psqtl{\lambda\!-\!u}
\ee
The face operator and elementary 2-boxes can be viewed as acting from any two adjacent nodes (in-states) to the remaining two adjacent nodes (out-states). In this manner, these operators can act in the four diagonal directions on distinct vector spaces spanned by link diagrams enumerating the allowed planar connectivities of the relevant nodes. 

The elementary 2-boxes satisfy the simple relations
\psset{unit=.075in}
\setlength{\unitlength}{.075in}
\bea
\begin{pspicture}[.45](0,0)(8,8)
\psline[linestyle=dashed,dash=.5 .5,linewidth=.25pt](2.4,0)(6.4,0)
\psline[linestyle=dashed,dash=.5 .5,linewidth=.25pt](2.4,8)(6.4,8)
\rput(0,4){\pdiamondb}
\rput(4,4){\pdiamondb}
\psarc[linewidth=1pt](4.4,4){2.236}{63.4}{116.5}
\psarc[linewidth=1pt](4.4,4){2.236}{-116.5}{-63.4}
\end{pspicture}
\;=\;
\begin{pspicture}[.45](0,0)(4,8)
\rput(0,4){\pdiamondb}
\end{pspicture}\ ,\qquad\qquad
\begin{pspicture}[.45](0,0)(8,8)
\psline[linestyle=dashed,dash=.5 .5,linewidth=.25pt](2.4,0)(6.4,0)
\psline[linestyle=dashed,dash=.5 .5,linewidth=.25pt](2.4,8)(6.4,8)
\rput(0,4){\pdiamonda}
\rput(4,4){\pdiamonda}
\psarc[linewidth=1pt](4.4,4){2.236}{63.4}{116.5}
\psarc[linewidth=1pt](4.4,4){2.236}{-116.5}{-63.4}
\end{pspicture}
\;=\;\beta\;
\begin{pspicture}[.45](0,0)(4,8)
\rput(0,4){\pdiamonda}
\end{pspicture}
\label{simprelations}
\eea
and similar relations where the dashed lines indicate that the corners and associated incident edges are identified. 
Viewed as acting horizontally, these are the standard relations $I\, I=I$ and $e_j^2=\beta e_j$ of the linear TL algebra as in Section~4. 
In the planar algebra, however, the relations (\ref{simprelations}) are valid for action in any direction, horizontally or vertically. In physical terms, the planar TL algebra is interpreted as a loop gas with fugacity
\be
\beta=2\cos\lambda=x+x^{-1},\qquad x=e^{i\lambda},\qquad \beta\in (-2,2)
\ee
assigned to each closed loop.

\subsection{Inversion and Yang-Baxter Equations}
\psset{unit=.075in}
\setlength{\unitlength}{.075in}
Let us prove the inversion and Yang-Baxter relations in the planar TL algebra. Diagrammatically, the inversion relation is
\bea
&&\begin{pspicture}[.45](0,0)(0,8)
\pspolygon[linewidth=.25pt](0,4)(2,0)(4,4)(2,8)
\psarc(0,4){.35}{-63.4}{63.4}
\rput(2,4){\small $u$}
\psline[linestyle=dashed,dash=.5 .5,linewidth=.25pt](2,0)(6,0)
\psline[linestyle=dashed,dash=.5 .5,linewidth=.25pt](2,8)(6,8)
\end{pspicture}\punit4
\begin{pspicture}[.45](0,0)(0,8)
\pspolygon[linewidth=.25pt](0,4)(2,0)(4,4)(2,8)
\psarc(0,4){.35}{-63.4}{63.4}
\rput(2,4){\small $\,-u$}
\end{pspicture}\punit4
\;=\;s_1(-u)s_1(u)\;
\begin{pspicture}[.45](0,0)(8,8)
\psline[linestyle=dashed,dash=.5 .5,linewidth=.25pt](2.4,0)(6.4,0)
\psline[linestyle=dashed,dash=.5 .5,linewidth=.25pt](2.4,8)(6.4,8)
\rput(0,4){\pdiamondb}
\rput(4,4){\pdiamondb}
\psarc[linewidth=1pt](4.4,4){2.236}{63.4}{116.5}
\psarc[linewidth=1pt](4.4,4){2.236}{-116.5}{-63.4}
\end{pspicture}\;+\;s_1(-u)s_0(-u)\;
\begin{pspicture}[.45](0,0)(8,8)
\psline[linestyle=dashed,dash=.5 .5,linewidth=.25pt](2.4,0)(6.4,0)
\psline[linestyle=dashed,dash=.5 .5,linewidth=.25pt](2.4,8)(6.4,8)
\rput(0,4){\pdiamondb}
\rput(4,4){\pdiamonda}
\psarc[linewidth=1pt](4.4,4){2.236}{63.4}{116.5}
\psarc[linewidth=1pt](4.4,4){2.236}{-116.5}{-63.4}
\end{pspicture}\nonumber\\[12pt]
&&\mbox{}+s_0(u)s_1(u)\;
\begin{pspicture}[.45](0,0)(8,8)
\psline[linestyle=dashed,dash=.5 .5,linewidth=.25pt](2.4,0)(6.4,0)
\psline[linestyle=dashed,dash=.5 .5,linewidth=.25pt](2.4,8)(6.4,8)
\rput(0,4){\pdiamonda}
\rput(4,4){\pdiamondb}
\psarc[linewidth=1pt](4.4,4){2.236}{63.4}{116.5}
\psarc[linewidth=1pt](4.4,4){2.236}{-116.5}{-63.4}
\end{pspicture}\;+\;s_0(u)s_0(-u)\;
\begin{pspicture}[.45](0,0)(8,8)
\psline[linestyle=dashed,dash=.5 .5,linewidth=.25pt](2.4,0)(6.4,0)
\psline[linestyle=dashed,dash=.5 .5,linewidth=.25pt](2.4,8)(6.4,8)
\rput(0,4){\pdiamonda}
\rput(4,4){\pdiamonda}
\psarc[linewidth=1pt](4.4,4){2.236}{63.4}{116.5}
\psarc[linewidth=1pt](4.4,4){2.236}{-116.5}{-63.4}
\end{pspicture}
\;\;=\;s_1(u)s_1(-u)
\begin{pspicture}[.45](0,0)(4,8)
\rput(0,4){\pdiamondb}
\end{pspicture}
\eea
The cancellation of the three omitted terms follows from the trigonometric identity between their weights
\be
[s_1(-u)s_0(-u)+s_0(u)s_1(u)+\beta\,s_0(u)s_0(-u)]\;
\begin{pspicture}[.45](0,0)(4,8)
\rput(0,4){\pdiamonda}
\end{pspicture}\;=\;0
\ee

The Yang-Baxter equations express the equality of two planar tangles
\psset{unit=.11in}
\setlength{\unitlength}{.11in}
\be
\phexbl u\punit{2}\phextl v\punit{6.5}\pdiamond{v\!-\!u}\punit4\;=\;
\pdiamond{v\!-\!u}\punit{2}\phexbr v\punit{2}\phextr u\punit8\mbox{}\\[2pt]
\ee
Setting $w=v-u$ and allowing for the five possible connections of the external nodes, this reduces to the diagrammatic equations\\[-2pt]
\psset{unit=.075in}
\setlength{\unitlength}{.075in}
\be
s_1(-u)s_1(-v)s_1(-w)\;\phexblb\punit{2}\phextla\punit{6.5}\pdiamondb\punit4\;=\;s_1(-u)s_1(-v)s_1(-w)\;
\pdiamondb\punit{2}\phexbra\punit{2}\phextrb\punit8\mbox{}\\[12pt]
\ee
\bea
&s_1(-u)s_0(v)s_1(-w)\;\phexblb\punit{2}\phextlb\punit{6.5}\pdiamondb\punit4\;=\;
s_0(u)s_1(-v)s_0(w)\;\pdiamonda\punit{2}\phexbra\punit{2}\phextra\punit8\;+
s_0(u)s_1(-v)s_1(-w)\;\pdiamondb\punit{2}\phexbra\punit{2}\phextra\punit8&\nonumber\\[12pt]
&+\;
s_1(-u)s_1(-v)s_0(w)\;\pdiamonda\punit{2}\phexbra\punit{2}\phextrb\punit8\;+\;
s_0(u)s_0(v)s_0(w)\;\pdiamonda\punit{2}\phexbrb\punit{2}\phextra\punit8&
\eea
The first equation, which is a trivial identity, occurs three times under rotations through $120^\circ$. The second equation occurs twice under rotations through $180^\circ$ and follows from the trigonometric identity
\bea
s_1(-u)s_0(v)s_1(-w)&=&\beta\, s_0(u)s_1(-v)s_0(w)+
s_0(u)s_1(-v)s_1(-w)\nonumber\\
&&\mbox{}+
s_1(-u)s_1(-v)s_0(w)+
s_0(u)s_0(v)s_0(w)
\eea

\subsection{Boundary Triangles and Boundary YBE}

To incorporate boundaries, we introduce 1-triangles. An elementary 1-triangle with no internal degrees of freedom is defined by
\bea
K(u)\;=\;K(u,\xi)\;=\;\ptri {u,\xi}\;=\;\ptriarc{}
\eea
where $\xi$ is a fixed boundary parameter which is often suppressed. The 1 in 1-triangle refers to the fact that there is 1 connectivity in and 1 connectivity out. This boundary condition, which we label by $(r,s)=(1,1)$, will play the role of the {\em vacuum} boundary condition. For given boundary 1-triangles, the Boundary Yang-Baxter Equations (BYBEs) express the equality of the two boundary tangles
\psset{unit=.085in}
\setlength{\unitlength}{.085in}
\def\lbybe#1#2#3#4{
\begin{pspicture}[.45](0,-1)(12,17)
\pspolygon[linewidth=.25pt](12,16)(0,4)(4,0)(12,8)(8,12)(4,8)(12,0)(12,16)
\rput(4,4){\small $#1$}
\rput(8,8){\small $#2$}
\rput(10.3,4){\small $#3$}
\rput(10.3,12){\small $#4$}
\psarc(0,4){.35}{-45}{45}
\psarc(4,8){.35}{-45}{45}
\psline[linestyle=dashed,dash=.5 .5,linewidth=.25pt](4,0)(12,0)
\end{pspicture}}
\def\rbybe#1#2#3#4{
\begin{pspicture}[.45](0,-1)(12,17)
\pspolygon[linewidth=.25pt](12,0)(0,12)(4,16)(12,8)(8,4)(4,8)(12,16)(12,0)
\rput(4,12){\small $#1$}
\rput(8,8){\small $#2$}
\rput(10.3,4){\small $#4$}
\rput(10.3,12){\small $#3$}
\psarc(0,12){.35}{-45}{45}
\psarc(4,8){.35}{-45}{45}
\psline[linestyle=dashed,dash=.5 .5,linewidth=.25pt](4,16)(12,16)
\end{pspicture}}
\bea
\lbybe{u\!-\!v}{\lambda\!-\!u\!-\!v}{u,\xi}{v,\xi}\ \ =\ \ \rbybe{u\!-\!v}{\lambda\!-\!u\!-\!v}{u,\xi}{v,\xi}
\eea
For the elementary 1-triangle, for example, this follows from the following four identities where 
$\omega_1=s_1(v-u)s_0(u+v)$, $\omega_2=s_0(u-v)s_0(u+v)$, $\omega_3=s_1(v-u)s_1(-u-v)$, 
$\omega_4=s_0(u-v)s_1(-u-v)$ and equality applies to connectivities as well as weights
\psset{unit=.06in}
\setlength{\unitlength}{.06in}
\bea
\begin{array}{c}
\omega_1\;
\begin{pspicture}[.45](0,-1)(12,17)
\rput(5.5,8){\lbybe{}{}{}{}}
\psarc(8,4){2.8}{-135}{135}
\psarc(8,12){2.8}{-135}{45}
\psarc(4,0){2.8}{45}{135}
\psarc(4,8){2.8}{-135}{-45}
\end{pspicture}
\ \ =\ \ 
\omega_1\;
\begin{pspicture}[.45](0,-1)(12,17)
\rput(5.5,8){\rbybe{}{}{}{}}
\psarc(8,12){2.8}{-135}{135}
\psarc(8,4){2.8}{-45}{135}
\psarc(4,16){2.8}{-135}{-45}
\psarc(4,8){2.8}{45}{135}
\end{pspicture}
\ \ ,\qquad\qquad
\omega_2\;
\begin{pspicture}[.45](0,-1)(12,17)
\rput(5.5,8){\lbybe{}{}{}{}}
\psarc(8,4){2.8}{0}{360}
\psarc(8,12){2.8}{-135}{45}
\psarc(0,4){2.8}{-45}{45}
\end{pspicture}
\ \ =\ \ 
\omega_2\;
\begin{pspicture}[.45](0,-1)(12,17)
\rput(5.5,8){\rbybe{}{}{}{}}
\psarc(8,12){2.8}{0}{360}
\psarc(8,4){2.8}{-45}{135}
\psarc(0,12){2.8}{-45}{45}
\end{pspicture}\\ \\
\omega_3\;
\begin{pspicture}[.45](0,-1)(12,17)
\rput(5.5,8){\lbybe{}{}{}{}}
\psarc(8,4){2.8}{-135}{45}
\psarc(8,12){2.8}{-45}{45}
\psarc(4,0){2.8}{45}{135}
\psarc(4,8){2.8}{-135}{45}
\psarc(12,8){2.8}{135}{225}
\end{pspicture}
\ \ =\ \ 
\omega_3\;
\begin{pspicture}[.45](0,-1)(12,17)
\rput(5.5,8){\rbybe{}{}{}{}}
\psarc(8,12){2.8}{-45}{135}
\psarc(8,4){2.8}{-45}{45}
\psarc(4,16){2.8}{-135}{-45}
\psarc(4,8){2.8}{-45}{135}
\psarc(12,8){2.8}{135}{225}
\end{pspicture}
\ \ ,\qquad\qquad
\omega_4\;
\begin{pspicture}[.45](0,-1)(12,17)
\rput(5.5,8){\lbybe{}{}{}{}}
\psarc(8,4){2.8}{-225}{45}
\psarc(8,12){2.8}{-45}{45}
\psarc(0,4){2.8}{-45}{45}
\psarc(4,8){2.8}{-45}{45}
\psarc(12,8){2.8}{135}{225}
\end{pspicture}
\ \ =\ \ 
\omega_4\;
\begin{pspicture}[.45](0,-1)(12,17)
\rput(5.5,8){\rbybe{}{}{}{}}
\psarc(8,12){2.8}{-45}{225}
\psarc(0,12){2.8}{-45}{45}
\psarc(4,8){2.8}{-45}{45}
\psarc(12,8){2.8}{135}{225}
\psarc(8,4){2.8}{-45}{45}
\end{pspicture}
\end{array}
\eea
Further solutions to the BYBEs are constructed in Section~3.4.

\subsection{Braids}

The planar TL algebra extends to a planar braid-monoid (tangle) algebra by adding braid 2-boxes.
The braid 2-boxes are defined by {\it braid limits} of the face operators
\psset{unit=.12in}
\setlength{\unitlength}{.12in}
\bea
b=k\lim_{u\to -i\infty} {X(u)\over s_1(-u)}\ =\ 
\begin{pspicture}[.45](0,0)(4,4)
\pspolygon[linewidth=.25pt](0,0)(4,0)(4,4)(0,4)
\psline[linewidth=1pt](0,2)(4,2)
\psline[linewidth=1pt](2,0)(2,1.5)
\psline[linewidth=1pt](2,2.5)(2,4)
\psarc(0,0){.35}{0}{90}
\end{pspicture}\;,
\qquad b^{-1}=k^{-1}\lim_{u\to i\infty} {X(u)\over s_1(-u)}\ =\ 
\begin{pspicture}[.45](0,0)(4,4)
\pspolygon[linewidth=.25pt](0,0)(4,0)(4,4)(0,4)
\psline[linewidth=1pt](2,0)(2,4)
\psline[linewidth=1pt](0,2)(1.5,2)
\psline[linewidth=1pt](2.5,2)(4,2)
\psarc(0,0){.35}{0}{90}
\end{pspicture}
\label{braidLimit}
\eea
Although the constant $k$ is arbitrary, the choice $k=-ie^{-i\lambda/2}=-i x^{-1/2}$ is compatible with crossing symmetry since then
\bea
b\;=\;-i\big(x^{-1/2}X(0) -x^{1/2} X(\lambda)\big),\qquad b^{-1}\;=\;-i\big(x^{-1/2}X(\lambda) -x^{1/2} X(0)\big)
\eea
and
\bea
X(u)\;=\;{x^{-1/2}e^{iu}b+x^{1/2}e^{-iu}b^{-1}\over i(x-x^{-1})},\qquad
X(\lambda-u)\;=\;{x^{-1/2}e^{iu}b^{-1}+x^{1/2}e^{-iu}b\over i(x-x^{-1})}
\label{sHalfAnsatz} 
\eea

\def\pbraida{
\begin{pspicture}[.45](0,0)(0,8)
\pspolygon[linewidth=.25pt](0,4)(2,0)(4,4)(2,8)
\psline[linewidth=1pt](1,6)(3,2)
\psline[linewidth=1pt](1,2)(1.75,3.5)
\psline[linewidth=1pt](2.25,4.5)(3,6)
\end{pspicture}}
\def\pbraidb{
\begin{pspicture}[.45](0,0)(0,8)
\pspolygon[linewidth=.25pt](0,4)(2,0)(4,4)(2,8)
\psline[linewidth=1pt](1,2)(3,6)
\psline[linewidth=1pt](1,6)(1.75,4.5)
\psline[linewidth=1pt](2.25,3.5)(3,2)
\end{pspicture}}
There are many relations that hold in the planar TL braid-monoid algebra. The braid relations
\be
\psset{unit=5mm}
\setlength{\unitlength}{5mm}
\begin{pspicture}[.475](4,4)
%\psgrid
\pspolygon[linewidth=.25pt](0,2)(1,4)(3,4)(2,2)(0,2)
\pspolygon[linewidth=.25pt](2,2)(3,4)(4,2)(3,0)(2,2)
\pspolygon[linewidth=.25pt](1,0)(0,2)(2,2)(3,0)(1,0)
\psarc(1,0){.15}{0}{119}
\psarc(0,2){.15}{0}{59}
\psarc(2,2){.15}{-59}{59}
\psline[linewidth=1pt](.5,1)(2.5,1)
\psline[linewidth=1pt](.5,3)(2.5,3)
\psline[linewidth=1pt](2.5,3)(3.5,1)
\psline[linewidth=1pt](1,2)(1.4,2.8)
\psline[linewidth=1pt](1.6,3.2)(2,4)
\psline[linewidth=1pt](2,0)(1.6,.8)
\psline[linewidth=1pt](1.4,1.2)(1,2)
\psline[linewidth=1pt](2.5,1)(2.9,1.8)
\psline[linewidth=1pt](3.1,2.2)(3.5,3)
\end{pspicture}\ =\ 
\begin{pspicture}[.475](4,4)
%\psgrid
\pspolygon[linewidth=.25pt](0,2)(1,4)(2,2)(1,0)(0,2)
\pspolygon[linewidth=.25pt](2,2)(1,4)(3,4)(4,2)(2,2)
\pspolygon[linewidth=.25pt](1,0)(2,2)(4,2)(3,0)(1,0)
\psarc(1,0){.15}{0}{59}
\psarc(0,2){.15}{-59}{59}
\psarc(2,2){.15}{0}{119}
\psline[linewidth=1pt](1.5,1)(3.5,1)
\psline[linewidth=1pt](1.5,3)(3.5,3)
\psline[linewidth=1pt](.5,3)(1.5,1)
\psline[linewidth=1pt](2,0)(2.4,.8)
\psline[linewidth=1pt](2.6,1.2)(3,2)
\psline[linewidth=1pt](3,2)(2.6,2.8)
\psline[linewidth=1pt](2.4,3.2)(2,4)
\psline[linewidth=1pt](.5,1)(.9,1.8)
\psline[linewidth=1pt](1.1,2.2)(1.5,3)
\end{pspicture}
\ee
for example, follow immediately by taking the limit $u,v,v-u\to \pm i\infty$ in the YBE. The inverse relation
\psset{unit=.075in}
\setlength{\unitlength}{.075in}
\bea
\begin{pspicture}[.45](0,0)(8,8)
\psline[linestyle=dashed,dash=.5 .5,linewidth=.25pt](2.4,0)(6.4,0)
\psline[linestyle=dashed,dash=.5 .5,linewidth=.25pt](2.4,8)(6.4,8)
\rput(0,4){\pbraida}
\rput(4,4){\pbraidb}
\psarc[linewidth=1pt](4.4,4){2.236}{63.4}{116.5}
\psarc[linewidth=1pt](4.4,4){2.236}{-116.5}{-63.4}
\end{pspicture}
\;=\;
\begin{pspicture}[.45](0,0)(4,8)
\rput(0,4){\pdiamondb}
\end{pspicture}
\label{planarinverse}
\eea
follows by taking $u\to i\infty$ in the inversion relation.
The twist relation is
\bea
\begin{pspicture}[.45](0,0)(8,8)
\psline[linestyle=dashed,dash=.5 .5,linewidth=.25pt](2.4,0)(6.4,0)
\psline[linestyle=dashed,dash=.5 .5,linewidth=.25pt](2.4,8)(6.4,8)
\rput(0,4){\pbraida}
\rput(4,4){\pdiamonda}
\psarc[linewidth=1pt](4.4,4){2.236}{63.4}{116.5}
\psarc[linewidth=1pt](4.4,4){2.236}{-116.5}{-63.4}
\end{pspicture}
\;=\;\omega\;
\begin{pspicture}[.45](0,0)(4,8)
\rput(0,4){\pdiamonda}
\end{pspicture}\;,\qquad\quad \omega=ix^{3/2}
\label{planartwist}
\eea
Another relation, which we will need later, is the rotated partner of (\ref{planarinverse}) with a spectator 1-triangle
\bea
\begin{pspicture}[.45](0,0)(8,8)
\psline[linewidth=.25pt](0,0)(4,0)
\psline[linewidth=.25pt](0,4)(4,4)
\psline[linewidth=.25pt](0,8)(4,8)
\psline[linewidth=.25pt](0,0)(0,8)
\psline[linewidth=.25pt](4,0)(4,8)
\psline[linewidth=.25pt](4,4)(8,8)
\psline[linewidth=.25pt](4,4)(8,0)
\psline[linewidth=.25pt](8,0)(8,8)
\psarc[linewidth=1pt](4,4){2}{-90}{90}
\psline[linewidth=1pt](0,2)(4,2)
\psline[linewidth=1pt](0,6)(4,6)
\psline[linewidth=1pt](2,0)(2,1.5)
\psline[linewidth=1pt](2,2.5)(2,5.5)
\psline[linewidth=1pt](2,6.5)(2,8)
\psline[linestyle=dashed,dash=.5 .5,linewidth=.25pt](4,0)(8,0)
\psline[linestyle=dashed,dash=.5 .5,linewidth=.25pt](4,8)(8,8)
\end{pspicture}\;=\;
\begin{pspicture}[.45](0,0)(8,8)
\psline[linewidth=.25pt](0,0)(4,0)
\psline[linewidth=.25pt](0,4)(4,4)
\psline[linewidth=.25pt](0,8)(4,8)
\psline[linewidth=.25pt](0,0)(0,8)
\psline[linewidth=.25pt](4,0)(4,8)
\psline[linewidth=.25pt](4,4)(8,8)
\psline[linewidth=.25pt](4,4)(8,0)
\psline[linewidth=.25pt](8,0)(8,8)
\psarc[linewidth=1pt](4,4){2}{-90}{90}
\psarc[linewidth=1pt](0,4){2}{-90}{90}
\psarc[linewidth=1pt](4,0){2}{90}{180}
\psarc[linewidth=1pt](4,8){2}{-180}{-90}
\psline[linestyle=dashed,dash=.5 .5,linewidth=.25pt](4,0)(8,0)
\psline[linestyle=dashed,dash=.5 .5,linewidth=.25pt](4,8)(8,8)
\end{pspicture}
%\ =\ \pdiamonda
\label{planarsbdy}
\eea

\subsection{Integrable and Conformal Boundary Conditions}

In this section, we start with the vacuum boundary condition and use the fusion construction of Behrend and Pearce~\cite{BP01} to build an infinite family of solutions to the BYBE labelled by Kac labels $(r,s)$ with $r,s=1,2,\ldots$. The $(r,s)$  integrable boundary condition leads to the $(r,s)$ conformal boundary condition in the continuum scaling limit. The construction process is valid in the planar Temperley-Lieb algebra.  
Since the arguments are formally the same as in \cite{BP01}, we just summarize the relevant results.

The $(r,s)$ solution is built in a two-stage process as the fusion product $(r,1)\otimes_f (1,s)$ of integrable seams acting on the vacuum $(1,1)$ 1-triangle. It is represented by the 1-triangle
\setlength{\unitlength}{15mm}
\psset{unit=15mm}
\bea
\raisebox{-1.4\unitlength}[1.6\unitlength][1.2\unitlength]
{\begin{pspicture}(4.5,2.0)
\psline[linewidth=2pt](.8,1)(1,1)
\psline[linewidth=2pt](.8,2)(1,2)
\psline[linewidth=2pt](-.55,1)(-.35,1)
\psline[linewidth=2pt](-.55,2)(-.35,2)
\psarc[linewidth=2pt](6,1.5){.5}{63}{90}
\psarc[linewidth=2pt](6,1.5){.5}{-90}{-63}
\put(0.45,1.5){\makebox(0,0)[]{$=$}}
\put(-.1,2.6){\spos{bc}{(r,s)}}
\put(3.5,2.6){\spos{bc}{(r,1)}}
\put(6.5,2.6){\spos{bc}{(1,s)}}
\multiput(-0.1,0.5)(6.6,0){2}{\line(0,1){2}}
\multiput(1,0.5)(1,0){4}{\line(0,1){2}}
\multiput(5,0.5)(1,0){2}{\line(0,1){2}}
\multiput(1,0.5)(0,1){3}{\line(1,0){5}}
\put(-0.6,1.5){\line(1,2){0.5}}
\put(-0.6,1.5){\line(1,-2){0.5}}
\put(6,1.5){\line(1,2){0.5}}
\put(6,1.5){\line(1,-2){0.5}}
\put(5,.5){\oval(.2,.2)[tr]}
\put(5,1.5){\oval(.2,.2)[tr]}
\multiput(1,.5)(1,0){2}{\oval(.2,.2)[tr]}
\multiput(1,1.5)(1,0){2}{\oval(.2,.2)[tr]}
\put(1.5,1){\spos{}{u\!-\!\xi_{r\!-\!1}}}
\put(2.5,1){\spos{}{u\!-\!\xi_{r\!-\!2}}}
\put(5.5,1){\spos{}{u\!-\!\xi_1}}
\put(1.5,2){\spos{}{-\!u\!-\!\xi_{r\!-\!2}}}
\put(2.5,2){\spos{}{-\!u\!-\!\xi_{r\!-\!3}}}
\put(5.5,2){\spos{}{-\!u\!-\!\xi_0}}
\put(-0.31,1.5){\spos{}{u,\xi}}
\put(6.3,1.5){\spos{}{\pm i\infty}}
\multiput(6,0.5)(0,2){2}{\makebox(0.5,0){\dotfill}}
\multiput(2,0.5)(1,0){2}{\spos{}{\bullet}}
\multiput(5,0.5)(1,0){1}{\spos{}{\bullet}}
\multiput(2,2.5)(1,0){2}{\spos{}{\bullet}}
\multiput(5,2.5)(1,0){1}{\spos{}{\bullet}}
\put(1,.35){$\underbrace{\hspace{5\unitlength}}_{\mbox{$r-1$ columns}}$}
\end{pspicture}}
\qquad\qquad\qquad\\[3pt]
\nonumber
\eea
Here there are $r-1$ double columns of faces, the column inhomogeneities are 
\bea
\xi_k=\xi+k\lambda
\label{xik}
\eea
The solid dots indicate that a projector $P^r$, defined below, is applied along the bottom (or equivalently top and bottom) edges of the right-hand side.
Any residual degrees of freedom (connectivities) on these edges are regarded as internal to the boundary. 

The projectors $P^r$, which act on the top and the bottom, are given by
\renewcommand{\text}[6]{\begin{picture}(#1,#2)\put(#3,#4){\p{#5}{\disp#6}}\end{picture}}
\renewcommand{\l}{\lambda}
\setlength{\unitlength}{6mm}
\psset{unit=6mm}
\bea
\label{PE}\raisebox{-5.75\unitlength}[5.75\unitlength][5.75\unitlength]{
\text{4.1}{11.5}{4.08}{6}{r}{P^r\ \ \propto\ \ }\,
\begin{pspicture}(6,11.5)\multiput(0,6)(1,1){2}{\line(1,-1){5}}
\put(2,8){\line(1,-1){4}}\put(3,9){\line(1,-1){3}}
\put(4,10){\line(1,-1){2}}\put(5,11){\line(1,-1){1}}
\multiput(0,6)(1,-1){2}{\line(1,1){5}}
\put(2,4){\line(1,1){4}}\put(3,3){\line(1,1){3}}
\put(4,2){\line(1,1){2}}\put(5,1){\line(1,1){1}}
\put(1,6){\pp{}{\!\mi(r\mi2)\l}}
\put(4,3){\pp{}{-2\l}}\put(4,9){\pp{}{-2\l}}
\put(5,2){\pp{}{-\l}}\put(5,4){\pp{}{-\l}}
\put(5,8){\pp{}{-\l}}\put(5,10){\pp{}{-\l}}
\multiput(0,0.5)(0,0.25){45}{\pp{}{.}}
\multiput(1,0.5)(0,0.25){18}{\pp{}{.}}
\multiput(1,7.25)(0,0.25){18}{\pp{}{.}}
\multiput(5,0.5)(0,0.25){2}{\pp{}{.}}
\multiput(5,11.25)(0,0.25){2}{\pp{}{.}}
\multiput(6,0.5)(0,0.25){45}{\pp{}{.}}
\psarc(5,1){.25}{45}{135}
\end{pspicture}
\text{3.38}{11.5}{2.37}{6}{r}{=\;}\,
\begin{pspicture}(6,11.5)\multiput(6,6)(-1,1){2}{\line(-1,-1){5}}
\put(4,8){\line(-1,-1){4}}\put(3,9){\line(-1,-1){3}}
\put(2,10){\line(-1,-1){2}}\put(1,11){\line(-1,-1){1}}
\multiput(6,6)(-1,-1){2}{\line(-1,1){5}}
\put(4,4){\line(-1,1){4}}\put(3,3){\line(-1,1){3}}
\put(2,2){\line(-1,1){2}}\put(1,1){\line(-1,1){1}}
\put(5,6){\pp{}{\!\mi(r\mi2)\l}}
\put(2,3){\pp{}{-2\l}}\put(2,9){\pp{}{-2\l}}
\put(1,2){\pp{}{-\l}}\put(1,4){\pp{}{-\l}}
\put(1,8){\pp{}{-\l}}\put(1,10){\pp{}{-\l}}
\multiput(0,0.5)(0,0.25){45}{\pp{}{.}}
\multiput(1,0.5)(0,0.25){2}{\pp{}{.}}
\multiput(1,11.25)(0,0.25){2}{\pp{}{.}}
\multiput(5,0.5)(0,0.25){18}{\pp{}{.}}
\multiput(5,7.25)(0,0.25){18}{\pp{}{.}}
\multiput(6,0.5)(0,0.25){45}{\pp{}{.}}
\psarc(1,1){.25}{45}{135}
\end{pspicture}}\qquad\qquad
\eea
These projectors are normalized to satisfy $(P^r)^2=P^r$ and act on $r-1$ strings to kill any diagram with closed half-arcs, that is, where any two of the $r-1$ strings are connected. For $\lambda/\pi$ irrational, there are an infinite number of projectors labelled by $r=1,2,3,\ldots$. However, for $\lambda/\pi$ rational, some projectors may diverge as is clear from (\ref{normalizedprojectors}). For the principal series, we restrict to $1\le r\le m$ ensuring the existence of the normalized projectors.

\psset{unit=.1in}
\setlength{\unitlength}{.1in}
For $1\le s\le m$, the $(1,s)$ solution is given by the braid limit of the $(s,1)$ solution
\bea
\begin{pspicture}[.45](0,-1)(4,9.5)
\pspolygon[linewidth=.25pt](0,4)(4,0)(4,8)(0,4)
\rput(2.3,4){\small $- i\infty$}
\rput(2.3,9){\small $(1,s)$}
\end{pspicture}&=&\lim_{\xi\to -i\infty}
\begin{pspicture}[.45](0,-1)(4,9.5)
\pspolygon[linewidth=.25pt](0,4)(4,0)(4,8)(0,4)
\rput(2.3,4){\small $u,\xi$}
\rput(2.3,9){\small $(s,1)$}
\end{pspicture}\ \ =\ \  
\begin{pspicture}[.45](-8,0)(8,8)
\rput(8,9){\small $(1,1)$}
\psline[linewidth=.25pt](-8,0)(4,0)
\psline[linewidth=.25pt](-8,4)(4,4)
\psline[linewidth=.25pt](-8,8)(4,8)
\psline[linewidth=.25pt](-8,0)(-8,8)
\psline[linewidth=.25pt](-4,0)(-4,8)
\psline[linewidth=.25pt](0,0)(0,8)
\psline[linewidth=.25pt](4,0)(4,8)
\psline[linewidth=.25pt](4,4)(8,8)
\psline[linewidth=.25pt](4,4)(8,0)
\psline[linewidth=.25pt](8,0)(8,8)
\psarc[linewidth=1pt](4,4){2}{-90}{90}
\psline[linewidth=1pt](-8,2)(4,2)
\psline[linewidth=1pt](-8,6)(4,6)
\psline[linewidth=1pt](2,0)(2,1.5)
\psline[linewidth=1pt](2,2.5)(2,5.5)
\psline[linewidth=1pt](2,6.5)(2,8)
\psline[linewidth=1pt](-2,0)(-2,1.5)
\psline[linewidth=1pt](-2,2.5)(-2,5.5)
\psline[linewidth=1pt](-2,6.5)(-2,8)
\psline[linewidth=1pt](-6,0)(-6,1.5)
\psline[linewidth=1pt](-6,2.5)(-6,5.5)
\psline[linewidth=1pt](-6,6.5)(-6,8)
\psline[linestyle=dashed,dash=.5 .5,linewidth=.25pt](4,0)(8,0)
\psline[linestyle=dashed,dash=.5 .5,linewidth=.25pt](4,8)(8,8)
\rput(-6,0){\spos{}{\bullet}}
\rput(-2,0){\spos{}{\bullet}}
\rput(2,0){\spos{}{\bullet}}
\rput(-6,8){\spos{}{\bullet}}
\rput(-2,8){\spos{}{\bullet}}
\rput(2,8){\spos{}{\bullet}}
\put(-8,-.5){$\underbrace{\hspace{12\unitlength}}_{\mbox{$s-1$ columns}}$}
\end{pspicture}\\
\nonumber
\eea
and a similar expression with underpasses replacing overpasses on the right side of the equation for  $\xi\to +i\infty$. The solid dots indicate that a projector $P^s$ is applied along the row. The limits exist provided the face operators are suitably normalized. 
Repeatedly applying (\ref{planarsbdy}) to either braid limit gives
\bea
\begin{pspicture}[.45](0,-1)(4,9.5)
\pspolygon[linewidth=.25pt](0,4)(4,0)(4,8)(0,4)
\rput(2.3,4){\small $\pm i\infty$}
\rput(2.3,9){\small $(1,s)$}
\end{pspicture}\ \ =\ \  
\begin{pspicture}[.45](-8,0)(8,8)
\psline[linewidth=.25pt](-8,0)(4,0)
\psline[linewidth=.25pt](-8,4)(4,4)
\psline[linewidth=.25pt](-8,8)(4,8)
\psline[linewidth=.25pt](-4,0)(-4,8)
\psline[linewidth=.25pt](-8,0)(-8,8)
\psline[linewidth=.25pt](0,0)(0,8)
\psline[linewidth=.25pt](4,0)(4,8)
\psline[linewidth=.25pt](4,4)(8,8)
\psline[linewidth=.25pt](4,4)(8,0)
\psline[linewidth=.25pt](8,0)(8,8)
\psarc[linewidth=1pt](4,4){2}{-90}{90}
\psarc[linewidth=1pt](0,4){2}{-90}{90}
\psarc[linewidth=1pt](4,0){2}{90}{180}
\psarc[linewidth=1pt](4,8){2}{-180}{-90}
\psarc[linewidth=1pt](-4,4){2}{-90}{90}
\psarc[linewidth=1pt](0,0){2}{90}{180}
\psarc[linewidth=1pt](0,8){2}{-180}{-90}
\psarc[linewidth=1pt](-8,4){2}{-90}{90}
\psarc[linewidth=1pt](-4,0){2}{90}{180}
\psarc[linewidth=1pt](-4,8){2}{-180}{-90}
\psline[linestyle=dashed,dash=.5 .5,linewidth=.25pt](4,0)(8,0)
\psline[linestyle=dashed,dash=.5 .5,linewidth=.25pt](4,8)(8,8)
\rput(-6,0){\spos{}{\bullet}}
\rput(-2,0){\spos{}{\bullet}}
\rput(2,0){\spos{}{\bullet}}
\rput(-6,8){\spos{}{\bullet}}
\rput(-2,8){\spos{}{\bullet}}
\rput(2,8){\spos{}{\bullet}}
\put(-8,-.5){$\underbrace{\hspace{12\unitlength}}_{\mbox{$s-1$ columns}}$}
\end{pspicture}\label{generalsbdy}\\
\nonumber
\eea
showing that the $s-1$ rightmost strings pass straight through the boundary tangle. For $s>m$, we define the $(1,s)$ boundary condition by the right side of (\ref{generalsbdy}) with $s-1$ columns with no projector but the action is restricted to the vector space of link states $\calV^{(s)}$ as explained in the next section. For $s\le m$, the two definitions are equivalent. The effect of the $(1,s)$ boundary condition is to close the $\ell=s-1$ defects on the right boundary as indicated in Figure~2 and discussed in the following.

\begin{figure}[thbp]
\psset{unit=1.1cm}
\setlength{\unitlength}{1.1cm}
\begin{center}
\begin{pspicture}(10,7)
%\psgrid
\pspolygon[fillstyle=solid,fillcolor=lightlightblue,linewidth=.25pt](0,0)(8,0)(8,4)(0,4)(0,0)
\pspolygon[fillstyle=solid,fillcolor=lightblue,linewidth=.25pt](8,0)(10,0)(10,4)(8,4)(8,0)
\rput(0,5){$\color{black}(r',s')=(1,1)$}
\rput(10.6,5){$\color{black}(r,s)=(1,3)$}
\rput[bl](8,0){\emptysquare}
\rput[bl](9,0){\emptysquare}
\rput[bl](8,1){\emptysquare}
\rput[bl](9,1){\emptysquare}
\rput[bl](8,2){\emptysquare}
\rput[bl](9,2){\emptysquare}
\rput[bl](8,3){\emptysquare}
\rput[bl](9,3){\emptysquare}
\psline[linecolor=blue,linestyle=dashed,dash=.25 .25,linewidth=2pt](0,4)(10,4)
\psline[linecolor=blue,linewidth=1.5pt](8.5,0)(8.5,4)
\psline[linecolor=blue,linewidth=1.5pt](9.5,0)(9.5,4)
\psline[linecolor=blue,linewidth=1.5pt](8,.5)(8.4,.5)
\psline[linecolor=blue,linewidth=1.5pt](8.6,.5)(9.4,.5)
\psline[linecolor=blue,linewidth=1.5pt](9.6,.5)(10,.5)
\psline[linecolor=blue,linewidth=1.5pt](8,1.5)(8.4,1.5)
\psline[linecolor=blue,linewidth=1.5pt](8.6,1.5)(9.4,1.5)
\psline[linecolor=blue,linewidth=1.5pt](9.6,1.5)(10,1.5)
\psline[linecolor=blue,linewidth=1.5pt](8,2.5)(8.4,2.5)
\psline[linecolor=blue,linewidth=1.5pt](8.6,2.5)(9.4,2.5)
\psline[linecolor=blue,linewidth=1.5pt](9.6,2.5)(10,2.5)
\psline[linecolor=blue,linewidth=1.5pt](8,3.5)(8.4,3.5)
\psline[linecolor=blue,linewidth=1.5pt](8.6,3.5)(9.4,3.5)
\psline[linecolor=blue,linewidth=1.5pt](9.6,3.5)(10,3.5)
\rput[bl](0,0){\loopb}
\rput[bl](1,0){\loopb}
\rput[bl](2,0){\loopa}
\rput[bl](3,0){\loopa}
\rput[bl](4,0){\loopa}
\rput[bl](5,0){\loopb}
\rput[bl](6,0){\loopa}
\rput[bl](7,0){\loopa}
\rput[bl](0,1){\loopa}
\rput[bl](1,1){\loopa}
\rput[bl](2,1){\loopa}
\rput[bl](3,1){\loopa}
\rput[bl](4,1){\loopa}
\rput[bl](5,1){\loopa}
\rput[bl](6,1){\loopb}
\rput[bl](7,1){\loopb}
\rput[bl](0,2){\loopb}
\rput[bl](1,2){\loopb}
\rput[bl](2,2){\loopb}
\rput[bl](3,2){\loopb}
\rput[bl](4,2){\loopb}
\rput[bl](5,2){\loopb}
\rput[bl](6,2){\loopb}
\rput[bl](7,2){\loopa}
\rput[bl](0,3){\loopa}
\rput[bl](1,3){\loopa}
\rput[bl](2,3){\loopa}
\rput[bl](3,3){\loopa}
\rput[bl](4,3){\loopa}
\rput[bl](5,3){\loopa}
\rput[bl](6,3){\loopa}
\rput[bl](7,3){\loopa}
\psarc[linecolor=blue,linewidth=1.5pt](0,1){.5}{90}{270}
\psarc[linecolor=blue,linewidth=1.5pt](0,3){.5}{90}{270}
\psarc[linecolor=blue,linewidth=1.5pt](10,1){.5}{-90}{90}
\psarc[linecolor=blue,linewidth=1.5pt](10,3){.5}{-90}{90}
\psarc[linecolor=purple,linewidth=1.5pt](1,4){.5}{0}{180}
\psarc[linecolor=purple,linewidth=1.5pt](7,4){.5}{0}{180}
\psarc[linecolor=purple,linewidth=1.5pt](4,4){.5}{0}{180}
\psarc[linecolor=purple,linewidth=1.5pt](7,4){1.5}{0}{180}
\psarc[linecolor=purple,linewidth=1.5pt](6,3.105){3.62}{14}{166}
\end{pspicture}
\caption{A typical configuration on the strip showing connectivities. The action on the link state is explained in the next section. The boundary condition is of type $(r',s')=(1,1)$ on the left and type $(r,s)=(1,3)$ on the right so there are $\ell=s\!-\!1=2$ defects in the bulk. The strings propagating along the right boundary are spectators connected to the defects.}
\end{center}
\end{figure}
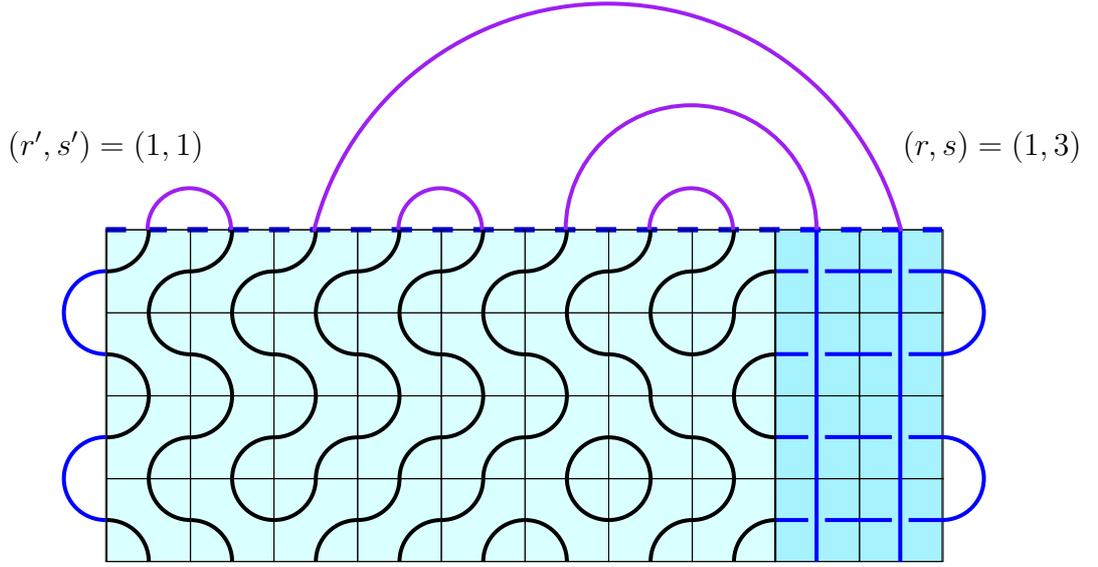

\subsection{Boundary Crossing}
\psset{unit=2.25mm}
\setlength{\unitlength}{2.25mm}
Since all closed half-arcs are projected out by the fusion projector, it follows that the normalized $(r,1)$ boundary tangle is
\bea
\begin{pspicture}[.45](0,-1)(4,9.5)
\pspolygon[linewidth=.25pt](0,4)(4,0)(4,8)(0,4)
\rput(2.3,4){\small $u,\xi$}
\rput(2.3,9){\small $(r,1)$}
\end{pspicture}&=&\ 
\begin{pspicture}[.45](-8,0)(8,8)
\psline[linewidth=.25pt](-8,0)(4,0)
\psline[linewidth=.25pt](-8,4)(4,4)
\psline[linewidth=.25pt](-8,8)(4,8)
\psline[linewidth=.25pt](-4,0)(-4,8)
\psline[linewidth=.25pt](-8,0)(-8,8)
\psline[linewidth=.25pt](0,0)(0,8)
\psline[linewidth=.25pt](4,0)(4,8)
\psline[linewidth=.25pt](4,4)(8,8)
\psline[linewidth=.25pt](4,4)(8,0)
\psline[linewidth=.25pt](8,0)(8,8)
\psarc[linewidth=1pt](4,4){2}{-90}{90}
\psarc[linewidth=1pt](0,4){2}{-90}{90}
\psarc[linewidth=1pt](4,0){2}{90}{180}
\psarc[linewidth=1pt](4,8){2}{-180}{-90}
\psarc[linewidth=1pt](-4,4){2}{-90}{90}
\psarc[linewidth=1pt](0,0){2}{90}{180}
\psarc[linewidth=1pt](0,8){2}{-180}{-90}
\psarc[linewidth=1pt](-8,4){2}{-90}{90}
\psarc[linewidth=1pt](-4,0){2}{90}{180}
\psarc[linewidth=1pt](-4,8){2}{-180}{-90}
\psline[linestyle=dashed,dash=.5 .5,linewidth=.25pt](4,0)(8,0)
\psline[linestyle=dashed,dash=.5 .5,linewidth=.25pt](4,8)(8,8)
\rput(-6,0){\spos{}{\bullet}}
\rput(-2,0){\spos{}{\bullet}}
\rput(2,0){\spos{}{\bullet}}
\rput(-6,8){\spos{}{\bullet}}
\rput(-2,8){\spos{}{\bullet}}
\rput(2,8){\spos{}{\bullet}}
\rput(-2.1,-1.9){$\underbrace{\hspace{12\unitlength}}_{\mbox{$r-1$ columns}}$}
\end{pspicture}
-{s_{r-1}(0)s_0(2u)\over s_0(\xi+u)s_r(\xi-u)}\ 
\begin{pspicture}[.45](-8,0)(8,8)
\psline[linewidth=.25pt](-8,0)(4,0)
\psline[linewidth=.25pt](-8,4)(4,4)
\psline[linewidth=.25pt](-8,8)(4,8)
\psline[linewidth=.25pt](-4,0)(-4,8)
\psline[linewidth=.25pt](-8,0)(-8,8)
\psline[linewidth=.25pt](0,0)(0,8)
\psline[linewidth=.25pt](4,0)(4,8)
\psline[linewidth=1pt](-2,0)(-2,8)
\psline[linewidth=1pt](2,0)(2,8)
\psline[linewidth=.25pt](4,4)(8,8)
\psline[linewidth=.25pt](4,4)(8,0)
\psline[linewidth=.25pt](8,0)(8,8)
\psarc[linewidth=1pt](-8,0){2}{0}{90}
\psarc[linewidth=1pt](-8,8){2}{-90}{0}
\psline[linestyle=dashed,dash=.5 .5,linewidth=.25pt](4,0)(8,0)
\psline[linestyle=dashed,dash=.5 .5,linewidth=.25pt](4,8)(8,8)
\rput(-6,0){\spos{}{\bullet}}
\rput(-2,0){\spos{}{\bullet}}
\rput(2,0){\spos{}{\bullet}}
\rput(-6,8){\spos{}{\bullet}}
\rput(-2,8){\spos{}{\bullet}}
\rput(2,8){\spos{}{\bullet}}
\rput(-2.1,-1.9){$\underbrace{\hspace{12\unitlength}}_{\mbox{$r-1$ columns}}$}
\end{pspicture}\qquad\mbox{}\label{explicitr1}\\[6pt]
\nonumber
\eea
This is a combination of equations (2.29) and (2.30) of \cite{BP01}. The closed loop which is implicitly present in the second term has been cancelled against a factor $\beta$ in the prefactor. 
It is noted that only the first term, which is an $s$-type boundary condition of type $(1,r)$, 
survives in the braid limit $\xi\to\pm i\infty$. 

The boundary crossing relation follows readily
\bea
&&\psset{unit=2mm}
{1\over s_0(2u)}\;
\begin{pspicture}[.45](0,-1)(8,9.5)
\pspolygon[linewidth=.25pt](0,4)(4,0)(8,4)(4,8)(0,4)
\rput(4,4){\small $2u-\lambda$}
\psarc(0,4){.6}{-45}{45}
\end{pspicture}
\begin{pspicture}[.45](0,-1)(4,9.5)
\pspolygon[linewidth=.25pt](0,4)(4,0)(4,8)(0,4)
\rput(2.3,4){\small $u,\xi$}
\rput(2.3,9){\small $(r,1)$}
\end{pspicture}\;=\;
\psset{unit=2mm}\Bigg({s_2(-2u)\over s_0(2u)}\;
\begin{pspicture}[.45](0,-1)(8,9.5)
\pspolygon[linewidth=.25pt](0,4)(4,0)(8,4)(4,8)(0,4)
\psarc(0,4){.6}{-45}{45}
\psarc[linewidth=1pt](4,0){2.82}{45}{135}
\psarc[linewidth=1pt](4,8){2.82}{-135}{-45}
\end{pspicture}
\;+\;{s_{-1}(2u)\over s_0(2u)}\;
\begin{pspicture}[.45](0,-1)(8,9.5)
\pspolygon[linewidth=.25pt](0,4)(4,0)(8,4)(4,8)(0,4)
\psarc(0,4){.6}{-45}{45}
\psarc[linewidth=1pt](0,4){2.82}{-45}{45}
\psarc[linewidth=1pt](8,4){2.82}{135}{225}
\end{pspicture}\Bigg)\;
\begin{pspicture}[.45](0,-1)(4,9.5)
\pspolygon[linewidth=.25pt](0,4)(4,0)(4,8)(0,4)
\rput(2.3,4){\small $u,\xi$}
\rput(2.3,9){\small $(r,1)$}
\end{pspicture}\nonumber
\eea
\bea
&=&\!s_1(\xi-u)s_{r-1}(\xi+u)\ 
\begin{pspicture}[.45](-8,0)(8,8)
\psline[linewidth=.25pt](-8,0)(4,0)
\psline[linewidth=.25pt](-8,4)(4,4)
\psline[linewidth=.25pt](-8,8)(4,8)
\psline[linewidth=.25pt](-4,0)(-4,8)
\psline[linewidth=.25pt](-8,0)(-8,8)
\psline[linewidth=.25pt](0,0)(0,8)
\psline[linewidth=.25pt](4,0)(4,8)
\psline[linewidth=.25pt](4,4)(8,8)
\psline[linewidth=.25pt](4,4)(8,0)
\psline[linewidth=.25pt](8,0)(8,8)
\psarc[linewidth=1pt](4,4){2}{-90}{90}
\psarc[linewidth=1pt](0,4){2}{-90}{90}
\psarc[linewidth=1pt](4,0){2}{90}{180}
\psarc[linewidth=1pt](4,8){2}{-180}{-90}
\psarc[linewidth=1pt](-4,4){2}{-90}{90}
\psarc[linewidth=1pt](0,0){2}{90}{180}
\psarc[linewidth=1pt](0,8){2}{-180}{-90}
\psarc[linewidth=1pt](-8,4){2}{-90}{90}
\psarc[linewidth=1pt](-4,0){2}{90}{180}
\psarc[linewidth=1pt](-4,8){2}{-180}{-90}
\psline[linestyle=dashed,dash=.5 .5,linewidth=.25pt](4,0)(8,0)
\psline[linestyle=dashed,dash=.5 .5,linewidth=.25pt](4,8)(8,8)
\rput(-6,0){\spos{}{\bullet}}
\rput(-2,0){\spos{}{\bullet}}
\rput(2,0){\spos{}{\bullet}}
\rput(-6,8){\spos{}{\bullet}}
\rput(-2,8){\spos{}{\bullet}}
\rput(2,8){\spos{}{\bullet}}
\rput(-2.1,-1.9){$\underbrace{\hspace{12\unitlength}}_{\mbox{$r-1$ columns}}$}
\end{pspicture}
-s_{r-1}(0)s_2(-2u)\;
\begin{pspicture}[.45](-8,0)(8,8)
\psline[linewidth=.25pt](-8,0)(4,0)
\psline[linewidth=.25pt](-8,4)(4,4)
\psline[linewidth=.25pt](-8,8)(4,8)
\psline[linewidth=.25pt](-4,0)(-4,8)
\psline[linewidth=.25pt](-8,0)(-8,8)
\psline[linewidth=.25pt](0,0)(0,8)
\psline[linewidth=.25pt](4,0)(4,8)
\psline[linewidth=1pt](-2,0)(-2,8)
\psline[linewidth=1pt](2,0)(2,8)
\psline[linewidth=.25pt](4,4)(8,8)
\psline[linewidth=.25pt](4,4)(8,0)
\psline[linewidth=.25pt](8,0)(8,8)
\psarc[linewidth=1pt](-8,0){2}{0}{90}
\psarc[linewidth=1pt](-8,8){2}{-90}{0}
\psline[linestyle=dashed,dash=.5 .5,linewidth=.25pt](4,0)(8,0)
\psline[linestyle=dashed,dash=.5 .5,linewidth=.25pt](4,8)(8,8)
\rput(-6,0){\spos{}{\bullet}}
\rput(-2,0){\spos{}{\bullet}}
\rput(2,0){\spos{}{\bullet}}
\rput(-6,8){\spos{}{\bullet}}
\rput(-2,8){\spos{}{\bullet}}
\rput(2,8){\spos{}{\bullet}}
\rput(-2.1,-1.9){$\underbrace{\hspace{12\unitlength}}_{\mbox{$r-1$ columns}}$}
\end{pspicture}\nonumber\\[20pt]
&=&\psset{unit=2mm}
\;
\begin{pspicture}[.45](0,-1)(4,9.5)
\pspolygon[linewidth=.25pt](0,4)(4,0)(4,8)(0,4)
\rput(2.3,4){\small $\;\;\;\lambda\!-\!u,\xi$}
\rput(2.3,9){\small $(r,1)$}
\end{pspicture}
\label{bdyCrossing}
\eea
Here we used the identities
\bea
s_2(-2u)+\beta s_{-1}(2u)&=&s_0(2u)\\
s_0(\xi+u)s_r(\xi-u)-s_{r-1}(0)s_{-1}(2u)&=&s_1(\xi-u)s_{r-1}(\xi+u)
\eea

\setlength{\unitlength}{.08in}
\section{Linear Temperley-Lieb Algebra}

The linear Temperley-Lieb algebra~\cite{TempLieb,MartinBook} ${\cal T}={\cal T}(n,\lambda)$, with
$n\in{\Bbb Z}_{\ge0}$ and $\lambda\in{\Bbb R}$, is obtained by fixing the in- and out-states (or direction of transfer) of the planar TL algebra. The linear TL algebra thus acts on a fixed (distinguished) vector space and 
is generated by the identity $I$ and
the operators $e_1,\ldots,e_{n-1}$ satisfying for $j,k=1,2,\ldots,n\!-\!1$
\begin{eqnarray}
\begin{array}{ll}
e_j^2\;=\;\beta\:e_j,\quad &\beta=2\cos\!\lambda\\[4pt]
e_j\:e_k\:e_j\;=\;e_j,\quad &|j\!-\!k|=1\\[4pt]
e_j\:e_k\;=\;e_k\:e_j,\quad &|j\!-\!k|>1
\end{array}
\label{TL}
\end{eqnarray}
Here we work in the dense loop representation of the TL algebra
and represent the TL generators $e_j$ graphically by monoids~\cite{Kauf} acting on
$n$ strings
\bea
e_j\;=\punit1\vert 0{-2}1\vert 4{-2}2\punit6\ldots\punit2
  \vert 0{-2}{j\!-\!1}\monoid 4{-2}{j}{j\!+\!1}
  \vert {12}{-2}{j\!+\!2}\punit{14}\ldots\punit2\vert 0{-2}{n\!-\!1}\vert 4{-2}{n
\rule{0pt}{7pt}}\punit8
\eea
\bea
e_j^2\;=\;
\monoid 0{-4}{j}{j\!+\!1}\monoid 00{}{}\punit5=\punit1\ \beta \ \;\monoid 0{-2}{j}{j\!+\!1}\punit4
\;=\;\beta\,e_j
\eea
\bea
\rule[-7\unitlength]{0in}{14\unitlength}
e_je_{j+1}e_j\;=\;
\monoid 0{-6}{j}{j\!+\!1}\vert 0{-2}{}\monoid 02{}{}\punit4
\monoid 0{-2}{}{}\punit4\vert 0{-6}{j\!+\!2}\vert 02{}\punit2=\punit2
\vert 0{-6}{j}\monoid 0{-2}{}{}\vert 02{}\punit4\vert 0{-6}{j\!+\!1}\vert 02{}\punit4
\vert 0{-6}{j\!+\!2}\vert 0{-2}{}\vert 02{}\punit1
\;=\;e_j\\[-4pt]
\nonumber
\eea
For $\beta\ne 0$, $\beta^{-1}e_j$ and $I-\beta^{-1}e_j$ are orthogonal projectors. 

The number $C_n$ of independent words
$w\in {\cal T}(n,\lambda)$  is given by the Catalan numbers 
\begin{equation}
C_n={2n\choose n}-{2n\choose n\!-\!1}
={1\over n\!+\!1}{2n\choose n}=1,2,5,14,\ldots; \quad  n=1,2,3,4,\ldots
\end{equation}
The words of the linear TL algebra are divided into equivalence classes by the
number of strings or defects $\ell$ passing from the bottom to the top of the monoid
diagrams. For $n=4$, for example, we have ${\cal T}={\cal S}_0\cup{\cal S}_2\cup{\cal S}_4$ with
\begin{eqnarray}
\begin{array}{ll}
\ell=0:\quad&{\cal S}_0=\{e_1e_3,e_1e_3e_2,e_2e_1e_3,e_2e_1e_3e_2\}\\[6pt]
\ell=2:&{\cal S}_2=\{e_1,e_2,e_3,e_1e_2,e_2e_1,e_2e_3,e_3e_2,e_1e_2e_3,e_3e_2e_1\}\\[6pt]
\ell=4:&{\cal S}_4=\{I\}
\end{array}
\end{eqnarray}
Under the action of the generators of the TL algebra, the defects can hop by
two sites ($e_1\mapsto e_2e_1$ for example) or adjacent defects can be annihilated in pairs 
($e_3\mapsto e_1e_3$ for example). It follows that the action of the TL algebra
is block triangular on the classes ${\cal S}_\ell$ and that ${\cal T}={\cal T}(n,\lambda)$ admits
the subalgebras
\begin{equation}
{\cal T}_\ell=
\bigcup_{\mbox{\scriptsize $\ell'\le \ell$, $\ell-\ell'=0$ mod $2$}} {\cal S}_{\ell'},
\qquad 0\le \ell\le n
\end{equation}

The linear TL algebra ${\cal T}(n,\lambda)$ is semisimple for $\lambda/\pi$ irrational and not semisimple for $\lambda/\pi$ rational~\cite{MartinBook,GoodmanWenzl}. This means that Hamiltonians and transfer matrices constructed from the generators of this algebra are necessarily diagonalizable for $\lambda/\pi$ irrational. To see indecomposable representations, we therefore need to consider the case where $\lambda/\pi$ is rational. This general structure theorem tells us that indecomposable representations {\em exist\/} but gives no hint as to  how to construct them or relate them to boundary conditions of a physical system.

\subsection{Link Diagrams}
\setlength{\unitlength}{.04in}
The fixed vector space of states of the linear TL algebra is described by connectivities. However, arbitrary
connectivities cannot occur. Referring to the top edge of Figure~1, it is seen that
connectivity in neighbouring pairs
$\punit2\downmonoid00{}{}\punit8\downmonoid00{}{}\punit8\downmonoid00{}{}
\punit6\ldots\ $ is always allowed. This distinguished connectivity state will play the
role of {\it vacuum} for our theories. Other allowed connectivities
are generated by the action of the TL algebra on the vacuum state and are
described algebraically by right ideals and diagrammatically by planar link diagrams.

Let us consider the action of the TL algebra on the vector space of right
ideals
\begin{equation}
{\cal V}={\cal V}(n,\lambda)=\langle w{\cal T}: w\in{\cal T}
={\cal T}(n,\lambda)\rangle,\qquad  w{\cal T}=\{w t: t\in{\cal T}\}
\end{equation}
where $\langle\ldots\rangle$ denotes the linear span. In the loop representation, each right ideal admits a graphical representation as
a (planar) link diagram. For $n=4$, for example, there are six right ideals organized by the number of defects $\ell$:
\begin{eqnarray}
\begin{array}{ll}
\rule{0pt}{16pt}
\ell=0:\quad&e_2e_1e_3{\cal T}=\;\{e_2e_1e_3,e_2e_1e_3e_2\}\punit0=\punit6
\down2monoid 00{1}{4}\punit8\downmonoid {-8}0{2}{3}\\[8pt]
\ell=0:&\phantom{e_2}e_1e_3{\cal T}=\;\{e_1e_3,e_1e_3e_2\}\punit0=\punit2
\downmonoid00{1}{2}\punit8\downmonoid00{3}{4}\\[8pt]
\ell=2:&\phantom{e_2e_3}e_1{\cal T}
=\;\{e_1,e_1e_2,e_1e_2e_3,e_1e_3,e_1e_3e_2\}\punit0=\punit2
\downmonoid00{1}{2}\punit8\svert0{0}{3}\punit4\svert0{0}{4}\\[8pt]
\ell=2:&\phantom{e_1e_3}e_2{\cal T}=
\;\{e_2,e_2e_1,e_2e_3,e_2e_1e_3,e_2e_1e_3e_2\}\punit0=\punit2
\svert001\punit4\downmonoid00{2}{3}\punit8\svert004\\[8pt]
\ell=2:&\phantom{e_1e_2}e_3{\cal T}
=\;\{e_3,e_3e_2,e_1e_3,e_3e_2e_1,e_1e_3e_2\}\punit0=\punit2
\svert001\punit4\svert002\punit4\downmonoid00{3}{4}\\[8pt]
\ell=4:&\phantom{e_2\,e_3}{I\,\cal T}
=\;{\cal T}\punit0=\punit2
\svert001\punit4\svert002\punit4\svert003\punit4\svert004\punit4
\end{array}
\end{eqnarray}
The TL generators act on these link diagrams from below. We denote by ${\cal V}_\ell$ the vector space of right
ideals with exactly $\ell=s-1$ defects. Defects occur with a fixed parity given by $n-\ell=0$ \mbox{mod $2$}. Since defects can be annihilated in pairs but not created by the TL generators, the action of the TL generators is upper block triangular on the vector spaces ${\cal V}_\ell$. The dimension of the space ${\cal V}_\ell$ is 
\be
\mbox{dim}\,{\cal V}_\ell=\chi_{1,s}^{(n)}(1)=\binom n{{n-\ell\over 2}}-\binom n{{n-\ell-2\over 2}}
\ee

It is often convenient to encode the right ideals by Restricted
Solid-On-Solid (RSOS or Dyck) paths
$\ket a=(a_0,a_1,\ldots,a_n)$ where $a_0=a_n=0$ and $a_j$ is the number of
half-loops above the midpoint between strings $j$ and $j\!+\!1$ and
$|a_{j+1}-a_j|=1$ for each $j=0,1,\ldots,n-1$. For $n=6$, we have
\begin{eqnarray}
\mbox{}\hspace{-12pt}
{\cal V}_0=\mbox{\small$\langle(0,1,2,3,2,1,0),(0,1,2,1,2,1,0),
(0,1,0,1,2,1,0),(0,1,2,1,0,1,0),(0,1,0,1,0,1,0)\rangle$}
\label{V0basis}
\end{eqnarray}
The action of a single TL generator $e_j$ maps one right ideal into a scalar
multiple of another right ideal. It is instructive to write down matrices representing the
action of TL generators on the basis of right ideals. For 
$n=6$, the action of $e_1$ and $e_2$ on ${\cal V}_0$, for example, is given by
\begin{equation}
e_1=\mbox{$\small\left(\begin{array}{ccccc}
0&0&0&0&0\\
0&0&0&0&0\\
0&1&\beta&0&0\\
0&0&0&0&0\\
1&0&0&1&\beta
\end{array}\right)$},
\qquad 
e_2=\mbox{$\small\left(\begin{array}{ccccc}
0&0&0&0&0\\
1&\beta&1&0&0\\
0&0&0&0&0\\
0&0&0&\beta&1\\
0&0&0&0&0
\end{array}\right)$}
\end{equation}
where the order of the basis is as given in (\ref{V0basis}). In general, these matrices
are real but not symmetric with eigenvalues $0$ or $\beta$ so that $\beta^{-1}e_j$ are projectors for $\beta\ne 0$. We stress that these matrices are {\it non-local\/} in the sense that the action on all link states must be considered to write the matrix representative of a given $e_j$. Despite the graphical depiction of the TL generators, these matrices are not (time-reversal) symmetric. This results from the action on link states which encode the history from time $-\infty$ and explicitly breaks the time-reversal symmetry associated with local representations of TL. 

Later it will be useful to restrict the action of the TL generators onto {\it spin-$(s-1)/2$ subspaces} defined by
\bea
\calV^{(s)}=\big\langle\, |a\rangle\in\calV_0:
\{a_0,a_1,\ldots,a_n\}=\{\ldots,s\!-\!1,s\!-\!2,\ldots,1,0\}\big\rangle,\quad s=1,2,3,\ldots
\label{sStates}
\eea
where precisely the last $s$ heights are fixed and $s-1=\ell$ is the number of defects.

\subsection{Face Operators and Local Relations}

\setlength{\unitlength}{6mm}
\psset{unit=6mm}
A solution of
the YBE~\cite{BaxBook} is obtained by taking the local face operators of the planar TL algebra and fixing the direction of transfer
\begin{equation}
X_j(u)\;=\raisebox{-1.75\unitlength}{
\begin{pspicture}(2,3.5)
\multiput(0,2)(1,-1){2}{\line(1,1){1}}
\multiput(0,2)(1,1){2}{\line(1,-1){1}}
%\put(.8,1.2){\line(1,0){.4}}
\psarc(1,1){.2}{45}{135}
\put(1,2){\pp{}{u}}
\multiput(0,0.5)(0,0.25){13}{\pp{}{.}}
\multiput(1,0.5)(0,0.25){2}{\pp{}{.}}
\multiput(1,3.25)(0,0.25){2}{\pp{}{.}}
\multiput(2,0.5)(0,0.25){13}{\pp{}{.}}
\put(0,0){\pp{b}{j-1}}
\put(1,0){\pp{b}{j}}
\put(2,0){\pp{b}{j+1}}
\end{pspicture}}
\;=\;s_1(-u)\:I\;+\;s_0(u)\:e_j
\end{equation}
These operators act from below on the fixed vector space ${\cal V}(n,\lambda)$ between string $j$ and $j+1$. It follows that the $X_j(u)$ satisfy the
operator form of the YBE
\bea
X_j(u)\,X_{j+1}(u\pl v)\,X_j(v)\;=\;X_{j+1}(v)\,X_j(u\pl v)\,X_{j+1}(u)
\eea
depicted graphically by
\begin{equation}
\begin{array}{r@{}c@{}l}
\begin{pspicture}(3,6.3)
\multiput(0,2)(1,-1){2}{\line(1,1){2}}
\multiput(0,4)(1,1){2}{\line(1,-1){2}}
\psarc(1,1){.2}{45}{135}
\psarc(1,3){.2}{45}{135}
\psarc(2,2){.2}{45}{135}
\put(0,2){\line(1,-1){1}}\put(0,4){\line(1,1){1}}
\put(1,2){\pp{}{u}}\put(2,3){\pp{}{u+v}}\put(1,4){\pp{}{v}}
\multiput(0,0.5)(0,0.25){21}{\pp{}{.}}
\multiput(1,0.5)(0,0.25){2}{\pp{}{.}}
\multiput(1,5.25)(0,0.25){2}{\pp{}{.}}
\multiput(2,0.5)(0,0.25){6}{\pp{}{.}}
\multiput(2,4.25)(0,0.25){6}{\pp{}{.}}
\multiput(3,0.5)(0,0.25){21}{\pp{}{.}}
\put(0,0){\pp{b}{j-1}}
\put(1,0){\pp{b}{j}}
\put(2,0){\pp{b}{j+1}}
\put(3,0){\pp{b}{j+2}}
\end{pspicture}
&\ptext{2.6}{5.5}{1.3}{3}{}{=}&
\begin{pspicture}(3,5.5)
\multiput(0,3)(1,1){2}{\line(1,-1){2}}
\multiput(0,3)(1,-1){2}{\line(1,1){2}}
\psarc(2,1){.2}{45}{135}
\psarc(2,3){.2}{45}{135}
\psarc(1,2){.2}{45}{135}
\put(2,1){\line(1,1){1}}\put(2,5){\line(1,-1){1}}
\put(2,2){\pp{}{v}}\put(1,3){\pp{}{u+v}}
\put(2,4){\pp{}{u}}
\multiput(0,0.5)(0,0.25){21}{\pp{}{.}}
\multiput(1,0.5)(0,0.25){6}{\pp{}{.}}
\multiput(1,4.25)(0,0.25){6}{\pp{}{.}}
\multiput(2,0.5)(0,0.25){2}{\pp{}{.}}
\multiput(2,5.25)(0,0.25){2}{\pp{}{.}}
\multiput(3,0.5)(0,0.25){21}{\pp{}{.}}
\put(0,0){\pp{b}{j-1}}
\put(1,0){\pp{b}{j}}
\put(2,0){\pp{b}{j+1}}
\put(3,0){\pp{b}{j+2}}
\end{pspicture}
\end{array}\qquad
\label{TLYBE}
\end{equation}
The local face operators also satisfy the single-site commutation relation
\begin{equation}
X_j(u)\:X_j(v)\;=\;s_1(-u)s_1(-v)I+s_0(u+v)e_j\;=\;X_j(v)\:X_j(u)
\label{Xjcommute}
\end{equation}
and hence the inversion relation
\begin{equation}
X_j(-u)\:X_j(u)\;=\;s_1(-u)\:s_1(u)\:I
\label{inversion}
\end{equation}

The triangle
boundary weights on the right 
\setlength{\unitlength}{10mm}
\bea
K_j(u,\xi)\;=\;\raisebox{-1.85\unitlength}{
\bpic(2,3.2)
\multiput(0,2)(1,-1){1}{\line(1,1){1}}
\multiput(0,2)(1,1){1}{\line(1,-1){1}}
\put(1,1){\line(0,1){2}}
\put(0.6,2){\spos{c}{u,\xi}}
\multiput(0,0.75)(0,0.15){18}{\pp{}{.}}
\multiput(1,0.75)(0,0.15){18}{\pp{}{.}}
\put(0,0.3){\pp{b}{j-1}}
\put(1.0,0.3){\pp{b}{j}}
\epic}\\[-26pt]\nonumber
\eea
must satisfy the Boundary
Yang-Baxter Equation (BYBE)
\bea
X_j(u\mi v)K_{j+1}(u,\xi)X_j(u\pl v)K_{j+1}(v,\xi)=
K_{j+1}(v,\xi)X_j(u\pl v)K_{j+1}(u,\xi)X_j(u\mi v)
\label{BYBE}
\eea
depicted graphically by
\setlength{\unitlength}{18mm}
\psset{unit=18mm}
\begin{eqnarray}
\rule{0pt}{1.0in}
\raisebox{-\unitlength}{
\begin{pspicture}(2.5,2)
\put(0.5,0.5){\botrightrefnodots{u,\xi}{v,\xi}{u\!-\!v}{u\!+\!v}}
\multiput(1,0)(0,0.1){25}{\pp{}{.}}
\multiput(2,0)(0,0.1){6}{\pp{}{.}}
\multiput(1.5,2)(0,0.1){5}{\pp{}{.}}
\put(1,-0.05){\spos{t}{j\!-\!1}}
\put(1.5,-0.05){\spos{t}{j}}
\put(2,-0.05){\spos{t}{j\!+\!1}}
\psarc(1.5,0){.075}{45}{135}
\psarc(1.5,1){.075}{45}{135}
\end{pspicture}
\ptext{0.4}{2}{0.5}{1.25}{}{=}
\begin{pspicture}(2.5,2)
\put(0.5,0){\toprightrefnodots{v,\xi}{u,\xi}{u\!+\!v}{u\!-\!v}}
\multiput(1,0)(0,0.1){25}{\pp{}{.}}
\multiput(2,2)(0,0.1){6}{\pp{}{.}}
\multiput(1.5,0)(0,0.1){5}{\pp{}{.}}
\put(1,-0.05){\spos{t}{j\!-\!1}}
\put(1.5,-0.05){\spos{t}{j}}
\put(2,-0.05){\spos{t}{j\!+\!1}}
\psarc(1.5,0.5){.075}{45}{135}
\psarc(1.5,1.5){.075}{45}{135}
\end{pspicture}}\qquad\quad\\[0pt]\nonumber
\end{eqnarray}
A similar relation holds on the left boundary. Here $\xi\in{\Bbb C}$ is an arbitrary parameter. Physically, $\xi$ is a thermodynamic
variable governing the boundary interactions. More specifically, it is a generalized boundary magnetic field. From the single-site commutation relation (\ref{Xjcommute}), it follows
immediately that a fundamental solution of the BYBE (\ref{BYBE}) is given by
\begin{equation}
K_{j+1}(u)=I
\end{equation}
It is the {\it vacuum} solution which is labelled by $(r,s)=(1,1)$. 

Explicitly, in the linear TL algebra, the first few normalized fusion projectors are
\be
P^1_j=P^2_j=I,\quad P^3_j=I-\frac{1}{s_2(0)}\,e_j,\quad
P^4_j=I-\frac{s_2(0)}{s_3(0)}\,(e_j+e_{j+1})+
\frac{1}{s_3(0)}\,(e_je_{j+1}+e_{j+1}e_j)
\label{normalizedprojectors}
\ee
It is noted that the denominators vanish for certain rational values of $\lambda/\pi$. However, this does not occur for the principal series with $\lambda=\pi/(m+1)$ in the cases under consideration where $r=1,2,\ldots, m$.
Using the TL algebra and the fusion projectors, it is possible~\cite{BP01} to systematically
build further solutions  $K^{(r,s)}_j$ of the BYBE labelled by arbitrary Kac labels
$(r,s)$ with $r,s=1,2,\ldots$
\bea
K^{(r,s)}_{j+1}(u,\xi)\ =\ P^r_{j+2} P^s_{j+r+1} 
\prod_{k=1}^{r-1}X_{j+k}(u-\xi_{r-k})\prod_{\ell=1}^{s-1}X_{j+\ell+r-1}(-i\infty)\nonumber\\
\times\prod_{\ell=s-1}^{1}X_{j+\ell+r-1}(i\infty)\prod_{k=r-1}^{1}X_{j+k}(u+\xi_{r-k})
P^r_{j+2} P^s_{j+r+1}
\label{Krs}
\eea
depicted graphically by
\psset{unit=.9cm}
\setlength{\unitlength}{.9cm}
\bea
K^{(r,s)}_{j+1}(u,\xi)\ =\ \raisebox{-1.85\unitlength}{
\bpic(2,3.0)
\multiput(0,2)(1,-1){1}{\line(1,1){1}}
\multiput(0,2)(1,1){1}{\line(1,-1){1}}
\put(1,1){\line(0,1){2}}
\put(0.6,2){\spos{c}{u,\xi}}
\multiput(0,0.75)(0,0.15){18}{\pp{}{.}}
\multiput(1,0.75)(0,0.15){18}{\pp{}{.}}
\put(0,0.3){\pp{b}{j}}
\put(1.0,0.3){\pp{b}{j+1}}
\put(0,3.25){$(r,s)$}
\epic}\hspace{-.2in}=\;
\begin{pspicture}[.5](0,-1)(6,10.5)
%\psgrid
\psline[linewidth=.5pt](0,1)(1,0)
\psline[linewidth=.5pt](0,1)(4,5)
\psline[linewidth=.5pt](1,0)(5,4)
\psline[linewidth=.5pt](4,3)(3,4)
\psline[linewidth=.5pt](1,2)(2,1)
\psline[linewidth=.5pt](2,3)(3,2)
\psline[linewidth=.5pt](4,3)(3,4)
\psline[linewidth=.5pt](4,5)(5,4)
\psline[linewidth=.5pt](4,5)(0,9)
\psline[linewidth=.5pt](5,6)(1,10)
\psline[linewidth=.5pt](5,4)(5,6)
\psline[linewidth=.5pt](2,3)(3,2)
\psline[linewidth=.5pt](4,5)(5,6)
\psline[linewidth=.5pt](3,6)(4,7)
\psline[linewidth=.5pt](2,7)(3,8)
\psline[linewidth=.5pt](1,8)(2,9)
\psline[linewidth=.5pt](0,9)(1,10)
\rput(1,1){$u\!-\!\xi_{r-1}$}
\rput(2,2){$u\!-\!\xi_{r-2}$}
\rput(4,4){$u\!-\!\xi_1$}
\rput(4,6){$u\!+\!\xi_1$}
\rput(2,8){$u\!+\!\xi_{r-2}$}
\rput(1,9){$u\!+\!\xi_{r-1}$}
\rput(4.65,5){$i\infty$}
\rput(5.7,5){$(1,s)$}
\psarc(1,0){.15}{45}{135}
\psarc(2,1){.15}{45}{135}
\psarc(4,3){.15}{45}{135}
\psarc(4,5){.15}{45}{135}
\psarc(2,7){.15}{45}{135}
\psarc(1,8){.15}{45}{135}
\psline[linewidth=1pt](.5,-.5)(.5,.5)
\psline[linewidth=1pt](.5,1.5)(.5,8.5)
\psline[linewidth=1pt](1.5,2.5)(1.5,7.5)
\psline[linewidth=1pt](2.5,3.5)(2.5,6.5)
\psline[linewidth=1pt](3.5,4.5)(3.5,5.5)
\psline[linewidth=1pt](.5,9.5)(.5,10.5)
\psline[linewidth=1pt](1.5,-.5)(1.5,.5)
\psline[linewidth=1pt](2.5,-.5)(2.5,1.5)
\psline[linewidth=1pt](3.5,-.5)(3.5,2.5)
\psline[linewidth=1pt](4.5,-.5)(4.5,3.5)
\psline[linewidth=1pt](1.5,9.5)(1.5,10.5)
\psline[linewidth=1pt](2.5,8.5)(2.5,10.5)
\psline[linewidth=1pt](3.5,7.5)(3.5,10.5)
\psline[linewidth=1pt](4.5,6.5)(4.5,10.5)
\rput(0,-.75){$j$}
\rput(1,-.75){$j\!+\!1$}
\rput(2,-.75){$j\!+\!2$}
\rput(3,-.75){$\ldots$}
\rput(5.3,-.75){$j\!+\!r$}
\multiput(1.5,.5)(1,1){4}{\spos{}{\bullet}}
\multiput(1.5,9.5)(1,-1){4}{\spos{}{\bullet}}
\end{pspicture}
\label{Krsgraph}
\eea
where $\xi_k=\xi+k\lambda$ and the solid dots indicate the action of the fusion projectors. The $(1,s)$ boundary triangle $K_{j+1}^{(1,s)}(\xi=i\infty)$ occurring on the right side of (\ref{Krsgraph}) has itself a similar graphical depiction but with $\xi_k=\pm i\infty$, $r$ replaced by $s$ and the boundary triangle omitted or equivalently acting as the identity. As in Section~3.4, for $s>m$, the fusion projectors are omitted and the action is simply restricted to the vector space of link states $\calV^{(s)}$.

After suitable normalization, it follows from (\ref{explicitr1}) that the boundary weights are given in terms of projectors by
\bea
K^{(r,s)}_{j}(u,\xi)\;=\; P_{j+1}^rP_{j+r}^s-
{s_{r-1}(0)s_0(2u)\over s_0(\xi+u)s_r(\xi-u)}\,P_{j+1}^re_jP_{j+1}^rP_{j+r}^s
\eea
{}From the recursive definition of the projectors~\cite{BP01}, it also follows that at $u=\xi$
\bea
K^{(r,s)}_{j}(\xi,\xi)\;=\;P_{j}^{r+1}P_{j+r}^s
\eea

\section{Commuting Transfer Matrices and Hamiltonians}

\subsection{Double-Row Transfer Matrices on a Strip}

The YBEs, supplemented by the additional local relations, are sufficient to imply
commuting transfer matrices and integrability.  To work on a strip with fixed boundary conditions on the
right and left, we need to work with $N$ column Double-row Transfer Matrices
(DTMs)~\cite{Sklyanin,BPO96} represented schematically in the planar TL algebra by the $N$-tangle
\setlength{\unitlength}{13mm}
\psset{unit=13mm}
\bea
\vec D(u)\;
=\quad
\raisebox{-1.3\unitlength}[1.3\unitlength][
1.1\unitlength]{\begin{pspicture}(6.4,2.4)(0.4,0.1)
\multiput(0.5,0.5)(6,0){2}{\line(0,1){2}}
\multiput(1,0.5)(1,0){3}{\line(0,1){2}}
\multiput(5,0.5)(1,0){2}{\line(0,1){2}}
\multiput(1,0.5)(0,1){3}{\line(1,0){5}}
\put(1,1.5){\line(-1,2){0.5}}\put(1,1.5){\line(-1,-2){0.5}}
\put(6,1.5){\line(1,2){0.5}}\put(6,1.5){\line(1,-2){0.5}}
\multiput(1.5,1)(1,0){2}{\sposb{}{u}}\put(5.5,1){\sposb{}{u}}
\multiput(1.5,2)(1,0){2}{\sposb{}{\lambda\!-\!u}}
\put(5.5,2){\sposb{}{\lambda\!-\!u}}
\put(0.71,1.5){\sposb{}{\lambda\!-\!u\ \ \ }}\put(6.29,1.5){\sposb{}{u}}
\multiput(0.5,0.5)(0,2){2}{\makebox(0.5,0){\dotfill}}
\multiput(6,0.5)(0,2){2}{\makebox(0.5,0){\dotfill}}
\psarc(1,.5){.125}{0}{90}
\psarc(1,1.5){.125}{0}{90}
\psarc(2,.5){.125}{0}{90}
\psarc(2,1.5){.125}{0}{90}
\psarc(5,.5){.125}{0}{90}
\psarc(5,1.5){.125}{0}{90}
\psarc[linewidth=1pt](1,1.5){.5}{245}{270}
\psarc[linewidth=1pt](1,1.5){.5}{90}{115}
\psarc[linewidth=1pt](6,1.5){.5}{65}{90}
\psarc[linewidth=1pt](6,1.5){.5}{270}{295}
\end{pspicture}}\qquad
\label{DTM}
\eea
As we will explain later, this schematic representation in the {\it planar} TL algebra needs to be interpreted
appropriately to write $\vec D(u)$ in terms of the generators of the {\it linear} TL 
algebra and to write down its associated matrix. 

Following the diagrammatic proof of \cite{BPO96}, which is valid in the planar TL algebra, the DTMs
$\vec D(u)$ form a commuting family with $[\vec D(u),\vec D(v)]=0$.
Similarly, using boundary crossing (\ref{bdyCrossing}) and following the diagrammatic proof of \cite{BPO96} yields the crossing symmetry $\vec D(u)=\vec D(\lambda-u)$. It also follows, at least for $(r,s)=(1,s)$, that $\vec D(u)$ is invariant under reflections about the vertical. Hence the eigenvectors of $\vec D(u)$ are either odd or even under the action of the chiral operator $\vec C$ that implements the left-right reflection on link states and the eigenvalues of $\vec D(u)$ are labelled by the quantum number $C=\pm 1$. 

In contrast to the situation for RSOS models, the DTMs $\vec D(u)$ here are not transpose symmetric and are not normal so there is no guarantee that they are diagonalizable. Nevertheless, we conjecture that for the one-boundary cases (one non-vacuum boundary) the DTMs $\vec D(u)$ are diagonalizable. This is supported by all of our numerics. The situation is very different, however, for the two-boundary cases (two non-vacuum boundaries) where in certain cases, as in Section~8, the transfer matrices are not diagonalizable and admit a Jordan cell structure.

\subsection{Hamiltonian Limits}

One way to take the Hamiltonian limit is to write  $\vec D(u)$ in terms of the linear TL algebra. Given a solution $K_j(u)$ of the right BYBE and assuming $\beta\ne 0$, we define a DTM acting on 
${\cal T}(N\!+\!2,\lambda)$ by
\begin{equation}
\vec D(u)=\beta^{-1}\,
e_{-1}\Big(\prod_{j=0}^{N-1} X_j(u)\Big) K_{N}(u) \Big(\prod_{j=N-1}^0 X_j(u)\Big)\beta^{-1}e_{-1}
\label{TLDTM}
\end{equation}
where the products are ordered and we have assumed the vacuum boundary condition on the
left. This is the appropriate interpretation of (\ref{DTM}) with the projectors
$\beta^{-1}e_{-1}$ enforcing closure on the left. As is clear in the diagram (\ref{expandedD}), the $(1,1)$ boundary triangle on the left is replaced by a connection generated by the two TL generators $e_{-1}$. 
\psset{unit=.9cm}
\setlength{\unitlength}{.9cm}
\bea
\vec D(u)\ \ \;=\ \ 
\begin{pspicture}[.5](-1,-1)(6,10.5)
%\psgrid
\psline[linewidth=.5pt](0,1)(1,0)
\psline[linewidth=.5pt](0,1)(4,5)
\psline[linewidth=.5pt](1,0)(5,4)
\psline[linewidth=.5pt](4,3)(3,4)
\psline[linewidth=.5pt](1,2)(2,1)
\psline[linewidth=.5pt](2,3)(3,2)
\psline[linewidth=.5pt](4,3)(3,4)
\psline[linewidth=.5pt](4,5)(5,4)
\psline[linewidth=.5pt](4,5)(0,9)
\psline[linewidth=.5pt](5,6)(1,10)
\psline[linewidth=.5pt](5,4)(5,6)
\psline[linewidth=.5pt](2,3)(3,2)
\psline[linewidth=.5pt](4,5)(5,6)
\psline[linewidth=.5pt](3,6)(4,7)
\psline[linewidth=.5pt](2,7)(3,8)
\psline[linewidth=.5pt](1,8)(2,9)
\psline[linewidth=.5pt](0,9)(1,10)
\rput(1,1){$u$}
\rput(2,2){$u$}
\rput(4,4){$u$}
\rput(4,6){$u$}
\rput(2,8){$u$}
\rput(1,9){$u$}
\rput(4.575,5){$u,\xi$}
\psarc(1,0){.15}{45}{135}
\psarc(2,1){.15}{45}{135}
\psarc(4,3){.15}{45}{135}
\psarc(4,5){.15}{45}{135}
\psarc(2,7){.15}{45}{135}
\psarc(1,8){.15}{45}{135}
\psline[linewidth=1pt](-.5,.5)(-.5,9.5)
\psline[linewidth=1pt](.5,1.5)(.5,8.5)
\psline[linewidth=1pt](1.5,2.5)(1.5,7.5)
\psline[linewidth=1pt](2.5,3.5)(2.5,6.5)
\psline[linewidth=1pt](3.5,4.5)(3.5,5.5)
\psline[linewidth=1pt](1.5,-.5)(1.5,.5)
\psline[linewidth=1pt](2.5,-.5)(2.5,1.5)
\psline[linewidth=1pt](3.5,-.5)(3.5,2.5)
\psline[linewidth=1pt](4.5,-.5)(4.5,3.5)
\psline[linewidth=1pt](1.5,9.5)(1.5,10.5)
\psline[linewidth=1pt](2.5,8.5)(2.5,10.5)
\psline[linewidth=1pt](3.5,7.5)(3.5,10.5)
\psline[linewidth=1pt](4.5,6.5)(4.5,10.5)
\psbezier[linewidth=1pt](-.5,-.5)(-.5,0)(.5,0)(.5,-.5)
\psbezier[linewidth=1pt](-.5,.5)(-.5,0)(.5,0)(.5,.5)
\psbezier[linewidth=1pt](-.5,9.5)(-.5,10)(.5,10)(.5,9.5)
\psbezier[linewidth=1pt](-.5,10.5)(-.5,10)(.5,10)(.5,10.5)
\rput(-1,-.75){$j=$}
\rput(0,-.75){$-1$}
\rput(1,-.75){$0$}
\rput(2,-.75){$1$}
\rput(3,-.75){$\ldots$}
\rput(4,-.75){$N\!-\!1$}
\rput(5,-.75){$N$}
\end{pspicture}
\label{expandedD}
\eea
The form in (\ref{TLDTM}) and (\ref{expandedD}) is the form used in our numerics.

For $\lambda\ne 0$, a suitably normalized Hamiltonian $\calH$ is defined by
\bea
\calH=-\half\sin\lambda\,\deriv{}u \log \vec D(u)\Big|_{u=0}
\eea
so that
\bea
\vec D(u)=\vec D(0)\,e^{-2u\calH/\sin\lambda+O(u^2)},\qquad\qquad [\vec D(0),\vec D(u)]=0
\eea
\newcommand{\nn}{\nonumber\\}

\psset{unit=.6cm}
\setlength{\unitlength}{.6cm}
\def\emptysquarea{\hspace{-.2\unitlength}
\begin{pspicture}(1,1)
\pspolygon[linewidth=.25pt](0,0)(1,0)(1,1)(0,1)(0,0)
\end{pspicture}}
\def\loopa{\hspace{-.2\unitlength}
\begin{pspicture}(1,1)
\pspolygon[linewidth=.25pt](0,0)(1,0)(1,1)(0,1)(0,0)
\psarc[linewidth=1pt](1,0){.5}{90}{180}
\psarc[linewidth=1pt](0,1){.5}{-90}{0}
\end{pspicture}}
\def\loopb{\hspace{-.2\unitlength}
\begin{pspicture}(1,1)
\pspolygon[linewidth=.25pt](0,0)(1,0)(1,1)(0,1)(0,0)
\psarc[linewidth=1pt](0,0){.5}{0}{90}
\psarc[linewidth=1pt](1,1){.5}{180}{270}
\end{pspicture}}
%%%%%%%
Here we derive the Hamiltonian by expanding $\vec D(u)$ in (\ref{DTM}) to first order in $u$. Initially omitting the projectors, this can be carried out diagrammatically in the planar algebra:
\bea
&\!\!\!\!&\! \beta\vec D(u)\;=\;
\begin{pspicture}[.4](-.5,0)(6,2)
\rput(.5,.5){\small $u$}
\rput(.5,1.5){\small $\lambda\!-\!u$}
\rput(3.5,.5){\small $u$}
\rput(3.5,1.5){\small $\lambda\!-\!u$}
\rput[bl](0,0){\emptysquarea}
\rput[bl](0,1){\emptysquarea}
\rput[bl](1,0){\emptysquarea}
\rput[bl](1,1){\emptysquarea}
\rput[bl](2,0){\emptysquarea}
\rput[bl](2,1){\emptysquarea}
\rput[bl](3,0){\emptysquarea}
\rput[bl](3,1){\emptysquarea}
\psline[linecolor=black,linewidth=1pt](5,0)(5,2)
\psline[linecolor=black,linewidth=1pt](6,0)(6,2)
\psarc(0,1){.5}{90}{270}
\psarc(4,1){.5}{-90}{90}
\end{pspicture}
\;-\;{s_{r-1}(0)s_0(2u)\over s_0(\xi+u)s_{r}(\xi-u)}\;
\begin{pspicture}[.4](-.5,0)(6,2)
\rput(.5,.5){\small $u$}
\rput(.5,1.5){\small $\lambda\!-\!u$}
\rput(3.5,.5){\small $u$}
\rput(3.5,1.5){\small $\lambda\!-\!u$}
\rput[bl](0,0){\emptysquarea}
\rput[bl](0,1){\emptysquarea}
\rput[bl](1,0){\emptysquarea}
\rput[bl](1,1){\emptysquarea}
\rput[bl](2,0){\emptysquarea}
\rput[bl](2,1){\emptysquarea}
\rput[bl](3,0){\emptysquarea}
\rput[bl](3,1){\emptysquarea}
\psline[linecolor=black,linewidth=1pt](5,0)(5,2)
\psline[linecolor=black,linewidth=1pt](6,0)(6,2)
\psarc(0,1){.5}{90}{270}
\psarc(4,0){.5}{0}{90}
\psarc(4,2){.5}{-90}{0}
\end{pspicture}\qquad\mbox{}\nonumber\\[16pt]
&\!\!=\!\!&\!
\beta s_1(-u)^{2N}\Bigg(
\begin{pspicture}[.4](-.5,0)(6,2)
\rput[bl](0,0){\loopa}
\rput[bl](0,1){\loopb}
\rput[bl](1,0){\loopa}
\rput[bl](1,1){\loopb}
\rput[bl](2,0){\loopa}
\rput[bl](2,1){\loopb}
\rput[bl](3,0){\loopa}
\rput[bl](3,1){\loopb}
\psline[linecolor=black,linewidth=1pt](5,0)(5,2)
\psline[linecolor=black,linewidth=1pt](6,0)(6,2)
\psarc(0,1){.5}{90}{270}
\psarc(4,1){.5}{-90}{90}
\end{pspicture}
\;-\;
{s_{r-1}(0)s_0(2u)\over s_0(\xi+u)s_{r}(\xi-u)}\;
\begin{pspicture}[.4](-.5,0)(6,2)
\rput[bl](0,0){\loopa}
\rput[bl](0,1){\loopb}
\rput[bl](1,0){\loopa}
\rput[bl](1,1){\loopb}
\rput[bl](2,0){\loopa}
\rput[bl](2,1){\loopb}
\rput[bl](3,0){\loopa}
\rput[bl](3,1){\loopb}
\psline[linecolor=black,linewidth=1pt](5,0)(5,2)
\psline[linecolor=black,linewidth=1pt](6,0)(6,2)
\psarc(0,1){.5}{90}{270}
\psarc(4,0){.5}{0}{90}
\psarc(4,2){.5}{-90}{0}
\end{pspicture}\ \Bigg)
\qquad\mbox{}\nonumber\\[16pt]
&\!\!+\!\!&\!
\beta s_0(u)s_1(-u)^{2N-1}\Bigg(
\begin{pspicture}[.4](-.5,0)(6,2)
\rput[bl](0,0){\loopa}
\rput[bl](0,1){\loopb}
\rput[bl](1,0){\loopb}
\rput[bl](1,1){\loopb}
\rput[bl](2,0){\loopa}
\rput[bl](2,1){\loopb}
\rput[bl](3,0){\loopa}
\rput[bl](3,1){\loopb}
\psline[linecolor=black,linewidth=1pt](5,0)(5,2)
\psline[linecolor=black,linewidth=1pt](6,0)(6,2)
\psarc(0,1){.5}{90}{270}
\psarc(4,1){.5}{-90}{90}
\end{pspicture}
\;+\;\cdots\;+\;
\begin{pspicture}[.4](-.5,0)(6,2)
\rput[bl](0,0){\loopa}
\rput[bl](0,1){\loopb}
\rput[bl](1,0){\loopa}
\rput[bl](1,1){\loopb}
\rput[bl](2,0){\loopa}
\rput[bl](2,1){\loopb}
\rput[bl](3,0){\loopb}
\rput[bl](3,1){\loopb}
\psline[linecolor=black,linewidth=1pt](5,0)(5,2)
\psline[linecolor=black,linewidth=1pt](6,0)(6,2)
\psarc(0,1){.5}{90}{270}
\psarc(4,1){.5}{-90}{90}
\end{pspicture}\ \Bigg)\qquad\mbox{}\\[16pt]
&\!\!+\!\!&\!
\beta s_0(u)s_1(-u)^{2N-1}\Bigg(
\begin{pspicture}[.4](-.5,0)(6,2)
\rput[bl](0,0){\loopa}
\rput[bl](0,1){\loopb}
\rput[bl](1,0){\loopa}
\rput[bl](1,1){\loopa}
\rput[bl](2,0){\loopa}
\rput[bl](2,1){\loopb}
\rput[bl](3,0){\loopa}
\rput[bl](3,1){\loopb}
\psline[linecolor=black,linewidth=1pt](5,0)(5,2)
\psline[linecolor=black,linewidth=1pt](6,0)(6,2)
\psarc(0,1){.5}{90}{270}
\psarc(4,1){.5}{-90}{90}
\end{pspicture}\;+\;\cdots\;+\;
\begin{pspicture}[.4](-.5,0)(6,2)
\rput[bl](0,0){\loopa}
\rput[bl](0,1){\loopb}
\rput[bl](1,0){\loopa}
\rput[bl](1,1){\loopb}
\rput[bl](2,0){\loopa}
\rput[bl](2,1){\loopb}
\rput[bl](3,0){\loopa}
\rput[bl](3,1){\loopa}
\psline[linecolor=black,linewidth=1pt](5,0)(5,2)
\psline[linecolor=black,linewidth=1pt](6,0)(6,2)
\psarc(0,1){.5}{90}{270}
\psarc(4,1){.5}{-90}{90}
\end{pspicture}\ \Bigg)
\;\mbox{}\nonumber\\[16pt]
&\!\!+\!\!&\!
s_0(u)s_1(-u)^{2N-1}\Bigg(
\begin{pspicture}[.4](-.5,0)(6,2)
\rput[bl](0,0){\loopa}
\rput[bl](0,1){\loopa}
\rput[bl](1,0){\loopa}
\rput[bl](1,1){\loopb}
\rput[bl](2,0){\loopa}
\rput[bl](2,1){\loopb}
\rput[bl](3,0){\loopa}
\rput[bl](3,1){\loopb}
\psline[linecolor=black,linewidth=1pt](5,0)(5,2)
\psline[linecolor=black,linewidth=1pt](6,0)(6,2)
\psarc(0,1){.5}{90}{270}
\psarc(4,1){.5}{-90}{90}
\end{pspicture}
\;+\;
\begin{pspicture}[.4](-.5,0)(6,2)
\rput[bl](0,0){\loopb}
\rput[bl](0,1){\loopb}
\rput[bl](1,0){\loopa}
\rput[bl](1,1){\loopb}
\rput[bl](2,0){\loopa}
\rput[bl](2,1){\loopb}
\rput[bl](3,0){\loopa}
\rput[bl](3,1){\loopb}
\psline[linecolor=black,linewidth=1pt](5,0)(5,2)
\psline[linecolor=black,linewidth=1pt](6,0)(6,2)
\psarc(0,1){.5}{90}{270}
\psarc(4,1){.5}{-90}{90}
\end{pspicture}\ \Bigg)
+\;\mbox{O}(u^2)\mbox{}\nonumber
\eea
Collecting connectivities together gives
\bea
\vec D(u)
&\!\!=\!\!&
\Big((1-u\cot\lambda)^{2N}+2\beta^{-1}\,{u\over\sin\lambda}\Big)\;
\begin{pspicture}[.4](0,0)(6,2)
\rput[bl](0,0){\emptysquarea}
\rput[bl](0,1){\emptysquarea}
\rput[bl](1,0){\emptysquarea}
\rput[bl](1,1){\emptysquarea}
\rput[bl](2,0){\emptysquarea}
\rput[bl](2,1){\emptysquarea}
\rput[bl](3,0){\emptysquarea}
\rput[bl](3,1){\emptysquarea}
\psline[linecolor=black,linewidth=1.3pt](.5,0)(.5,2)
\psline[linecolor=black,linewidth=1.3pt](1.5,0)(1.5,2)
\psline[linecolor=black,linewidth=1.3pt](2.5,0)(2.5,2)
\psline[linecolor=black,linewidth=1.3pt](3.5,0)(3.5,2)
\psline[linecolor=black,linewidth=1pt](5,0)(5,2)
\psline[linecolor=black,linewidth=1pt](6,0)(6,2)
\end{pspicture}\nonumber\\[16pt]
&\!\!-\!\!&
{2u\over\sin\lambda}\,{s_{r-1}(0)\over s_0(\xi)s_{r}(\xi)}\;
\begin{pspicture}[.4](0,0)(6,2)
\rput[bl](0,0){\emptysquarea}
\rput[bl](0,1){\emptysquarea}
\rput[bl](1,0){\emptysquarea}
\rput[bl](1,1){\emptysquarea}
\rput[bl](2,0){\emptysquarea}
\rput[bl](2,1){\emptysquarea}
\rput[bl](3,0){\emptysquarea}
\rput[bl](3,1){\emptysquarea}
\psline[linecolor=black,linewidth=1.3pt](.5,0)(.5,2)
\psline[linecolor=black,linewidth=1.3pt](1.5,0)(1.5,2)
\psline[linecolor=black,linewidth=1.3pt](2.5,0)(2.5,2)
\psline[linecolor=black,linewidth=1pt](5,0)(5,2)
\psline[linecolor=black,linewidth=1pt](6,0)(6,2)
\psarc(4,0){.5}{0}{180}
\psarc(4,2){.5}{-180}{0}
\end{pspicture}\\[16pt]
&\!\!+\!\!&
{2u\over\sin\lambda}\Bigg(\;
\begin{pspicture}[.4](0,0)(6,2)
\rput[bl](0,0){\emptysquarea}
\rput[bl](0,1){\emptysquarea}
\rput[bl](1,0){\emptysquarea}
\rput[bl](1,1){\emptysquarea}
\rput[bl](2,0){\emptysquarea}
\rput[bl](2,1){\emptysquarea}
\rput[bl](3,0){\emptysquarea}
\rput[bl](3,1){\emptysquarea}
\psline[linecolor=black,linewidth=1.3pt](2.5,0)(2.5,2)
\psline[linecolor=black,linewidth=1.3pt](3.5,0)(3.5,2)
\psline[linecolor=black,linewidth=1pt](5,0)(5,2)
\psline[linecolor=black,linewidth=1pt](6,0)(6,2)
\psarc(1,0){.5}{0}{180}
\psarc(1,2){.5}{-180}{0}
\end{pspicture}
\;+\;\cdots\;+\;
\begin{pspicture}[.4](0,0)(6,2)
\rput[bl](0,0){\emptysquarea}
\rput[bl](0,1){\emptysquarea}
\rput[bl](1,0){\emptysquarea}
\rput[bl](1,1){\emptysquarea}
\rput[bl](2,0){\emptysquarea}
\rput[bl](2,1){\emptysquarea}
\rput[bl](3,0){\emptysquarea}
\rput[bl](3,1){\emptysquarea}
\psline[linecolor=black,linewidth=1.3pt](.5,0)(.5,2)
\psline[linecolor=black,linewidth=1.3pt](1.5,0)(1.5,2)
\psline[linecolor=black,linewidth=1pt](5,0)(5,2)
\psline[linecolor=black,linewidth=1pt](6,0)(6,2)
\psarc(3,0){.5}{0}{180}
\psarc(3,2){.5}{-180}{0}
\end{pspicture}\;\Bigg)+\;\mbox{O}(u^2)\qquad\mbox{}\nonumber
\eea
In summary, we find
\bea
\vec D(u)\;=\;I-{2u\over\sin\lambda}\,\calH\;+\;\mbox{O}(u^2)\;=\;
I+{2u\over\sin\lambda}\Big((\beta^{-1}-N\cos\lambda)I-\calH^{(r,s)}\Big)\;+\;\mbox{O}(u^2)
\eea
where, reinstating the projectors,
\bea
\calH^{(r,s)}=-\sum_{j=1}^{N-1} e_j+{s_{r-1}(0)\over s_0(\xi)s_r(\xi)}\,P^r_{N+1}e_{N}P^r_{N+1}
\label{rsHams}
\eea
This operator is understood to be acting on the vector space ${\cal V}^{(s)}$. 
Each Hamiltonian in the infinite hierarchy $\calH^{(r,s)}$ with $r=1,2,\ldots,m$ and $s=1,2,\ldots$ 
is integrable and can be solved, for example, by Bethe ansatz. This statement is true for either sign of the Hamiltonian and for arbitrary complex values of $\xi$.

In the continuum scaling limit, the finite-size corrections to the Hamiltonian yield the dilatation Virasoro generator $L_0$ as in (\ref{HtoL0}). For the principal series with fixed real $\xi$, we find that the continuum limit depends on the range of $\xi$ but is otherwise independent of $\xi$. 
For $0<\xi<\pi-r\lambda$, the Hamiltonian $\calH^{(r,s)}$ converges to the representation of $L_0$ labelled by $(r,s)$ whereas, for $-r\lambda<\xi<0$, the Hamiltonian $\calH^{(r,s)}$ converges to the representation of $L_0$ labelled by $(r-1,s)$. 
In the sequel, and in particular in the numerics in Section~7, we assume that $0<\xi<\pi-r\lambda$. 
By scaling the imaginary part of $\xi$ appropriately with $\log N$, it is also possible to induce a boundary renormalization group flow between these two boundary conditions labelled by $(r,s)$ and $(r-1,s)$.

It is noted that, for $r=m+1$, the allowed range of $\xi$ for convergence to the representation of $L_0$ labelled by $(r,s)$ vanishes. This is the reason, although the required projector exists, our current construction fails beyond the first $m$ columns in the Kac table. It is also noted that, for $r=m+2$, the projector does not  exist thereby preventing the construction of the $(r,s)=(m+1,s)$ representation of $L_0$ by choosing $\xi<0$.

\section{Relation to Six-Vertex Model: Bethe Ansatz and Functional Equations}

\subsection{Six-Vertex Model}

In principle, it is possible to derive the Bethe ansatz and functional equations of the logarithmic minimal models directly using non-local connectivities. However, it is more expedient to consider the related faithful  representation~\cite{MartinBook,PaSa90,GoodmanWenzl,Kulish} of the linear Temperley-Lieb algebra given by the six-vertex model with vertex weights
\psset{unit=.1in}
\setlength{\unitlength}{.1in}
\bea
\begin{array}{rl}
W\left(\verta\right)\;=\;W\left(\vertb\right)\;=\;s_1(-u),&\quad W\left(\vertc\right)\;=\;W\left(\vertd\right)\;=\;s_0(u)\qquad\\[20pt]
 W\left(\verte\right)\;=\;e^{iu},&\quad W\left(\vertf\right)\;=\;e^{-iu}
 \end{array}
 \label{sixvertex}
\eea
In the usual six-vertex model, the last two vertex weights are both 1, the model is arrow reversal symmetric and the central charge is $c=1$. In contrast, the assignment of weights here preserves conservation of arrows at a vertex but breaks the arrow reversal symmetry and moves the central charge away from the fixed value $c=1$ to the value (\ref{ctsc}). 

The elementary face weights acting on $({\Bbb C}^2)^{\otimes N}$ are given by
\psset{unit=6mm}
\setlength{\unitlength}{6mm}
\begin{equation}
X_j(u)\;=\raisebox{-1.75\unitlength}{
\bpic(2,3.5)
\multiput(0,2)(1,-1){2}{\line(1,1){1}}
\multiput(0,2)(1,1){2}{\line(1,-1){1}}
\put(.8,1.2){\line(1,0){.4}}
\put(1,2){\pp{}{u}}
\multiput(0,0.5)(0,0.25){13}{\pp{}{.}}
\multiput(1,0.5)(0,0.25){2}{\pp{}{.}}
\multiput(1,3.25)(0,0.25){2}{\pp{}{.}}
\multiput(2,0.5)(0,0.25){13}{\pp{}{.}}
\put(.5,0){\pp{b}{j}}
\put(1.5,0){\pp{b}{j+1}}
\epic}
\;=\;s_1(-u)\:I\;+\;s_0(u)\:e_j
\end{equation}
where
\bea
e_j\;=\;I\otimes I\otimes\cdots \otimes I\otimes \smat{0&0&0&0\cr 0&x&1&0\cr 0&1&x^{-1}&0\cr 0&0&0&0}\otimes I\otimes
\cdots \otimes I\otimes I
\eea
with $x=e^{i\lambda}$ as above. The $4\times 4$ matrix acts at positions $j$ and $j+1$. The elementary boundary $K$ matrices acting from ${\Bbb C}^2$ to ${\Bbb C}^2$ are given by
\psset{unit=.06in}
\setlength{\unitlength}{.06in}
\bea
K\left(\!\vertbdya{}\,\right)\;=\;x^{1/2},\qquad K\left(\!\vertbdyb{}\,\right)\;=\;x^{-1/2}
\eea

For arbitrary $\lambda$, double-row transfer matrices $\vec T(u)$ for the six-vertex model with one non-trivial $(r,s)$ boundary condition can be built using the TL algebra and fusion projectors following the prescription given above. These matrices have similar properties to the logarithmic minimal models --- they form commuting families and are diagonalizable. The number $n$ of down arrows is related to the number of defects $\ell$ by $|N-2n|=\ell$. These two models differ, however, in one crucial aspect. The number of down arrows is a good quantum number for the six-vertex model but the number of defects is not conserved for the logarithmic minimal models since defects can be annihilated in pairs. Consequently, the six-vertex transfer matrices are block diagonal whereas the logarithmic minimal model transfer matrices are block triangular. It is precisely this block {\em triangular\/} structure that allows for the appearance of Jordan cells.

Since the six-vertex model (\ref{sixvertex}) gives a {\it faithful} representation of the linear TL algebra, it follows~\cite{MartinBook,GoodmanWenzl,Kulish} that all other representations, including the logarithmic minimal models, satisfy the same Bethe ansatz and functional equations. Moreover, the eigenvalues are necessarily a subset of the six-vertex eigenvalues possibly with different multiplicities. This has been confirmed by numerics on small systems.
For the purposes of calculating eigenvalue spectra, it therefore suffices to solve standard six-vertex Bethe ansatz equations~\cite{YangYang,ABBBQ}. At present, only the Bethe ansatz for the largest eigenvalues in the cases $(1,s)$ with $r=1$ have been worked out. The other cases are more complicated since they necessarily involve complex conjugate pairs of roots. Of course any mapping onto the six-vertex model will, of necessity, miss the indecomposable representations discussed in Section~8.

\subsection{Bulk and Boundary Free Energies}

As discussed in the previous subsection, for a given value of the crossing parameter $\lambda$, the largest eigenvalues of the DTMs $\vec D(u)$ in the vacuum sector with $(1,1)|(1,1)$ boundary conditions agree exactly at each finite size with the largest eigenvalues of the six-vertex model with open boundary conditions~\cite{ABBBQ,PaSa90}. It immediately follows that these models have the same bulk and boundary free energies. Through finite-size corrections, it also follows that these models have the same central charge. 

The bulk and boundary free energies can be obtained analytically by solving the relevant inversion relations~\cite{Baxt82,OP}. The boundary free energies are derived in \cite{NepP}. Here we just present the forms needed for the principal series with $\lambda=\pi/m$ with $m=3,4,\ldots$. These forms need to be modified for $\lambda>\pi/3$. 
The bulk free energy per face for $0<\lambda<\pi/2$ and $-\lambda/2<\mbox{Re}(u)<3\lambda/2$ is
\be
f_{bulk}(u,\lambda)=\int_{-\infty}^\infty 
{\cosh(\pi-2\lambda)t \sinh ut\sinh(\lambda-u)t\over t\,\sinh\pi t\,\cosh\lambda t}\,dt
\ee
For $-{\lambda\over 2}<\mbox{Re}(u)<{3\lambda\over 2}$ and $\lambda/2<\mbox{Re}(\xi)<3\lambda/2$, the $s$-independent boundary free energies are given by
\bea
f_{bdy}(u,\xi,\lambda)=f_{bdy}^{(r,s)}(u,\xi,\lambda)=f_0(u,\lambda)+f_r(u,\xi,\lambda)
\eea
Here 
\bea
f_0(u,\lambda)
=-2\!\int_{-\infty}^\infty 
{\sinh{(\pi-3\lambda)t\over 2}\sinh{\lambda t\over 2}\sinh ut\sinh(\lambda-u)t
\over t\,\sinh{\pi t\over 2}\cosh\lambda t}\,dt,\qquad 0<\lambda<{\pi\over 3}
\eea
\bea
f_r(u,\xi,\lambda)
=2\!\!\int_{-\infty}^\infty 
{\cosh{(\pi-2\xi-r\lambda)t}\cosh{r\lambda t}\sinh ut\sinh(\lambda-u)t
\over t\,\sinh{\pi t}\cosh\lambda t}\,dt,\qquad 2\xi+r\lambda<\pi
\eea
and
\bea
f_1(u,\xi,\lambda)&=&\log[s_0(\xi+u)s_1(\xi-u)]
\eea

In the Hamiltonian limit, the relevant expressions are given by minus the derivatives with respect to $u$ at $u=0$. The bulk free energy is
\bea
f_{bulk}(\lambda)=-\int_{-\infty}^\infty 
{\cosh(\pi-2\lambda)t \tanh\lambda t \over \sinh\pi t}\,dt
=\cot\lambda-\sin\lambda \int_{-\infty}^\infty {dt\over \cosh \pi t\,(\cosh 2\lambda t-\cos\lambda)}
\eea
For $-{\lambda\over 2}<\mbox{Re}(u)<{3\lambda\over 2}$ and $\lambda/2<\mbox{Re}(\xi)<3\lambda/2$, the boundary free energies are given by
\be
f_{bdy}(\lambda)=f_{bdy}^{(r,s)}(\lambda)=f_0(\lambda)+f_r(\xi,\lambda)
\ee
Here
\bea
f_0(\lambda)
=2\!\int_{-\infty}^\infty 
{\sinh{(\pi-3\lambda)t\over 2}\sinh{\lambda t\over 2}\tanh\lambda t
\over \sinh{\pi t\over 2}}\,dt,\qquad 0<\lambda<{\pi\over 3}
\eea
\bea
f_r(\xi,\lambda)
=-2\!\!\int_{-\infty}^\infty 
{\cosh{(\pi-2\xi-r\lambda)t}\cosh{r\lambda t}\tanh\lambda t
\over \sinh{\pi t}}\,dt,\qquad 2\xi+r\lambda<\pi
\eea
and
\bea
f_1(\xi,\lambda)=-{\sin\lambda\over \sin\xi\sin(\lambda+\xi)}
\eea
These explicit integrals are needed for numerics.

\section{Numerical Strip Partition Functions}

In this section, we report some numerical results for finite-size partition functions on the strip. These results are preliminary in the sense that we only consider the principal series and that the Bethe ansatz has not yet been implemented for $(r,s)$ boundary conditions with $r>1$. As we have already seen, the logarithmic minimal models are Yang-Baxter integrable so ultimately all of these results should be obtainable analytically. 

For $(r,s)=(1,s)$, finite-size sequences of numerical eigenvalues were obtained by solving the Bethe ansatz equations. These were generated for system sizes out to $N=40$ with $N$ of a definite parity. The numerical eigenvalues and numerical locations of the zeros of $Q$ and $T$ were checked against the values obtained by direct numerical diagonalization of the logarithmic minimal transfer matrices for system sizes out to $N=16$. For $(r,s)$ with $r>1$, finite-size sequences of numerical eigenvalues were obtained by direct numerical diagonalization of the logarithmic minimal transfer matrices and Hamiltonians for system sizes out to $N=16$. 
In these calculations, we fixed $u=\xi=\lambda/2$ for the transfer matrices and $\xi=(\pi -r \lambda)/2$ for the Hamiltonians. The precise choice for $\xi$ is not relevant since, in the appropriate interval, the limit is independent of $\xi$. In all cases, the numerical sequences were extrapolated using van den Broeck-Schwartz approximants~\cite{vBS} to extract the finite-size corrections. 

We present numerical results for both the isotropic lattice and Hamiltonian limit and show that these indeed agree. Typically, because there is no need to enforce closure on the left with a TL projector, the Hamiltonian calculation gives an extra digit of precision. In presenting numerical results, the numerical errors in the last digit (indicated in parenthesis) are a subjective indication of errors.

\subsection{Finite-Size Corrections}

The partition function for a $P\times N$ strip with one non-trivial boundary condition is
\be
 Z_{(1,1)|(r,s)}^{(P,N)}\;=\;{\rm Tr}\,\vec D(u)^P\;=\;\sum_n D(u)^P
  \;=\;\sum_ne^{-PE_n}
\ee
The general form of the finite-size corrections are by now standard~\cite{BCN,Aff}. 
For double-row transfer matrices, the finite-size corrections for the energies are
\be
 E_n\;=\;-\log D(u)\;=\;2Nf_{bulk}+f_{bdy}
  +\frac{2\pi\sin\vartheta}{N}\left(-\frac{c}{24}+\Delta+k\right)+\cdots, \quad k=0,1,2,\ldots
\ee
where $\Delta=\Delta_{r,s}$, $\vartheta={\pi u\over \lambda}$ is the anisotropy angle and $k$ labels the level in the conformal tower. 
Similarly, for the Hamiltonians $\calH^{(r,s)}$, the finite-size corrections for the energies are
\be
E_n=Nf_{bulk}+f_{bdy}
  +\frac{\pi v_s}{N}\left(-\frac{c}{24}+\Delta+k\right)+\cdots, \quad k=0,1,2,\ldots
\ee
where $\Delta=\Delta_{r,s}$ and $v_s={\pi\sin\lambda\over\lambda}$ is the ``velocity of sound". 
For the full matrices
\be
 {N\over \pi v_s}\Big(\calH^{(r,s)}-(Nf_{bulk}+f_{bdy})I\Big)\to L_0-{c\over 24}
\label{HtoL0}
\ee
Since the free energies are independent of $s$, the same is true for a Hamiltonian, say $\calH^{(1,s')|(1,s)}$, with two non-trivial boundaries. In this case, however, the matrices may exhibit a non-trivial Jordan canonical form as we demonstrate in Section~8.

\subsection{Critical Dense Polymers ($m=1$, $c=-2$)}

The first member ${\cal LM}(1,2)$ of the principal series is very interesting since it is a logarithmic CFT in the universality class of critical dense polymers~\cite{Sa87,Sa92}. This model is exceptional because $\lambda={\pi\over 2}$ implies the loop fugacity vanishes ($\beta=0$) and $e_j^2=0$ so that loops are forbidden.  Consequently, the two orthogonal projectors $\beta^{-1}e_j$ and $I-\beta^{-1}e_j$ no longer exist and the general fusion construction of integrable boundary conditions fails.  Consequently, we only consider $r=1$. Nevertheless, the model is still Yang-Baxter integrable and there exists an infinite family of integrable and conformal boundary conditions labelled by $s=1,2,3,\ldots$ corresponding to acting on different vector spaces of link states $\calV^{(s)}$ (\ref{sStates}). Remarkably, for this exceptional case, the limiting transfer matrices satisfy simple inversion identities, similar to those of the rational Ising model~\cite{BaxBook,OPW}, which enable the eigenvalue spectra to be calculated exactly on a finite lattice.  Specifically, for $(1,s)$ boundary conditions, we find in agreement with \cite{Sa92}
\be
c=-2;\qquad\Delta_{1,s}={(2-s)^2-1\over 8}, \quad\qquad s=1,2,3,\ldots
\ee
and obtain the complete set of associated finitized characters (\ref{finitizedchar}). We report these analytic results elsewhere~\cite{PRinPrep}. Since this case has been solved analytically, we omit any discussion of the numerics.

\subsection{Critical Percolation ($m=2$, $c=0$)}

The second member ${\cal LM}(2,3)$ of the principal series is also very interesting since it corresponds to critical percolation~\cite{Sa87,EberleFlohr06}. In this case, the (suitably normalized) transfer matrix $\vec D(u)$ is a stochastic matrix and its Hamiltonian limit ${\cal H}$ is an intensity matrix~\cite{PRGN}. As an aside, we point out that the entries of the Perron-Frobenius eigenvectors of these matrices are related~\cite{RasStroganov,PRGN} to the counting of fully packed loop configurations with connections to alternating sign matrices.

\noindent
{Isotropic Lattice:}
\bea
\begin{array}{rll}
(r,s)=(1,1):&Z_{(1,1)}(q)=q^{-c/24}(1+q^2+q^3+2q^4+2q^5+\cdots),& c=.0000000(1)\\[8pt]
(r,s)=(1,2):&Z_{(1,2)}(q)=q^{-c/24+\Delta}(1+q+q^2+2q^3+3q^4+\cdots),& \Delta=.0000000(1)\\[8pt]
(r,s)=(2,1):&Z_{(2,1)}(q)=q^{-c/24+\Delta}(1+q+q^2+2q^3+\cdots),& \Delta=.624(2)
\end{array}
\eea

\noindent
Hamiltonian Limit:
\bea
\begin{array}{rll}
(r,s)=(1,1):&Z_{(1,1)}(q)=q^{-c/24}(1+q^2+q^3+2q^4+\cdots),& c=.00000000(1)\\[8pt]
(r,s)=(1,2):&Z_{(1,2)}(q)=q^{-c/24+\Delta}(1+q+q^2+2q^3+3q^4+\cdots),& \Delta=.00000000(1)\\[8pt]
(r,s)=(1,3):&Z_{(1,3)}(q)=q^{-c/24+\Delta}(1+q+2q^2+2q^3+\cdots),& \Delta=.33333333(1)\\[8pt]
(r,s)=(1,4):&Z_{(1,4)}(q)=q^{-c/24+\Delta}(1+q+2q^2+\cdots),& \Delta=1.000000(3)\\[8pt]
(r,s)=(2,1):&Z_{(2,1)}(q)=q^{-c/24+\Delta}(1+q+q^2+2q^3+\cdots),& \Delta=.625(2)
\end{array}
\eea

\subsection{Logarithmic Ising Model ($m=3$, $c=\half$)}

\noindent
Isotropic Lattice:
\bea
\begin{array}{rll}
(r,s)=(1,1):&Z_{(1,1)}(q)=q^{-c/24}(1+q^2+q^3+2q^4+2q^5+\cdots),& c=.49999999(3)\\[8pt]
(r,s)=(1,2):&Z_{(1,2)}(q)=q^{-c/24+\Delta}(1+q+q^2+2q^3+3q^4+\cdots),& \Delta=.062499999(2)\\[8pt]
(r,s)=(1,3):&Z_{(1,3)}(q)=q^{-c/24+\Delta}(1+q+2q^2+2q^3+\cdots),& \Delta=.49999999(7)\\[8pt]
(r,s)=(1,4):&Z_{(1,4)}(q)=q^{-c/24+\Delta}(1+q+2q^2+\cdots),& \Delta=1.3125(1)\\[8pt]
(r,s)=(2,1):&Z_{(2,1)}(q)=q^{-c/24+\Delta}(1+q+q^2+2q^3+\cdots),& \Delta=.4999(2)
\end{array}
\eea

\noindent
Hamiltonian Limit:
\bea
\begin{array}{rll}
(r,s)=(1,1):&Z_{(1,1)}(q)=q^{-c/24}(1+q^2+q^3+2q^4+\cdots),& c=.499999999(2)\\[8pt]
(r,s)=(1,2):&Z_{(1,2)}(q)=q^{-c/24+\Delta}(1+q+q^2+2q^3+3q^4+\cdots),& \Delta=.062499999(2)\\[8pt]
(r,s)=(1,3):&Z_{(1,3)}(q)=q^{-c/24+\Delta}(1+q+2q^2+2q^3+\cdots),& \Delta=.5000000(1)\\[8pt]
(r,s)=(1,4):&Z_{(1,4)}(q)=q^{-c/24+\Delta}(1+\cdots),& \Delta=1.31249(2)\\[8pt]
(r,s)=(2,1):&Z_{(2,1)}(q)=q^{-c/24+\Delta}(1+q+q^2+2q^3+\cdots),& \Delta=.5001(2)
\end{array}
\eea

\subsection{Logarithmic Tricritical Ising Model ($m=4$, $c=7/10$)}

\noindent
Isotropic Lattice:
\bea
\begin{array}{rll}
(r,s)=(1,1):&Z_{(1,1)}(q)=q^{-c/24}(1+q^2+q^3+2q^4+2q^5+\cdots),& c=.69999(2)\\[8pt]
(r,s)=(1,2):&Z_{(1,2)}(q)=q^{-c/24+\Delta}(1+q+q^2+2q^3+3q^4+\cdots),& \Delta=.0999993(8)\\[8pt]
(r,s)=(1,3):&Z_{(1,3)}(q)=q^{-c/24+\Delta}(1+\cdots),& \Delta=.60007(8)\\[8pt]
(r,s)=(1,4):&Z_{(1,4)}(q)=q^{-c/24+\Delta}(1+\cdots),& \Delta=1.5002(3)\\[8pt]
(r,s)=(1,5):&Z_{(1,5)}(q)=q^{-c/24+\Delta}(1+\cdots),& \Delta=2.8001(2)\\[8pt]
(r,s)=(2,1):&Z_{(2,1)}(q)=q^{-c/24+\Delta}(1+q+q^2+2q^3+\cdots),& \Delta=.4374(1)
\end{array}
\eea

\noindent
Hamiltonian Limit:
\bea
\begin{array}{rll}
(r,s)=(1,1):&Z_{(1,1)}(q)=q^{-c/24}(1+q^2+q^3+2q^4+\cdots),& c=.70003(4)\\[8pt]
(r,s)=(1,2):&Z_{(1,2)}(q)=q^{-c/24+\Delta}(1+q+q^2+2q^3+3q^4+\cdots),& \Delta=.099998(3)\\[8pt]
(r,s)=(1,3):&Z_{(1,3)}(q)=q^{-c/24+\Delta}(1+\cdots),& \Delta=.59995(6)\\[8pt]
(r,s)=(1,4):&Z_{(1,4)}(q)=q^{-c/24+\Delta}(1+\cdots),& \Delta=1.50001(2)\\[8pt]
(r,s)=(1,5):&Z_{(1,5)}(q)=q^{-c/24+\Delta}(1+\cdots),& \Delta=2.80003(4)\\[8pt]
(r,s)=(2,1):&Z_{(2,1)}(q)=q^{-c/24+\Delta}(1+q+q^2+\cdots),& \Delta=.437(1)
\end{array}
\eea

\section{Examples of Indecomposable Representations}

\setlength{\unitlength}{.04in}
As already mentioned in Section~5.1, if the double-row transfer matrix has a $(1,s)$ boundary on one side and the vacuum on the other, then the transfer matrix appears to be diagonalizable. This is highly non-trivial because these transfer matrices are not normal matrices, but this observation is supported by numerical calculations for small sizes ($N\le 12$). Typically, the eigenvalues are distinct but this is not always the case, for example, in the Hamiltonian limit. We conjecture that in general these matrices are diagonalizable including in the Hamiltonian limit $u\to 0$ and assume this in the following discussion.

In general, the $s\ell(2)$ fusion rule
\be
(1,s_1)\otimes_f (1,s_2)=\mathop{\bigoplus}_{\mbox{\scriptsize $s_3=|s_1-s_2|+1$}\atop \mbox{\scriptsize $s_1+s_2-s_3=1\ \mbox{\scriptsize mod}\ 2$}}^{s_1+s_2-1} (1,s_3)
\ee
applies to the principal series whenever $\Delta_{(1,s_3)}-\Delta_{(1,s_3')} \notin {\Bbb Z}$ for any pair $s_3$, $s_3'$. If $\Delta_{(1,s_3)}-\Delta_{(1,s_3')} \in {\Bbb Z}$ for some pair $s_3$, $s_3'$, then there is the possibility to form an indecomposable representation. 
Looking numerically at many cases for different values of $m$, it seems that fusion yields an indecomposable representation in some but not all cases where $\Delta_{(1,s_3)}-\Delta_{(1,s_3')} \in {\Bbb Z}$. 
In this section, we present some typical examples to show that the fusion implied by taking non-vacuum boundary conditions on either side of the strip in these circumstances does lead to indecomposable representations of the Virasoro algebra. We hope to discuss the general fusion algebras in a future paper.

As a first example, consider the case of critical dense polymers ($m=1$) with $\lambda=\pi/2$ and $c=-2$ and consider the fusion
\bea
(1,2)\otimes_f(1,2)=(1,1)\oplus_i(1,3)\label{putative}
\eea
corresponding to having an $(r,s)=(1,2)$ boundary on both sides of the transfer matrix. These boundary conditions on the left and right each introduce a single defect for a total of two defects. Since the defects can be annihilated in pairs by the action of the TL algebra, the transfer matrix is upper block triangular with blocks labelled by the defect number $\ell=0,2$ corresponding to the $(1,1)$ and $(1,3)$ respectively. The finitized partition function reads
\bea
Z^{(N)}_{(1,2)|(1,2)}(q)=q^{-c/24}\,\mbox{Tr}\,q^{L_0^{(N)}}= \chi^{(N)}_{(1,1)}(q)+\chi^{(N)}_{(1,3)}(q)
\eea
but this is an indecomposable representation. To see what is going on, suppose $N=4$ and consider the Hamiltonian 
\bea
\calH=-\mbox{$\small\left(\begin{array}{ccccc}
0&1&0&0&0\\
2&0&1&0&1\\
0&0&0&1&0\\
0&0&1&0&1\\
0&0&0&1&0
\end{array}\right)$}+\sqrt{2}\,I
\eea
acting on the five states
\bea
\punit6
\down2monoid 00{1}{4}\punit8\downmonoid {-8}0{2}{3}\quad
\punit4
\downmonoid00{1}{2}\punit8\downmonoid00{3}{4}\quad
\punit8
\downmonoid00{1}{2}\punit8\svert0{0}{3}\punit4\svert0{0}{4}\quad
\punit4
\svert001\punit4\downmonoid00{2}{3}\punit8\svert004\quad
\punit4
\svert001\punit4\svert002\punit4\downmonoid00{3}{4}
\eea
A shift in the energy has been introduced to make the groundstate energy $E=0$. In accord with the imposed boundary conditions in the left side of (\ref{putative}), it is useful to interpret the left defect in these link states as being closed on the left $(\ell=1, s=2)$ and the right defect as being closed on the right $(\ell=1, s=2)$. Similarly, on the right side of (\ref{putative}), it is useful to interpret the two defects as closing on the right $(\ell=2, s=3)$. 
The $\ell=0$ and $\ell=2$ diagonal blocks are diagonalizable with eigenvalues $\{0,\sqrt{8}\}$ and $\{0,\sqrt{2},\sqrt{8}\}$ respectively. The Jordan canonical form for $\calH$ has rank-2 Jordan cells
\bea
\calH\sim \mbox{$\small\left(\begin{array}{ccccc}
0&0&1&0&0\\
0&\sqrt{8}&0&0&1\\
0&0&0&0&0\\
0&0&0&\sqrt{2}&0\\
0&0&0&0&\sqrt{8}
\end{array}\right)$}
\sim \mbox{$\small\left(\begin{array}{ccccc}
0&1&0&0&0\\
0&0&0&0&0\\
0&0&\sqrt{2}&0&0\\
0&0&0&\sqrt{8}&1\\
0&0&0&0&\sqrt{8}
\end{array}\right)$}\to 
\mbox{$\small\left(\begin{array}{ccccc}
0&1&0&0&0\\
0&0&0&0&0\\
0&0&1&0&0\\
0&0&0&2&1\\
0&0&0&0&2
\end{array}\right)$}
=L_0^{(4)}
\eea
This corresponds to the finitized partition function
\bea
Z^{(4)}_{(1,2)|(1,2)}(q)= \chi^{(4)}_{(1,1)}(q)+\chi^{(4)}_{(1,3)}(q)=q^{1/12}[(1+q^2)+(1+q+q^2)]
=q^{1/12}(2+q+2q^2)
\eea
Of course, the actual eigenvalues of $\calH$ only approach the integer energies indicated in the finitized Virasoro generator $L_0^{(N)}$ as $N\to\infty$. 
We see that every eigenvalue of the $(1,1)$ block has an exactly equal eigenvalue in the $(1,3)$ block and that together they form a rank-2 Jordan cell. This pattern continues for larger values of $N$ and has been checked numerically for $N\le 10$. The fact that this is possible is consistent with the identity~\cite{PRinPrep}
\bea
\chi^{(N)}_{(1,3)}(q)-\chi^{(N)}_{(1,1)}(q)=q^{1/12}\sum_{k=0}^{(N-4)/2} \Big\langle {{N-2\over 2}\atop k,k+1}\Big\rangle_{\! q}
\eea
where for $k\le n$ the generalized $q$-Narayana numbers
\bea
\Big\langle {M\atop k,n}\Big\rangle_{\! q}=q^{n-M+k(k+1)/2+n(n+1)/2} \Big(\gauss Mk_q \gauss {M+1}{n+1}_q-\gauss{M+1}k_q\gauss M{n+1}_q\Big)
\eea
are fermionic in the sense that they are polynomials with non-negative coefficients~\cite{PRinPrep}. 
We conjecture the exact form in the limit $N\to\infty$ is
\bea
L_0=\begin{pmatrix}
\mbox{Diag}(0,2,3,4,4,\ldots)&\vec J\\ \vec 0& \mbox{Diag}(0,1,2,2,3,3,4,4,4,4,\ldots)
\end{pmatrix}
\eea
This symbolic notation means that each energy $E$ in the expansion $\chi(q)=q^{-c/24}\sum_E q^E$ of the characters occurs on the matrix diagonal and a rank-2 Jordan cell is formed between as many coincident pairs as possible with an entry $1$ in $\vec J$. All other entries of $\vec J$ are $0$.

This indecomposable representation for critical dense polymers ($m=1$) is just the first in a sequence of indecomposable representations for the principal series
\bea
(1,2)\otimes_f(1,m+1)=(1,m)\oplus_i(1,m+2),\qquad m=1,2,3,\ldots
\eea
In this sequence, it seems that rank-2 Jordan cells are formed between as many coincident pairs as possible, that is, identical eigenvalues originating from distinct blocks. For small sizes, we have checked this explicitly for $m=1$ ($N=2,4,6,8$), $m=2$ ($N=3,5,7$), $m=3$ ($N=4,6,8$), $m=4$ ($N=5,7$) and $m=5,6,7,8,9$ (with $N=m+1$).

For critical dense polymers ($m=1$), we have also found the indecomposable representation
\be
(1,2)\otimes_f(1,4)=(1,3)\oplus_i(1,5)\label{extrafusion}
\ee
for $N=4,6,8$. In this case, a rank-2 Jordan cell is not always formed between coincident pairs but the Jordan form of the truncated Virasoro generator $L_0$ agrees with that of Gaberdiel and Kausch~\cite{GabK99} to the level calculated in their paper. Lastly, again for $m=1$ ($N=4,6$), we have confirmed the appearance of indecomposable representations resulting from fusion products involving indecomposable representations
\bea
(1,3)\otimes_f\big[(1,1)\oplus_i(1,3)\big]&=&\big[(1,1) \oplus_i(1,3)\big]\oplus\big[(1,3) \oplus_i(1,5)\big]\\
\big[(1,1) \oplus_i(1,3)\big]\otimes_f\big[(1,1) \oplus_i(1,3)\big]&=&2\big[(1,1) \oplus_i(1,3)\big]\oplus\big[(1,3) \oplus_i(1,5)\big]\qquad\mbox{}
\eea
We have not observed the appearance of higher rank Jordan cells in any of the cases studied.

\section{Discussion}

We have argued that the essential new physics in our logarithmic minimal 
theories derives from the non-local nature of the degrees of freedom  
in the form of connectivities. For these models, 
we have exhibited an infinite family of Yang-Baxter integrable lattice models on the strip which realize logarithmic CFTs in the continuum scaling limit. We have described the spectra of these theories on the strip for an infinite family of boundary conditions labelled by $(r,s)$ in an infinitely extended Kac table. Most importantly, we have shown how indecomposable representations arise in a consistent manner from within our lattice approach. 

The lattice approach to studying 
LCFTs opens up an alternative approach to this important class of problems while exposing the algebraic structures associated with integrability such as functional equations, Bethe ansatz, $T$-systems, $Y$-systems and Thermodynamic Bethe Ansatz.
We expect logarithmic lattice models to exist whenever there exists a braid-monoid algebra that can be extended to a planar algebra. We therefore expect that, from the lattice, it is possible to construct logarithmic dilute minimal models, logarithmic Wess-Zumino-Witten models as well as logarithmic models corresponding to higher fusions and higher rank.

Conventionally, to claim a consistent CFT, one must consider the system in other topologies, such as a cylinder or a torus~\cite{ReSa01,Torus}. This is particularly relevant to the question of comparing the logarithmic CFTs obtained in the scaling limit from our logarithmic minimal models with the logarithmic extensions of minimal models considered by other authors~\cite{FjelstadEtAl,EberleFlohr06,Torus}. At present, we can neither assert the equivalence nor inequivalence of these logarithmic CFTs. 
This task is particularly difficult because the precise equivalence of logarithmic CFTs may depend on the fine details of the fusion algebras and structure of the indecomposable representations. 
Obviously, there remains much work to be done.

\section*{Appendix A\ \ Decomposition of $Q_{r,s}$ into Irreducible Representations}
\renewcommand{\theequation}{A.\arabic{equation}}

In this appendix, we consider the quasi-rational quotient module
$Q_{r,s}:=V_{\Delta_{r,s}}/ V_{\Delta_{r,-s}}$ where $V_\Delta$ is the Verma module
of highest weight $\Delta$. 
In the celebrated work \cite{FF}, the embedding pattern of submodules 
of $V_{\Delta_{r,s}}$ is described based on which one can build the
irreducible quotient module $M_{\Delta_{r,s}}$ associated 
to $V_{\Delta_{r,s}}$. 
It is thus, in principle, a simple matter to determine how
the character of the quasi-rational module $Q_{r,s}$ decomposes into a finite number
of characters of irreducible modules. 
This decomposition is worked out explicitly in the following.

There are two possible embedding patterns. The typical one is conventionally
described by a diagram such as
\newcommand{\twoarrow}{\searrow\hspace{-1em}\nearrow}  
\def\irchi{\chi^{{\rm (irr)}}}
\begin{equation}
 \begin{array}{cccccccccccccc}
     & &  V_1  & \rightarrow & V_2 &\rightarrow&\cdots&\rightarrow&V_j&\rightarrow&\cdots\\   
     & \nearrow &  & & & & & & & &  \\
 V_0 & & & \twoarrow  &  & \twoarrow  & \cdots& \twoarrow& &\twoarrow& \cdots\\
      & \searrow & & & & & & & & & \\              
     & & V_1' & \rightarrow & V_2' &\rightarrow &\cdots&\rightarrow&V_j'&\rightarrow&\cdots
  \end{array}
\label{VV}
\end{equation}
where an arrow from module $A$ to
module $B$ indicates that $B$ is a submodule of $A$.
In this case, the irreducible modules associated to $V_j$ and $V_j'$
are $M_j=V_j/(V_{j+1}+V'_{j+1})$ and $M'_j=V'_j/(V_{j+1}+V'_{j+1})$, respectively, where we have used the unconventional $+$ instead of $\oplus$ since we reserve the notation $\oplus$ for {\it direct} sums.  
The decomposition of the character of the quotient module $V_m/V_{m+n}$, for example, 
into characters of irreducible modules reads
\bea
 \chi(V_m/V_{m+n})=\chi(M_m)+ \sum_{j=1}^{n-1}
   [ \chi (M_{m+j})+\chi(M'_{m+j})]+\chi(M'_{m+n})
\label{VonV}
\eea
Similar decompositions obviously apply to $V_m/V'_{m+n}$, $V'_m/V_{m+n}$
and $V'_m/V'_{m+n}$, and they all follow straightforwardly from (\ref{VV}).
The alternative embedding pattern is described by the diagram
\bea
 V_0\ \rightarrow\ V_1\ \rightarrow\ V_2\ \rightarrow\ \cdots\ \rightarrow\
  V_j\ \rightarrow\ \cdots
\label{V}
\eea
in which case the irreducible module associated to $V_j$ is
$M_j=V_j/V_{j+1}$, while the decomposition of the character of the quotient
module $V_m/V_{m+n}$ into characters of irreducible modules reads
\bea
 \chi(V_m/V_{m+n})= \sum_{j=0}^{n-1}\chi(M_{m+j})
\label{VonV1}
\eea
In either embedding pattern, we say that $V_j$ and $V'_j$ appear with rank $j$.

Now, for any pair $r,s$ of positive integers eventually labelling $Q_{r,s}$, let us write 
\begin{equation}
 r= r_0 + k p,\qquad s=s_0+ k' p',\ \qquad k,k'\ge 0 
\end{equation}
where $r_0,s_0$ are in the fundamental domain 
\begin{equation}
 1\le r_0 \le p,\qquad 1\le s_0 \le p'
\label{r0s0}
\end{equation}
If $r_0<p$ and $s_0<p'$, the embedding pattern associated to
$Q_{r,s}$ is of the type (\ref{VV}), while it is of type
(\ref{V}) if at least one of the upper bounds is saturated (\ref{r0s0}), 
that is, if $r_0=p$ or $s_0=p'$.
In general, the module $V_{\Delta_{r,s}}$ may be considered a submodule of 
$V_{\Delta_{r_0,s_0}}=V_{\Delta_{p-r_0,p'-s_0}}$ (if $|k-k'|$ is even)
or of $V_{\Delta_{r_0,p'-s_0}}=V_{\Delta_{p-r_0,s_0}}$ (if $|k-k'|$ is odd).
It is therefore a straightforward task to determine the decomposition of the character
of $Q_{r,s}$ into characters of irreducible modules: one merely has to identify
the locations of $V_{\Delta_{r,s}}$ and $V_{\Delta_{r,-s}}$
in the ambient embedding pattern. 
We will write the decompositions in terms of the characters
\bea
 \chi_{r,s}:=\chi(Q_{r,s})=\chi(V_{\Delta_{r,s}})-\chi(V_{\Delta_{r,-s}}),
  \ \ \ \ \ \ \ \ 
 \irchi_{\rho,\sigma}:=\chi(M_{\Delta_{\rho,\sigma}})
\label{chidefs}
\eea

We first consider the situation where
$1\leq r_0 < p,\ 1\leq s_0 < p'$, in which case the embedding pattern
headed by $V_{\Delta_{r,s}}$ looks like
\[\begin{array}{cccccccccc}
                    & \nearrow & {(r_1,s_1)}   & \rightarrow & {(r_2,s_2)} &
                                      \rightarrow  &      \cdots \\
            (r,s) & & & \twoarrow  &  & \twoarrow  & \\
                    & \searrow & {(r'_1,s'_1)} & \rightarrow & {(r'_2,s'_2)} &
                                   \rightarrow       &   \cdots
            \end{array} \]      
Suppose, for example, that $k\ge k'$, in which case
 $(r_{2n},s_{2n})=(r_0+(k-k'+2n)p,s_0)$, 
$(r'_{2n},s'_{2n})=(r_0,s_0+(k-k'+2n)p')$, 
$(r_{2n+1},s_{2n+1})=(r_0+(k-k'+2n+1)p,p'-s_0)$, 
$(r'_{2n+1},s'_{2n+1})=(r_0,p'-s_0+(k-k'+2n+1)p')$. In this chain,
the submodule $V_{\Delta_{r,-s}}$ appears with rank $2k'+1$ as it
corresponds to 
$(r,-s) \simeq (r_0 +(k+k'+1)p, p'-s_0)=(r_{2k'+1},s_{2k'+1})$.
The decomposition thus reads
\begin{equation} 
 \chi_{r,s}=\irchi_{r,s}+\sum_{i=1}^{2k'}\irchi_{r_i,s_i}
  +\sum_{i=1}^{2k'+1}\irchi_{r'_i,s'_i}
\label{hatchi}
\end{equation}
For general $r_0 < p,\ s_0 < p',\ 0\le k,k'$, we find 
\bea
 \chi_{r_0+kp,s_0+k'p'}&=&
  \irchi_{r_0+kp,s_0+k'p'}
  +\sum_{j=1}^{2\min(k,k')}\left(
  \irchi_{r_0+(|k-k'|+j)p,(-1)^js_0+(1-(-1)^j)p'/2}\right.\nn
  &&\left.+\irchi_{r_0,(-1)^js_0+(1-(-1)^j)p'/2+(|k-k'|+j)p'}\right)
 +\irchi_{r_0,p'-s_0+(k+k'+1)p'}
\eea

The linear embedding patterns (\ref{V}) corresponding
to exactly one saturated upper bound (\ref{r0s0}) may be analyzed
in a similar way. 
For general $1\leq r_0 < p,\ 1\leq s_0 < p',\ 0\le k,k'$, we thus find the 
decompositions
\bea
 \chi_{r_0+kp,(k'+1)p'}&=&\sum_{j=0}^{\min(2k,2k'+1)}
  \irchi_{(-1)^jr_0+(1-(-1)^j)p/2,(k+k'+1-j)p'}      \nn
 \chi_{(k+1)p,s_0+k'p'}&=&\sum_{j=0}^{\min(2k+1,2k')}
  \irchi_{(k+k'+1-j)p,(-1)^js_0+(1-(-1)^j)p'/2} 
\eea
Finally, if both upper bounds are saturated, 
that is $(r,s)=((k+1)p,(k'+1)p')$ where $k,k'\geq0$, 
the embedding pattern is linear and the decomposition reads
\begin{equation} 
 \chi_{(k+1)p,(k'+1)p'}=\sum_{j=0}^{\min(k,k')}\irchi_{(k+k'+1-2j)p,p'}
  =\sum_{j=0}^{\min(k,k')}\irchi_{p,(k+k'+1-2j)p'}
\end{equation}

It is observed that a quasi-rational quotient module $Q_{r,s}=V_{\Delta_{r,s}}/ V_{\Delta_{r,-s}}$ 
is irreducible if and only if it corresponds to a linear embedding pattern in which the 
submodule $V_{\Delta_{r,-s}}$ is the maximal proper submodule of $V_{\Delta_{r,s}}$. 
That is, the only irreducible quasi-rational quotient 
modules are $Q_{(k+1)p,s_0}$, $Q_{r_0,(k'+1)p'}$, $Q_{(k+1)p,p'}$ and
$Q_{p,(k'+1)p'}$ where $k,k'\geq0$.

\section*{Acknowledgements} PAP and JR are supported by the Australian Research Council. We
thank Michael Flohr for discussions at IPAM and for helpful correspondence and Hubert Saleur for useful discussions and help with the references at Saclay. We also thank Thomas Quella for comments.

%: Bibliography

\end{document}